\documentclass[12pt,a4paper,final]{iopart}

\usepackage{graphicx}
\pdfoutput=1
\usepackage{amsfonts,amsmath,amsthm,amssymb,amsopn}
\usepackage{dsfont,bbm}
\usepackage{mathrsfs}
\usepackage{leftidx}
\usepackage{cite}
\usepackage{booktabs}
\usepackage[retainorgcmds]{IEEEtrantools}
\usepackage{soul,xcolor}
\usepackage{color}
\definecolor{darkblue}{rgb}{0.,0.,0.5}
\definecolor{darkred}{rgb}{0.6,0.,0.}
\definecolor{darkgreen}{rgb}{0.,0.8,0.}
\definecolor{darkpink}{rgb}{1,0.08,0.5}

\usepackage[breaklinks=true,colorlinks=true,linkcolor=darkblue,urlcolor=magenta,citecolor=magenta]{hyperref}
\numberwithin{equation}{section}
\usepackage{cleveref}
\def\be{\begin{equation}}
\def\ee{\end{equation}}
\def\bra#1{\mathinner{\langle{#1}|}}
\def\ket#1{\mathinner{|{#1}\rangle}}

\def\ul#1{{\underline{#1}}}

\def\ol#1{\overline{#1}}
\def\rhoh{\ol{\rho}}

\def\vmbb#1{\mathbb{#1}}

\newcommand{\braket}[2]{\langle #1 \vert #2 \rangle}

\newcommand{\ua}{\uparrow}
\newcommand{\da}{\downarrow}
\newcommand{\ii}{ {\rm i} }
\newcommand{\dd}{ {\rm d} }

\newcommand{\ZZ}{\mathbb{Z}}
\newcommand{\NN}{\mathbb{N}}
\newcommand{\LL}{\mathbb{L}}

\newcommand{\RaR}{\mathbb{R}}
\newcommand{\CC}{\mathbb{C}}
\newcommand{\UU}{\mathbb{U}}
\newcommand{\y}{{\rm y}}
\newcommand{\x}{{\rm x}}
\newcommand{\z}{{\rm z}}

\newcommand{\half}{{\textstyle\frac{1}{2}}}

\newcommand{\mm}[1]{{\mathbf{#1}}}
\newcommand{\bb}[1]{{\mathbf{#1}}}

\def\Tr{{\,{\rm Tr}}}

\def\one{\mathbbm{1}}

\eqnobysec

\begin{document}
\setstcolor{red}

\title{Quasilocal charges in integrable lattice systems}
\author{Enej Ilievski$^1$, Marko Medenjak$^2$, Toma\v{z} Prosen$^2$, Lenart Zadnik$^2$}

\address{$^1$ Institute for Theoretical Physics,
University of Amsterdam, Science Park 904, Postbus 94485, 1090 GL Amsterdam, The Netherlands}
\address{$^2$ Department of physics, Faculty of Mathematics and Physics, University of Ljubljana,
Jadranska 19, SI-1000 Ljubljana, Slovenia}

\date{\today}

\begin{abstract}
We review recent progress in understanding the notion of locality in integrable quantum lattice systems. The central concept are the 
so-called quasilocal conserved quantities, which go beyond the standard perception of locality. Two systematic procedures to 
rigorously construct families of quasilocal conserved operators based on
quantum transfer matrices are outlined, specializing on anisotropic Heisenberg XXZ spin-1/2 chain. Quasilocal conserved operators stem 
from two distinct classes of representations of the auxiliary space algebra, comprised of unitary (compact) representations, which can 
be naturally linked to the fusion algebra and quasiparticle content of the model, and non-unitary (non-compact) representations giving 
rise to charges, manifestly orthogonal to the unitary ones.
Various condensed matter applications in which quasilocal conservation laws play an essential role are presented, with special 
emphasis on their implications for anomalous transport properties (finite Drude weight) and relaxation to non-thermal steady states in 
the quantum quench scenario.
\end{abstract}


\maketitle

\pagestyle{empty}
\tableofcontents
\pagestyle{headings}

\flushbottom
\clearpage
\section{Introduction}

Local conservation laws are amongst the most important fundamental concepts in theoretical physics. In generic
systems these usually comprise of energy, momentum, particle number, etc., and correspond to Noether charges connected to 
rather obvious physical symmetries. On the other hand, in systems which are exactly solvable, or integrable, the number of 
conservation laws and the corresponding conserved charges can be much larger and the underlying symmetries sometimes quite 
hidden. According to a widespread belief, integrability should provide us with a 1-to-1 correspondence between conserved charges and 
physical degrees of freedom. However, such a definition is only really applicable -- or unambiguous -- in classical deterministic 
(Hamiltonian) systems with a finite number of degrees of freedom where it amounts to the historical, Liouville--Arnold integrability.

Interacting quantum systems, where local degrees of freedom (quantum spins, fermions, or bosons) are arranged in a regular 1D 
lattice, are typically considered integrable in one of the following cases: Firstly, there may exist a canonical (Bogoliubov) 
transformation which maps the local degrees of freedom to non-interacting quasiparticles. Such is, for example, the situation with 
quantum transverse field Ising model, or $XY$ spin-$1/2$ chain \cite{LSM61}. These systems, which are reducible to a single particle 
picture and  are often referred to as quasi-free, shall not be of interest in this article, even though they allow for an illustration 
of some non-trivial many-body phenomena, such as area laws for entanglement~\cite{AreaLaws}.
Secondly, there exist systems exhibiting genuine interparticle interaction whose dynamics is representable in terms
of quasi-particles which undergo non-diffractive scattering without particle production. A central feature in such a case
is factorizability of an arbitrary multi-particle scattering process in terms of subsequent 2-particle scattering events, 
mathematically phrased in the form of the celebrated Yang--Baxter (or star-triangle) equation. One of the most remarkable physical 
consequences of that mechanism is the emergence of a macroscopic number of local integrals of motion (conservation laws).
One of these charges, usually the first one in the series, is considered as the Hamiltonian (with local interactions).
Here {\em locality} means that the densities of these charges act non-trivially only on a finite number of adjacent lattice sites.
Integrability in the sense of Yang and Baxter, which is universally understood within the framework of algebraic structures
known as quantum groups~\cite{Drinfeld89,FRT90,KasselBook,SierraBook}, is perhaps the most general widely acceptable definition of 
integrability known to date. Besides defining and describing integrability in closed quantum many-body systems in 1D 
\cite{KorepinBook}, it also covers 2D equilibrium classical statistical systems \cite{BaxterBook}, nonequilibrium classical driven 
diffusive 1D systems \cite{Schutz01}, as well as classical Hamiltonian systems~\cite{FaddeevBook,BBTBook}, and since more recently, 
also integrable nonequilibrium steady states of open quantum interacting systems \cite{Prosen_review}.

In recent years, a tremendous progress has been made in understanding a wide variety of nonequilibrium aspects of integrable systems,
a considerable part being covered by a series of review 
articles appearing in the present volume~\cite{Bernard_review,Calabrese_review,Caux_review,Cazalilla_review,DeLuca_review,Essler_review,
Vasseur_review,Vidmar_review,Schmiedmayer_review}.
However, interacting integrable quantum systems are for quite some time no longer only of pure mathematical or theoretical interest. 
In the last decade, a dramatic progress in experimental techniques for  manipulation of ultracold atoms enabled a few successful
experimental 
realizations~\cite{Bloch_review,Kinoshita04,Kinoshita06,Hofferberth_Nature07,Gring_Science12,Trotzky_Nature12,Cheneau_Nature12,Langen_Science15,Schmiedmayer_review}, some of which can directly probe
the nonequilibrium transport \cite{Hild,Ronzheimer,Schneider,Xia}.

The fact that certain integrable many-body systems can already be routinely controlled in a concrete experimental setup also underlies 
a remarkable degree of structural stability 
for some of their dynamical properties with respect to model imperfections (perturbations), in spite of the fact that strict
integrability technically requires precise (or fine-tuned) cancellations of most of generically allowed processes.
This may hint to an existence of a yet undisclosed quantum analogy of KAM (Kolmogorov--Arnold--Moser)
scenario~\cite{AABook}. To our opinion this is one of the potentially most exciting problems for future research~\cite{Prosen00,BCK15}.

As discussed above, Yang--Baxter integrability for a lattice system with $N$ sites, guarantees a macroscopic number $\propto N$ of 
local conservation laws and the corresponding local currents. By a local conservation law one understands an operator-valued 
continuity equation, involving a charge and a current density being operators supported on a finite number of, say $n \ll N$, physical 
sites. The summation of the local charge density over the whole volume of $N$ sites then defines an {\em extensive local conserved 
charge} of an integrable model. One might wonder whether  such local  conserved charges represent a complete set, meaning that any 
extensive conserved operator which scales linearly with $N$ can be represented as a linear combination of these local charges. Some formal completeness results for specific models have been put forward a while ago \cite{BabbittThomas79}, and one might have been tempted to conclude 
that local charges (derived from fundamental Yang--Baxter transfer matrix) are all the conserved operators needed to understand local 
physics. 
However, certain unconventional phenomena discovered later in studies of paradigmatic examples of interacting integrable systems
gave, in spite of a missing formal understanding, quite the opposite indications.
Firstly, it has been discovered~\cite{Zotos97,Zotos99} that the spin Drude weight in the integrable anisotropic Heisenberg chains ($XXZ$ model) 
is finite at finite temperature, despite the fact that contributions of all hitherto known local charges to spin current were zero. 
In more recent works it has been found~\cite{WoutersPRL,Pozsgay_GGE} that a Generalized Gibbs Ensemble (GGE) formed of 
the same standard set of local conserved charges fails to describe thermalization after a quantum quench in the gapped $XXZ$ model. 
These results hinted at the 
existence of additional {\em effectively} local conserved charges {\em linearly independent} from the strictly local ones. One should 
note that in studies of infinite quantum (and even classical) lattice systems, extensive observables form a vector space rather than 
the full algebra, so it is the linear independence and not functional independence that matters.

The first progress along the above lines came, unexpectedly, with the solution of an open $XXZ$ model~\cite{ProsenPRL106} driven
out of equilibrium with effective magnetic (particle) reservoirs at the boundary formulated in terms of Lindblad master equation. The 
steady state solution in the perturbative (weak-coupling) regime turned out to be tightly related to a novel effectively local (or quasilocal) conservation law which in turn explained the controversial problem of the ballistic conductivity by 
providing a rigorous non-trivial lower bound on the spin Drude weight and thus confirmed previous results of
several numerical studies~\cite{FHM03,HPZ11,Karrasch12,Karrasch13} and bosonization techniques~\cite{SPA09,SPA11}.
In a subsequent study~\cite{PI13}, a connection to certain non-standard solutions to Yang--Baxter equation has been uncovered, 
permitting a systematic construction of a large set of quasilocal conservation laws directly from commuting transfer 
matrices associated to complex-spin (non-unitarty) representations and yielding a further improved Mazur bound on the Drude 
weight. Generalizations of the results to periodic boundary conditions were simultaneously obtained
in Refs.~\cite{ProsenNPB14,Pereira14}.
A distinguished property of these so-called `non-unitary' quasilocal charges is that they do not exhibit the spin-reversal invariance 
of the $XXZ$ Hamiltonian and hence may have a nonvanishing overlap with observables which are odd with respect to spin reversal, such
as the spin current.
Very recently, even more exotic non-unitary quasilocal charges have been discovered where even the particle conservation
($U(1)$-symmetry) is broken~\cite{Zadnik16}. Similar constructions of quasilocal charges and consequent Drude weight bounds can be 
performed also in other gapless integrable quantum spin models, for example in spin-1 Fatteev--Zamolodchikov chain~\cite{PV16}.
We should remark, however, that it is the compactness of $q$-deformation rather than masslessness of the elementary excitations which 
plays the essential role in the construction of current carrying quasilocal charges which break the parity symmetry of the model (e.g., spin reversal).
This observation should make it possible to extend these concepts to massive integrable models like the sine-Gordon theory.

In spite of all rather profound implications mentioned above, the family of non-unitary quasilocal conserved operators could
not offer the answer to the puzzling findings of Refs.~\cite{WoutersPRL,Pozsgay_GGE,GA14} which cast doubts on the
applicability of the concept of a Generalized Gibbs Ensemble which was vividly debated about at the same time. 
In particular, it became clear that in a generic case the GGE has to be appropriately extended by incorporating
quasilocal conservation laws which are viable for the whole range of anisotropies,
invariant under spin-reversal transformation (i.e., of even parity), but still distinct from the canonical ones
obtained from expanding the fundamental transfer matrix.
Such quasilocal charges have been constructed (for the isotropic case) in Ref.~\cite{IMP15}, invoking transfer matrices built from 
unitary but non-fundamental spin representations of the auxiliary spin. Soon after, a study~\cite{Ilievski_GGE} confirmed that those 
charges exactly explain the GGE conundrum. 

\paragraph{Outline.}
The present review article aims at a coherent and pedagogical (i.e. non-
technical) introduction to the notion of quasilocal conserved charges and various physical
applications in which they take the center stage.
As the focus is primarily to elucidate the main ideas and their interrelations, a reader seeking for a more detailed
and rigorous exposition is referred to the cited literature. Sec.~\ref{sec2} consists of a minimal technical background for
getting familiar with the main concepts presented in this article. Sec.~\ref{sec3} is devoted to the construction of
what we call `unitary' quasilocal charges, namely conservation laws arising from the unitary representations of an
underlying symmetry group. In Sec.~\ref{sec:non-unitary} a more intricate case of `non-unitary' quasilocal charges which break the 
spin reversal (or, in general, some other $\ZZ_2$ parity) symmetry is presented.
Sec.~\ref{sec:applications} is dedicated to the exposition of several physical applications: Sec.~\ref{sec:mazur} discusses rigorous 
Mazur bounds on the spin Drude weight. Sec.~\ref{subsect:quenches} makes
a link to quantum quenches from spin-reversal symmetric initial states and highlights the duality between the spectra of quasilocal 
charges and Bethe root distributions which describe bound states in the formalism of the Thermodynamic Bethe Ansatz.
Sec.~\ref{sec:ness} illustrates the connection to integrable nonequilibrium steady states of boundary-driven quantum
master (Lindblad) equations. In this review, all the concepts are presented explicitly on a concrete example of
the $XXZ$ chain and the associated  $\mathcal{U}_q(\mathfrak{sl}(2))$ quantum symmetry. We conclude in Sec.~\ref{discussion} where 
certain possible generalizations to other integrable models and some questions which enter in the broader context are briefly 
discussed.

\section{Prerequisites}
\label{sec2}

In this section we introduce the framework and technical tools that shall be used in our paper.
In the Sec.~\ref{subsect:pseudolocal} we introduce the concepts of quantum spin systems on the lattice and the 
corresponding operator ($C^*$) algebra, and define the notions of locality, extensivity, pseudolocality and quasilocality. In  Sec.~\ref{subsect:YB} we define the main concepts of Yang--Baxter integrability: $R$-matrices, Lax 
matrices, transfer matrices, and fusion hierarchies which allow one to build unitary representations of these objects from the 
fundamental one. These concepts enable us to reformulate Bethe's original `coordinate ansatz'~\cite{Bethe31} in an entirely
algebraic language, a technique which is nowadays typically referred to as the \emph{quantum inverse scattering method} or
the \emph{algebraic Bethe ansatz}~\cite{KorepinBook,Faddeev_arxiv,Sklyanin_arxiv}.

The point of our review is to show that one can develop a new perspective on non-equilibrium quantum physics by combining the concepts from Yang--Baxter integrability with the notions of pseudo- and quasilocality of extended quantum lattice systems.

\subsection{Pseudolocal and quasilocal operators over quantum lattices}
\label{subsect:pseudolocal}

The main theme of this article are conserved charges of integrable lattice models which comply with a certain weaker version of 
locality. As such, they extend beyond the orthodox concept of local charges, derived from logarithmic derivatives of the fundamental 
transfer matrix~\cite{Sklyanin_arxiv,Faddeev_arxiv,KorepinBook,GaudinBook}, and exhibit physical relevance for computing 
time-averaged values of dynamical response functions.

Since we are only concerned with integrable systems, we can limit our discussion to a one-dimensional lattice $\Lambda=\ZZ$, although 
the concepts of this section can be readily extended to a $D$-dimensional lattice $\Lambda=\ZZ^D$. The total Hilbert 
space, formed by a tensor product of $d-$dimensional single-site Hilbert spaces, will be denoted by ${\cal H}$. The Hilbert space of a lattice 
subinterval between sites $x$ and $x'$, $x\le x'$, will be denoted by ${\cal H}_{[x,x']} \subset {\cal H}$ and the corresponding 
operator subalgebra by $\mathfrak{A}_{[x,x']}$. The entire quasilocal $C^*$ operator algebra $\mathfrak{A}$ is obtained as the limit 
of a sequence $\{\mathfrak{A}_{[-n,n]}; n=1,2,3\ldots\}$,  closed in the operator norm topology~\cite{BRBook}. We shall refer to an 
observable represented by an operator $a\in {\mathfrak A}$ as {\em local}, if it acts nontrivially only on a finite subinterval $[x,x']$,
\begin{align}
a=a_{[x,x']}\otimes\one_{\Lambda\setminus [x,x']}, \quad a_{[x,x']}\in\mathfrak{A}_{[x,x']}.
\end{align}
The {\em smallest} such interval is referred to as the {\em support} of $a$, and its length $r=x'-x+1$, as the {\em order} of 
locality. 
Denoting by ${\rm Tr}_{[x,x']}$ the trace over ${\cal H}_{[x,x']}$, one defines the {\em tracial state} $\omega_{0}$ as
\begin{equation}
\omega_{0}(a) = \frac{{\rm Tr}_{[x,x']} a_{[x,x']}}{{\rm Tr}_{[x,x']}\one_{[x,x']}},
\end{equation}
and extends it over an entire $\mathfrak{A}$ by continuity (of $\omega_0$). The tracial state can be interpreted as the infinite temperature Gibbs state, satisfying $\omega_{0}(ab)=\omega_{0}(ba)$ and having the strongest clustering property, namely being
separable: $\omega_{0}(ab)=\omega_{0}(a)\omega_{0}(b)$ for any pair of local observables $a,b$ with disjoint supports.

We define the Hilbert--Schmidt (HS) inner product as
\begin{align}
(a,b) &= \omega_{0}(a^\dagger b) - \omega_{0}(a^\dagger)\omega_{0}(b),
\label{eqn:HS}
\end{align}
and denote the corresponding HS norm\footnote{Note that, strictly speaking, $(a,b)$ and $\| a\|_{\rm HS}$ become a proper HS product 
and HS vector norm, respectively, only after one takes the identity operator $\one$ out of the algebra $\mathfrak{A}$. Otherwise 
they yield the HS product and HS norm of the corresponding `nearest' traceless observables. In other words, any operator 
of the form $c\one$, $c\in\CC$, has `zero length'.} by $\| a\|_{\rm HS}\equiv \sqrt{(a,a)}$. The latter satisfies the standard Cauchy-Schwartz 
inequality and a mixed inequality in relation to the operator norm $\| \bullet\|$,
\begin{equation}
|(a,b)| \le \| a\|_{\rm HS} \| b\|_{\rm HS},\qquad
\| ab \|_{\rm HS} \le \| a \|_{\rm HS} \| b \|.
\end{equation}
Equipped with these structures we can define an orthonormal basis of local observables. A choice of an on-site basis such that 
$(\sigma_x^\alpha,\sigma_x^{\alpha'})=\delta_{\alpha,\alpha'}$, induces the HS orthonormal basis of algebra $\mathfrak{A}_{[x,x']}$ 
consisting of elements of the form 
\begin{equation}
\sigma^{\ul{\alpha}}_{[x,x']} = \sigma^{\alpha_x}_x \otimes \sigma^{\alpha_{x+1}}_{x+1}\otimes \cdots \otimes \sigma^{\alpha_{x'}}_{x'}.
\label{clusterbasis}
\end{equation}
For example, in case of $2$-dimensional local Hilbert space, $\sigma^{\alpha\ge 1}_x$ are 
just the Pauli matrices, while for $3$-dimensional local space they are the Gell-Mann matrices, etc. In all cases we choose 
$\sigma^0_x = \one_x$.

We furthermore define a lattice shift automorphism by $\hat{\cal S}^y(a_{[x,x']}) = a_{[x+y,x'+y]}$ and associate to each element $a\in\mathfrak{A}$ a 
translationally invariant sum
\begin{equation}
A = \sum_{x} \hat{\cal S}^x(a),
\end{equation}
which represents an \emph{extensive} observable of a translationally invariant infinite quantum spin chain. Note that $A$ is not an 
element of quasilocal algebra $\mathfrak{A}$, but the above sum can still be attributed a precise mathematical meaning as a sequence of operators $\{A^{(N)}\}$ acting on finite lattices of increasing lengths $N$. For example, the Hamiltonian of locally interacting 
translationally invariant models, as well as other strictly local charges, are precisely of such form. In this sense, a local operator $a$ is called a {\em density} of an {\em extensive local} observable $A$.

The above sequences have the following properties: (i) volume scaling extensivity 
\begin{align}
&0 < \lim_{N\to\infty} \frac{1}{N} \left(A^{(N)},A^{(N)}\right) < \infty,
\label{eqn:volscaling}
\end{align}
and (ii) a finite overlap $\lim_{N\to\infty} (b,A^{(N)})\ne 0$ with at least one local operator $b$ (say $b=a$). In what follows, the upper index $N$ will be left out, since an extensive operator $A$ is always identified with the corresponding sequence.

By definition, any operator sequence $A$, satisfying extensivity (i) given by Eq.~\eqref{eqn:volscaling}, and the finite overlap criterion 
(ii), shall be referred to as {\em pseudolocal}. This relaxes the constraint on the strict locality of the densities and generalizes 
the concept in a physically meaningful way. As we shall argue later, pseudolocality of conserved charges is the decisive 
property responsible for ballistic (or non-ergodic~\cite{Prosen98,ProsenJPA98,ProsenPRE99}) scaling of dynamical response functions. 
Note that if the density $a$ can be written as a sum of mutually orthogonal terms $a_{[1,r]}$,
\begin{equation}
a=\sum_{r=1}^{N} a_{[1,r]},
\end{equation}
for which a stronger condition, known as {\em quasilocality}~\cite{PI13},
\begin{equation}
\| a_{[1,r]} \|_{\rm HS} < C e^{-\xi r}, \quad \xi >0,
\label{eqn:on1}
\end{equation}
holds, $A$ is automatically pseudolocal.

Here we have considered lattices with {\em open} boundaries. For systems with periodic or twisted boundary conditions, the same 
concepts can be introduced by making the shift operator $\hat{\cal S}$ periodic \cite{ProsenNPB14}. 

The definition of pseudolocality and quasilocality can be generalized (see Ref.~\cite{Doyon15}) to an arbitrary sufficiently strongly 
clustering state $\omega$ (say Gibbs, or generalized Gibbs state, etc.) simply by replacing the HS inner product by
\begin{equation}
(a,b) = \omega(a^{\dagger}b) - \omega(a^{\dagger})\omega(b),
\end{equation}
with the main conclusion that the set of all pseudolocal observables forms a Hilbert space.
 
\subsection{Yang--Baxter relation, quantum transfer matrices, and fusion hierarchies}
\label{subsect:YB}

A distinguished feature of integrable models is an existence of a \emph{macroscopic} number of conservation laws. They arise as a 
consequence of an exceptional amount of symmetry which is governed by algebraic structures known as
\emph{quantum groups}~\cite{Drinfeld89,FRT90,KasselBook,SierraBook}.
The central element in the story is the so-called quantum $\mathbf{R}$-matrix, an operator acting on a tensor product of a pair of vector spaces,
\begin{equation}
\mathbf{R}(\lambda):\mathcal{V}_{1}\otimes \mathcal{V}_{2} \rightarrow \mathcal{V}_{1}\otimes \mathcal{V}_{2},
\end{equation}
that can be considered as representations $\mathcal{V}_{1}$ and $\mathcal{V}_{2}$ of an underlying 
symmetry algebra, which we here for simplicity assume to be $\mathfrak{su}(2)$ or its quantum deformation. In addition, $\mathbf{R}
(\lambda)$ depends analytically on a spectral parameter $\lambda \in \CC$. The cornerstone equation of quantum integrability is 
obtained by embedding $\mathbf{R}$-matrices into a three-fold tensor product space $\mathcal{V}_{1}\otimes \mathcal{V}_{2}\otimes 
\mathcal{V}_{3}$, by making use of a suggestive notation $\mathbf{R}_{12}(\lambda)=\mathbf{R}(\lambda)\otimes \one$, and imposing the 
requirement
\begin{equation}
\mathbf{R}_{12}(\lambda-\mu)\mathbf{R}_{13}(\lambda)\mathbf{R}_{23}(\mu)=
\mathbf{R}_{23}(\mu)\mathbf{R}_{13}(\lambda)\mathbf{R}_{12}(\lambda-\mu),\quad \forall \lambda,\mu\in \CC,
\label{eqn:Yang--Baxter}
\end{equation}
where we have omitted the indices of vector spaces on which the operators act trivially.
This condition is the celebrated \emph{Yang--Baxter equation}~\cite{McGuire64,Yang67,BaxterBook} (YBE).
Physically speaking, YBE expresses equivalence of two distinct sequences of two-particle collisions which, as a consequence, give 
factorization property of the whole many-particle scattering process~\cite{McGuire64,ZZ79}. What is perhaps even more remarkable is, 
that such an equivalence automatically generates an infinite number of conserved quantities. The procedure is outlined below.

The simplest solution to YBE~\eqref{eqn:Yang--Baxter} is obtained when the $\mathbf{R}$-matrix acts in two fundamental spin representations
$\mathcal{V}_{1/2}\cong \CC^{2}$,
\begin{equation}
\mathbf{R}(\lambda):\quad \CC^{2}\otimes \CC^{2}\rightarrow \CC^{2}\otimes \CC^{2},\qquad \mathbf{R}(\lambda)=\lambda - \tfrac{\ii}{2} + \ii \mathbf{P},
\label{eqn:R_fundamental}
\end{equation}
where $\mathbf{P}$ is a permutation operator, $\mm{P}\ket{\psi_1}\otimes\ket{\psi_2} = \ket{\psi_2}\otimes\ket{\psi_1}$. Furthermore, 
we introduce the Lax operator $\mathbf{L}(\lambda)$ by interpreting one fundamental space of the $\mathbf{R}$-matrix as a local {\em 
physical} spin while the second fundamental space is referred to as an {\em auxiliary} space,
$\mathbf{L}_{12}(\lambda) \equiv \mathbf{R}_{12}(\lambda) =  \lambda + 2\ii \vec{\mathbf{s}}_{1}\cdot\vec{\mathbf{s}}_2$, or
\begin{equation}
\mathbf{L}(\lambda)=
\begin{pmatrix}
\lambda + \ii \mathbf{s}^{\z} & \ii \mathbf{s}^{-} \cr
\ii \mathbf{s}^{+} & \lambda - \ii \mathbf{s}^{\z}
\end{pmatrix}.
\label{eqn:Lax_isotropic}
\end{equation}
The spin generators fulfil the $\mathfrak{su}(2)$ algebraic relations, $[\mathbf{s}^{+},\mathbf{s}^{-}]=2\mathbf{s}^{\z}$ and $[\mathbf{s}^{\z},\mathbf{s}^{\pm}]=\pm \mathbf{s}^{\pm}$, and in terms of the Pauli matrices read $\mathbf{s}^{\z}=\tfrac{1}{2}\sigma^{\z}$ and $\mathbf{s}^{\pm}=\sigma^{\pm}=\tfrac{1}{2}(\sigma^{\x}\pm \ii \sigma^{\y})$. For clarity of notation, we shall 
here and below use bold-roman fonts to denote all operators which act nontrivially in auxiliary (non-physical) spaces. From 
YBE~\eqref{eqn:Yang--Baxter} it follows that the Lax operator Eq.~\eqref{eqn:Lax_isotropic} by construction obeys the
local \emph{fundamental commutation relation} (also known as the RLL relation~\cite{FRT90,Faddeev_arxiv}) over the auxiliary
vector space ${\cal H}_{\rm a}\otimes{\cal H}_{\rm a}$, ${\cal H}_{\rm a}\cong \CC^{2}$,
\begin{equation}
\mathbf{R}_{12}(\lambda-\mu)\mathbf{L}_{1}(\lambda)\mathbf{L}_{2}(\mu)=
\mathbf{L}_{2}(\mu)\mathbf{L}_{1}(\lambda)\mathbf{R}_{12}(\lambda-\mu),
\label{eqn:RLL_relation}
\end{equation}
which can be extended to the entire physical Hilbert space $\mathcal{H}_{\rm p}\cong (\CC^2)^{\otimes N}$ of the
$N$-spin lattice  
\begin{equation}
\mathbf{R}_{12}(\lambda-\mu)\mathbf{M}_{1}(\lambda)\mathbf{M}_{2}(\mu)=
\mathbf{M}_{2}(\mu)\mathbf{M}_{1}(\lambda)\mathbf{R}_{12}(\lambda-\mu),
\label{eqn:RTT_relation}
\end{equation}
by introducing the \emph{monodromy matrix} $\mathbf{M}(\lambda)$ acting over ${\cal H}_{\rm a}\otimes \mathcal{H}_{\rm p}$,
\begin{equation}
\mathbf{M}(\lambda)=\bb{L}(\lambda)^{\otimes N}.
\label{eqn:monodromy_matrix}
\end{equation}
Here and subsequently we use a compact notation of $\otimes^{N}$ to denote a `partial' tensor product, i.e. an operation where the tensor product only affects the physical components, whereas for the auxiliary components ordinary matrix multiplication applies. 
Finally, by tracing over the auxiliary space of Eq.~\eqref{eqn:monodromy_matrix} we produce the \emph{fundamental transfer matrix}
\begin{equation}
T(\lambda) = \Tr_{\rm a}\,\mathbf{M}(\lambda),
\label{eqn:transfer_matrix_def}
\end{equation}
acting over the spin chain Hilbert space $\mathcal{H}_{\rm p}$.

An infinite set of conservation laws is a consequence of commutativity property
\begin{equation}
[T(\lambda),T(\mu)]=0,\quad \forall \lambda,\mu \in \CC,
\label{eqn:commutation_fundamental}
\end{equation}
which follows directly from the definition~\eqref{eqn:monodromy_matrix} in combination with the YBE~\eqref{eqn:Yang--Baxter}. In fact, by considering higher-dimensional irreducible unitary representations of auxiliary spaces ($s> 1/2$), one sees that
the entire construction also holds  for \emph{higher-spin} transfer operators. These are constructed from Lax operators $\mathbf{L}_{s}(\lambda)$ 
associated with $(2s+1)$-dimensional auxiliary spaces $\mathcal{H}_{\rm a}={\cal V}_s\cong \CC^{2s+1}$ and satisfy
\begin{equation}
[T_{s}(\lambda),T_{s'}(\mu)]=0,
\qquad \forall s,s'\in \tfrac{1}{2}\ZZ_{+} \qquad {\rm and}\quad \lambda,\mu \in \CC.
\label{eqn:commutation_general}
\end{equation}

\subsubsection{Lax operator for the anisotropic Heisenberg model.}
\label{subsect:lax}

In this work we discuss the properties of quasilocal conservation laws in the anisotropic Heisenberg spin-$1/2$ chain ($XXZ$ model),
\begin{equation}
H=\sum_{x=0}^{N-1} 2\,\sigma_x^+  \sigma_{x+1}^- + 2\,\sigma_x^- \sigma_{x+1}^+ + \Delta\,\sigma_x^\z \sigma_{x+1}^\z,
\label{eqn:$XXZ$_Hamiltonian}
\end{equation}
where, unless otherwise stated, periodic boundary conditions are assumed.
Including the anisotropy requires employing a one-parametric deformation of the $\mathfrak{su}(2)$ symmetry algebra, which formally
gives rise to a quantum-deformed (quantized) enveloping algebra $\mathcal{U}_{q}(\mathfrak{sl}(2))$. The suitable deformation is
achieved through the deformation parameter $q=\exp(\eta)$, yielding the Lax operator of the following form (see e.g.~\cite{Faddeev_arxiv})
\begin{equation}
\mathbf{L}_{s}(\lambda)=\frac{1}{\sinh{(\eta)}}
\begin{pmatrix}
\sin{(\lambda+\ii\eta \mathbf{s}^{z})} & \ii \sinh{(\eta)} \mathbf{s}^{-} \cr
\ii \sinh{(\eta)}\mathbf{s}^{+} & \sin{(\lambda-\ii \eta \mathbf{s}^{z})}
\end{pmatrix}.
\label{eqn:Lax_operator_deformed}
\end{equation}
Three regimes are to be distinguished with respect to the anisotropy parameter $\Delta$:
\begin{itemize}
\item \emph{gapped regime}, corresponding to anisotropy $\Delta=\cosh{(\eta)} > 1$ with $\eta>0$,
\item \emph{gapless regime}, corresponding to $|\Delta| < 1$, which we shall write as $\Delta = \cos(\eta)$ 
with $q$-parameter lying on the unit circle $q=\exp{(\ii \eta)}$ for $\eta \in (0,\pi)$. In this regime, replacement $\eta \to -\ii \eta$ and $\lambda \to -\ii \lambda$ is needed in \eref{eqn:Lax_operator_deformed}
to restore the notation that is most often used (equivalent to exchanging $\sin$ and $\sinh$ in Eq.~(\ref{eqn:Lax_operator_deformed})), and that is used below.
\item \emph{isotropic point}, $\Delta=1$, is obtained from either of the regimes by taking the scaling limit, namely to write the spectral parameter as $\lambda \to \lambda \eta$ and then take $\eta\to 0$.
\end{itemize}

The Lax operator \eref{eqn:Lax_operator_deformed} is invariant under
the $q$-deformed quantum algebra $\mathcal{U}_{q}(\mathfrak{sl}(2))$.
By introducing $q$-deformation as $[x]_{q}=(q^{x}-q^{-x})/(q-q^{-1})$,
the $q$-deformed commutation relations read
\begin{equation}
[\mathbf{s}^{+},\mathbf{s}^{-}]=[2\mathbf{s}^{z}]_{q},\quad
q^{2\mathbf{s}^{z}}\mathbf{s}^{\pm}=q^{\pm 2}\mathbf{s}^{\pm}q^{2\mathbf{s}^{z}}.
\label{eqn:q-deformed_commutation}
\end{equation}
A family of irreducible unitary representations $\mathcal{V}_{s}$, $s\in\half\ZZ_+$, are spanned by basis vectors $\ket{n}$, $n=0,1,\ldots 2s$, writing
${\cal V}_{s} \simeq {\rm lsp}\{ \ket{n} \}$, on which $q$-deformed spin
generators act as
\begin{equation}
\begin{split}
\mathbf{s}^{\z} &= \sum_{n=0}^{2s} (s-n)\ket{n}\bra{n}, \\
\mathbf{s}^{+} &= \sum_{n=0}^{2s-1} \sqrt{[2s-n]_{q}[n+1]_{q}}\ket{n+1}\bra{n},\\
\mathbf{s}^{-} &= \sum_{n=0}^{2s-1} \sqrt{[2s-n]_{q}[n+1]_{q}}\ket{n}\bra{n+1}.
\end{split}
\label{eqn:compact_representation}
\end{equation}

In addition to finite-dimensional unitary representations of $\mathcal{U}_{q}(\mathfrak{sl}(2))$ algebra, YBE~\eqref{eqn:Yang--Baxter} in fact admits a much larger class of solutions which pertain to
generic complex-spin highest-weight representation $\mathcal{V}^{+}_{s}$, $s\in \CC$ (see e.g. Refs.\cite{FK95,Derkachov01,Karakh02}). These are of infinite dimension for a generic value of
$s$. 
For values of deformations corresponding to $\eta=\pi\,l/m$, with $l,m$, $l < m$, being co-prime positive integers -- or equivalently, 
for $q$ being a primitive root of unity -- we shall be interested in irreducible finite-dimensional sub-representations $\mathcal{V}^{(m)}_{s}$,
\begin{equation}
\begin{split}
\mathbf{s}^{\z}_{s} &= \sum_{n=0}^{m-1} (s-n)\ket{n}\bra{n}, \\
\mathbf{s}^{+}_{s} &= \sum_{n=0}^{m-2} [n+1]_q\ket{n}\bra{n+1}, \\
\mathbf{s}^{-}_{s} &= \sum_{n=0}^{m-2} [2s-n]_q\ket{n+1}\bra{n}.
\end{split}
\label{eqn:noncompact_irrep}
\end{equation}
Here the state $\ket{0}$ designates the highest-weight vector, alias the `vacuum', $\mm{s}^+_s \ket{0}=0$. Highest-weight transfer operators $T^{\rm hw}_{s}$ with 
$s \in \CC$ are defined according to the same prescription as in Eq.~\eqref{eqn:transfer_matrix_def}. Non-unitarity of
irreducible representations \eqref{eqn:noncompact_irrep} is reflected in the fact that
$\mathbf{s}^{+}_{s}\neq (\mathbf{s}^{-}_{s})^{\dagger}$.
Existence of an $\mathbf{R}$-matrix acting in a product of two different highest-weight spaces $\mathcal{V}_{s}\otimes 
\mathcal{V}_{s'}$ implies mutual commutations
\begin{equation}
[T^{\rm hw}_{s}(\lambda),T^{\rm hw}_{s'}(\mu)]=[T^{\rm hw}_{s}(\lambda),T_{s'}(\mu)]=0,
\end{equation}
for all distinct spin labels and pairs of spectral parameters $\lambda,\mu \in \CC$.

The standard set of local charges is generated by an expansion of $\log T_{\frac{1}{2}}(\lambda)$ around the so-called shift point,
\begin{equation}
H^{(k)}=-\ii \partial^{k-1}_{\lambda}\log T_{\frac{1}{2}}(\lambda+\tfrac{\ii \eta}{2})|_{\lambda=0},
\label{eqn:ultralocal}
\end{equation}
where $H^{(2)}\sim H$ is the Hamiltonian \eqref{eqn:$XXZ$_Hamiltonian}. The locality of conserved operators $H^{(k)}$ is manifested in 
the fact that each $H^{(k)}$ admits an expansion in terms of homogeneous sums of local densities $h^{(k)}$ of order $k$, i.e.
\begin{equation}
H^{(k)}= \sum_{x=0}^{N-1} \hat{\cal S}^x (h^{(k)}) \equiv \sum_{x=0}^{N-1}h^{(k)}_{x},
\label{eqn:ultralocality}
\end{equation}
for any finite length $N$.

Let us now switch the focus to the properties of higher-spin transfer matrices $T_s$ and their spectra, which play a vital role in
the construction of unitary quasilocal conserved charges. These properties will only be used later in the `fusion approach' (Sec. \ref{subsection:fusionapproach}) and
for obtaining closed-form results in the quantum quench problem (Sec.~\ref{subsect:closedform}).

\subsubsection{Quantum Hirota equation.}
\label{subsect:hirota}

The quantum Hirota equation~\cite{BLZII,KLWZ97,Zabrodin97,KSZ08,Gromov09,KL10}, also known as the $T$-system~\cite{KP92,KNS94},
is a bilinear difference equation which takes the form
\begin{equation}
T_{s}(\lambda + \tfrac{\ii \eta}{2})T_{s}(\lambda - \tfrac{\ii \eta}{2}) 
= \phi(\lambda + s\tfrac{\ii \eta}{2})\ol{\phi}(\lambda - s\tfrac{\ii \eta}{2})
+T_{s-\frac{1}{2}}(\lambda)T_{s+\frac{1}{2}}(\lambda),\quad s=\tfrac{1}{2}\mathbb{Z}_{+},
\label{eqn:Hirota}
\end{equation}
with bar denoting complex conjugation. This relation can be formally understood as the quantized version of Weyl's formula for 
characters of classical representations~\cite{BR90,KLWZ97}, while
physically it represents fusion rules on an underlying algebra in a covariant way.
Higher-spin transfer operators $T_{s}$ represent the \emph{canonical} solution to the Hirota equation. In this case, the scalar 
potentials  have to be identified as $\phi(\lambda)=T_{0}(\lambda + \tfrac{\ii \eta}{2})$ and
$\ol{\phi}(\lambda)=T_{0}(\lambda - \tfrac{\ii \eta}{2})$, where $T_{0}(\lambda)=(\sin{(\lambda)}/\sinh{(\eta)})^{N}$.

There exists some (gauge) freedom in choosing the operators $T_{s}$, which is the reason for defining their gauge-invariant
combinations known as the $Y$-operators. They are defined through the non-linear transformation
\begin{equation}
Y_{2s}=\frac{T_{s-\frac{1}{2}}T_{s+\frac{1}{2}}}{T_{0}^{[2s+1]}T_{0}^{[-2s-1]}}=
\frac{T^{+}_{s}T^{-}_{s}}{T_{0}^{[2s+1]}T_{0}^{[-2s-1]}}-\one,\qquad s=\tfrac{1}{2}\mathbb{Z}_{+},
\label{eqn:Y-functions}
\end{equation}
where the following compact notation is introduced:
$f^{[\pm k]}(\lambda)\equiv f(\lambda \pm k \tfrac{\ii \eta}{2}\mp \ii 0^{+})$ for $\eta\ne 0$, and
$f^{[\pm k]}(\lambda)\equiv f(\lambda\pm k \frac{\ii}{2}\mp \ii 0^{+})$ in the isotropic case (after applying a scaling limit 
$\lambda\to \lambda \eta$ and sending $\eta\to 0$). We shall write $f^{\pm}(\lambda)\equiv f^{[\pm 1]}(\lambda)$. The $Y$-operators 
obey the $Y$-system functional relations
\begin{equation}
Y^{+}_{j}Y^{-}_{j}=(\one+Y_{j-1})(\one+Y_{j+1}),\qquad j=1,2,\ldots
\label{eqn:Y-system}
\end{equation}
where the boundary condition $Y_{0}=0$ is assumed.

In this article, Hirota equation appears in two different (but related) contexts:
\begin{enumerate}
\item as the fusion relation among higher-spin transfer operators $T_{s}$ which is automatically inherited by their eigenvalues, and
\item as an analytic closed-form description of certain solutions of equilibrium states which typically arise in the
scope of quantum quench applications (cf. Sec.~\ref{subsect:quenches}).
\end{enumerate}

The Hirota equation~\eqref{eqn:Hirota}, can be understood as a discrete
integrable classical system of its own. A central relation in this regard is the Baxter's $TQ$-equation~\cite{Baxter73,shortcut,BFLM11,DM06}
\begin{equation}
T_{\frac{1}{2}}Q=T^{+}_{0}Q^{[-2]}+T^{-}_{0}Q^{[+2]},
\label{eqn:Baxter_TQ}
\end{equation}
which represents a discrete second-order difference equation for the fundamental transfer matrix $T_{1/2}$.
The operator $Q$ stands for Baxter's $Q$-operator. We do not derive it here explicitly (see e.g. Ref.~\cite{shortcut}),
but make use of its spectral representation which will provide the connection to Bethe eigenstates (cf. Eq.~\eqref{eqn:Q_spectrum}).

The $Q$-operator allows us to linearize Eq.~\eqref{eqn:Hirota}, i.e. enable us to express $T_{s}(\lambda)$ explicitly as a combination of $Q$-operators
\begin{equation}
\frac{T^{+}_{s}}{T^{[2s+1]}_{0}}=Q^{[2s+2]}Q^{[-2s]}
\sum_{k=0}^{2s}\frac{\zeta^{N}_{2s,k}}{Q^{[2(k-s)]}Q^{[2(k-s+1)]}},
\label{eqn:Hirota_solution}
\end{equation}
where the scalars are provided by
\begin{equation}
\zeta_{2s,k}(\lambda)=\frac{T^{[2(k-s)+1]}_{0}(\lambda)}{T^{[2s+1]}_{0}(\lambda)}.
\label{eqn:Hirota_scalar}
\end{equation}

Since the $TQ$-equation~\eqref{eqn:Baxter_TQ} is of a second order, it admits two (linearly) independent solutions, $Q$ and 
$\widetilde{Q}$, whose independence requires the Wronksian determinant to be non-degenerate,
\begin{equation}
T_{0}= Q^{+}\widetilde{Q}^{-} - Q^{-}\widetilde{Q}^{+}.
\label{eqn:quantum_Wronksian}
\end{equation}

By virtue of commutativity of $T_s(\lambda)$ and $Q(\mu)$, for all $s\in\half\ZZ_+,\lambda,\mu\in\CC$, all previously stated identities can 
be taken at the level of their eigenvalues. To distinguish commuting operators from their eigenvalues, we write the latter
with the calligraphic font. Bethe roots $\lambda_{j}$ are by definition zeros of eigenvalues of $Q$,
i.e. solutions of $\mathcal{Q}(\lambda)=0$. Bethe ansatz equations can be obtained algebraically by eliminating $\widetilde{Q}$ 
through the combination of Eq.~\eqref{eqn:Baxter_TQ} and the Wronskian condition~\eqref{eqn:quantum_Wronksian}, yielding an 
equation for the eigenvalues
\begin{equation}
\frac{T^{-}_{0}(\lambda_j)\mathcal{Q}^{[+2]}(\lambda_j)}{T^{+}_{0}(\lambda_j)\mathcal{Q}^{[-2]}(\lambda_j)}=-1.
\label{eqn:algebraic_Bethe_ equations}
\end{equation}

Similarly, Eq.~\eqref{eqn:Hirota_solution} turns out to be useful in studying the large-$N$ limit spectra of the
transfer operators $T_{s}$. We shall exploit this trick later on in Sec.~\ref{subsect:quenches}.

\section{Quasilocal charges from unitary representations}
\label{sec3}
In this section we construct quasilocal charges from half-integer representations of the auxiliary algebra \eref{eqn:compact_representation}, extending the standard family of local charges. In the first part we formulate the pseudolocality condition in terms of auxiliary transfer matrices and subsequently demonstrate its equivalence to the \emph{inversion identity}. Furthermore, the construction allows us to obtain a representation of conserved charges, which is useful for computation of 
their norms and subsequently performing orthogonalization procedure. Subsequently we present an alternative approach to 
obtain the inversion identity by resorting to previously discussed Hirota equation. The latter enables us to identify quasilocal 
charges which pertain to the gapless regime.
\subsection{Auxilliary transfer matrix approach}
\label{subsect:X}

Initially, we consider the $|\Delta| \ge 1$ regime of the $XXZ$ model and show that an infinite tower of conserved operators
\begin{equation}
X_{s}(\lambda) = -\ii \partial_{\lambda}\log\,\frac{T^{+}_{s}(\lambda)}{T^{[2s+1]}_{0}(\lambda)},\quad \lambda\in \RaR,
\quad s=\tfrac{1}{2},1,\tfrac{3}{2},\ldots
\label{eqn:X_definition}
\end{equation}
generated from the higher-spin transfer operators $T_{s}$ are indeed quasilocal conserved charges. The sketch of the proof given
below is based on establishing the inversion formula derived in Ref.~\cite{IMP15},
\begin{equation}
\frac{T^{+}_{s}(\lambda)T^{-}_{s}(\lambda)}{T^{[-2s-1]}_{0}(\lambda)T^{[2s+1]}_{0}(\lambda)} \stackrel{N\to \infty}{\longrightarrow} \one,
\label{eqn:inversion_formula}
\end{equation}
which allows for an alternative representation (or definition) of the charges Eq.~\eqref{eqn:X_definition} in a more
convenient product form
\begin{equation}
X_{s}(\lambda)=-\ii\partial_{\mu}\frac{T^{-}_{s}(\lambda)}{T^{[-2s-1]}_{0}(\lambda)}
\frac{T^{+}_{s}(\mu)}{T^{[2s+1]}_{0}(\mu)}\Big|_{\mu=\lambda},\qquad \lambda \in \RaR.
\label{eqn:X_product_from}
\end{equation}
Subsequently we will adopt Eq. \eqref{eqn:X_product_from} as a working definition when proving quasilocality property of operators $X_{s}(\lambda)$. Initially, we shall not rely on the apparatus of integrability but rather employ a direct technique using
auxiliary transfer matrices.

By doubling the auxiliary space the operator product on the left hand-side of Eq.~\eqref{eqn:inversion_formula} can be represented as 
\begin{equation}
\frac{T^{\mp}_{s}(\lambda)}{T^{[\mp 2s \mp 1]}_0(\lambda)}  \frac{ T^\pm_s(\mu)}{T^{[\pm 2s \pm 1]}_0(\mu)} =
\Tr_{\rm a}\left\{\vmbb{L}^\pm_s(\lambda,\mu)^{\otimes N}\right\},
\label{eqn:X_as_MPO}
\end{equation}
where the trace takes place in $\mathcal{V}_{s}\otimes \mathcal{V}_{s}$ and
$\mathbb{L}^{\pm}_{s}(\lambda,\mu)$ are composite Lax operators acting over
$\mathcal{V}_{s}\otimes \mathcal{V}_{s}\otimes \mathbb{C}^{2}$ given by
\begin{equation}
\mathbb{L}^{\pm}_{s}(\lambda,\mu) =
\mathcal{N}^{\pm}_{s}(\lambda,\mu)(\mathbf{L}^{\mp}_{s}(\lambda)\otimes \one_{s})(\one_{s} \otimes \mathbf{L}^{\pm}_{s}(\mu))
=\sum_{\alpha \in \mathcal{J}}\mathbb{L}^{\pm \alpha}_{s}(\lambda,\mu)\sigma^{\alpha},
\label{eqn:double_Lax}
\end{equation}
with the index set $\mathcal{J}=\{\x,\y,\z,0\}$. For later convenience, we have introduced the normalization factor
\begin{equation}
\mathcal{N}^{\pm}_{s}(\lambda,\mu) = \left(L^{[\mp(2s+1)]}_{0}(\lambda) L^{[\pm(2s+1)]}_{0}(\mu)\right)^{-1},
\label{eqn:normalization}
\end{equation}
where $L_{0}(\lambda)=\sin{(\lambda)}/\sinh{(\eta)}$ is the scalar Lax operator.

\begin{figure}[htb]
\centering
\includegraphics[width=0.8\textwidth]{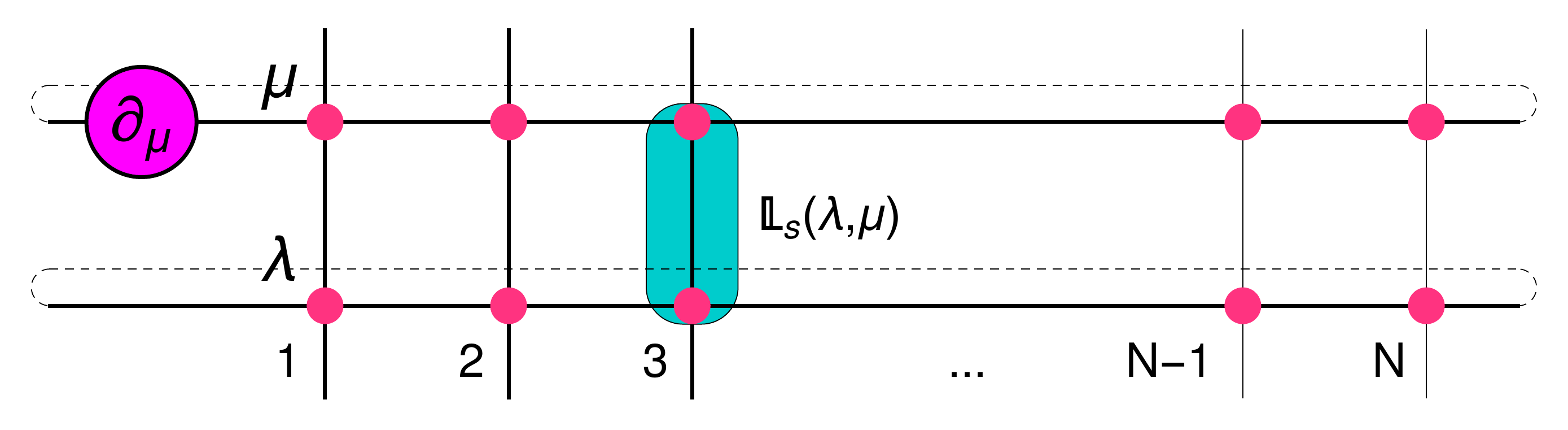}
\caption{Schematic depiction of a quasilocal charge $X_{s}(\lambda)$ for a spin chain composed of $N$ sites: Each vertex represents a 
copy of an irreducible spin-$s$ Lax operator $\mathbf{L}_{s}$. Each row represents one copy of an auxiliary space $\mathcal{V}_{s}$, 
carrying their own rapidity variables ($\lambda$ and $\mu$). Horizontal stacking pertains to tensor multiplication with respect to 
physical spaces $\mathcal{V}_{\frac{1}{2}}\cong \CC^{2}$, while vertical stacking should be understood as tensor multiplication with 
respect to auxiliary spin spaces (for physical components 
ordinary multiplication applies). The dashed lines denote partial tracing with respect to auxiliary spaces $\mathcal{V}_{s}$.
The upper row is acted upon by the derivative operation $\partial_{\mu}$ (magenta), where Leibniz chain rule should be assumed.
In addition, a reducible two-component Lax matrix $\mathbb{L}_{s}(\lambda,\mu)$
sits on every vertical rung (shown in blue only for the 3rd site). Notice that to generate a quasilocal charge $X_{s}(\lambda)$ one 
has to finally set $\mu=\lambda$.}
\label{fig:X-fig}
\end{figure}

The central object to establish pseudolocality of the family $X_s(\lambda)$ to be considered is the normalized
Hilbert--Schmidt kernel (HSK)
\begin{equation}
K_{s,s'}(\lambda,\mu) = \lim_{N\to\infty} \frac{1}{N} \left(X_s(\lambda),X_{s'}(\mu)\right).
\label{eqn:HSK}
\end{equation}
Evaluation of expression \eqref{eqn:HSK} requires the introduction of an auxiliary transfer operator over
$\mathcal{V}_{s}\otimes \mathcal{V}_{s}\otimes \mathcal{V}_{s^{\prime}}\otimes \mathcal{V}_{s^{\prime}}$, reading
\begin{equation}
\mathbb{T}_{s,s'}(\lambda,\lambda^{\prime},\mu,\mu^{\prime}) =
\frac{1}{2}\Tr_{\CC^{2}}\left(
(\mathbb{L}^{\mp}_{s}(\lambda,\lambda^{\prime})\otimes \one^{\otimes 2}_{s^{\prime}})
(\one^{\otimes 2}_{s}\otimes \mathbb{L}^{\pm}_{s^{\prime}}(\mu,\mu^{\prime}))\right).
\end{equation}
Equipped with this result, the quasilocality condition for $X_{s}(\lambda)$ is equivalent to demanding that
\begin{align}
K_{s,s'}(\lambda,\mu) &= \lim_{N\to \infty}\frac{1}{N}\Big\{\partial_{\lambda^{\prime}}\partial_{\mu^{\prime}}
\Tr\,\mathbb{T}_{s,s^{\prime}}(\lambda,\lambda^{\prime},\mu,\mu^{\prime})^{N}\vert_{\lambda^{\prime}=\lambda,\mu^{\prime}=\mu}
\nonumber \\
&-\big[\partial_{\lambda'}\Tr\,\mathbb{L}^{+0}_{s}(\lambda,\lambda^{\prime})^{N}\big]_{\lambda'=\lambda}
\big[\partial_{\mu'}\Tr\,\mathbb{L}^{-0}_{s'}(\mu,\mu^{\prime})^{N} \big]_{\mu'=\mu}\Big\},
\label{eqn:K-kernel_def}
\end{align}
is finite and non-zero. The goal is to obtain $K_{s,s'}(\lambda,\mu)$ by calculating the dominating (i.e. the largest in  
modulus) eigenvalues of auxiliary transfer matrices $\mathbb{T}_{s,s'}$ and $\mathbb{L}^{\pm 0}_{s}$.

Let $\tau^{j}_{s}(\lambda,\mu)$ denote the eigenvalues of $\mathbb{L}^{+ 0}_{s}(\lambda,\mu) = \mathbb{L}^{- 0}_{s}(\lambda,\mu) $,
while for coinciding parameters we put $\tau^{j}_{s}(\lambda)\equiv \tau^{j}_{s}(\lambda,\lambda)$ (and similarly
$\mathbb{L}^{\pm}_{s}(\lambda)\equiv \mathbb{L}^{\pm}_{s}(\lambda,\lambda)$, and
$\mathcal{N}^{\pm}_{s}(\lambda)\equiv \mathcal{N}^{\pm}_{s}(\lambda,\lambda)$). In the normalization we use, the dominating
eigenvalues $\tau^{0}_{s}(\lambda)$ of $\mathbb{L}^{\pm}_{s}(\lambda)$ are equal to $1$, while the rest of the spectrum is \emph{sub-unitary}, $|\tau^{j}_{s}(\lambda)|<1$ for $j\neq 0$. Moreover, by analyzing the spectra of matrices $\mathbb{L}^{\pm}_{s}$ one can
learn that the left/right eigenvector $\mathbb{L}^{\pm 0} \ket{\psi_0} =\ket{\psi_0}$, $\bra{\psi_0}\mathbb{L}^{\pm 0}  =\bra{\psi_0}$ 
(corresponding to the leading eigenvalue), is the \emph{spin-singlet} state
\begin{equation}
\ket{\psi_0} = (2s+1)^{-1/2}\sum_{k=0}^{2s} (-1)^{k} \ket{k}\otimes\ket{2s-k}.
\label{eqn:singlet_vacuum}
\end{equation}
The singlet vector $\ket{\psi_0}$ obeys $(\vec{\bf{S}}_{1}+\vec{\bf{S}}_{2})\ket{\psi_0}=0$ where (following Ref.~\cite{IMP15}) auxiliary spins are given by $\vec{\mathbf{S}}_{1}=(\vec{\mathbf{s}}\otimes \one_{s})$ and $\vec{\mathbf{S}}_{2}=\one_{s}\otimes \vec{\mathbf{s}}$ and act over $\mathcal{V}_{s}\otimes \mathcal{V}_{s}$.
For the remaining Pauli components $\mathbb{L}^{\pm \alpha}_{s}(\lambda)$, $\alpha\in\{\x,\y,\z\}$, we have
\begin{equation}
\vec{\vmbb{L}}^{-}_s(\lambda) \ket{\psi_0} = 0,\quad \bra{\psi_0}\vec{\vmbb{L}}^{+}_s(\lambda) = 0.
\label{eqn:hit}
\end{equation}
These relations imply that the product state
$\ket{\Psi_0} = \ket{\psi_0}\otimes \ket{\psi_0}\in\mathcal{V}^{\otimes 2}_{s}\otimes \mathcal{V}^{\otimes 2}_{s^{\prime}}$ is 
an eigenvector of $\mathbb{T}_{s,s^{\prime}}(\lambda,\lambda,\mu,\mu)$ with a unit eigenvalue 
\begin{equation}
\tau_{s,s^{\prime}}(\lambda,\lambda,\mu,\mu)=\tau_s^{0}(\lambda)\tau_{s'}^{0}(\mu)=1.
\label{eqn:factcond}
\end{equation}
The last step to perform in order to show that the kernel from Eq.~\eqref{eqn:K-kernel_def} is finite, is to rigorously show that $\tau_{s,s'}(\lambda,\lambda,\mu,\mu)=1$ is indeed the leading eigenvalue. This statement can be conveniently phrased by defining the operator
\begin{equation}
\mathbb{F}_{s,s^{\prime}}(\lambda,\mu) = \one - \mathbb{T}_{s,s'}(\lambda,\lambda,\mu,\mu),
\label{eqn:F-operator}
\end{equation}
and showing that it is a \emph{positive-definite} operator on the orthogonal complement of the singlet state $\ket{\Psi_0}$.

The $SU(2)$ symmetry of the isotropic point $\Delta=1$ makes the task of
demonstrating that the matrix~\eqref{eqn:F-operator} represents a contracting map much easier.
The scalar component of double Lax operator $\mathbb{L}^{+0}_{s}(\lambda,\mu)$ can be readily expressed in terms of the Casimir operator 
$\mathbf{C}=(\vec{\mathbf{S}}_{1}+\vec{\mathbf{S}}_{2})^{2}$,
\begin{equation}
\mathbb{L}^{+0}_{s}(\lambda,\mu)=\mathcal{N}^{+}_{s}(\lambda,\mu)\left((\lambda-\tfrac{\ii}{2})(\mu+\tfrac{\ii}{2}) \one-
\tfrac{1}{2}(\mathbf{C}-\vec{\mathbf{S}}^{2}_{1}-\vec{\mathbf{S}}^{2}_{2})\right),
\end{equation}
from where we conclude that the eigenvalues are
\begin{equation}
\tau^{j}_{s}(\lambda) = 1 - \tfrac{1}{2}\,\mathcal{N}_{s}(\lambda)j(j+1),\quad j=0,1,\ldots 2s,
\label{eqn:tau_j}
\end{equation}
while the dominating vector is clearly the spin singlet state $\ket{\Psi_0}$. A complete proof and further details on this part are 
presented in Ref.~\cite{IMP15} and the Supplementary material attached to it.

Note that factorizability of the leading eigenvalue, Eq.~\eqref{eqn:factcond}, in fact implies the inversion identity \eqref{eqn:inversion_formula}. Similar inversion formulae have been discussed earlier in the
literature~\cite{BaxterBook,Pearce87,Klumper89}. Quasilocality then follows essentially as a corollary of Eq.~\eqref{eqn:factcond}.
To finalize the proof it remains to be shown that the kernels 
$K_{s,s^{\prime}}(\lambda,\mu)$ given by Eq.~\eqref{eqn:K-kernel_def} are well-defined and can be evaluated directly by
accounting only for the contributions from the leading eigenvalues of auxiliary transfer matrices $\mathbb{T}_{s,s'}(\lambda,\lambda',\mu,\mu')$ and  $\mathbb{L}_{s}^{\pm 0}(\lambda,\mu)$.
Using arguments based on the first order perturbation theory in combination with factorizability of the leading 
eigenvalue results in
\begin{equation}
\begin{split}
K_{s,s^{\prime}}(\lambda,\mu) &= \big[\partial_{\lambda^{\prime}}\partial_{\mu^{\prime}}
\tau_{s,s^{\prime}}(\lambda,\lambda^{\prime},\mu,\mu^{\prime})\big]_{\mu^{\prime}=\mu,\lambda^{\prime}=\lambda}\\
&- \big[\partial_{\lambda'}\tau^{-0}_{s}(\lambda,\lambda^{\prime})\big]_{\lambda'=\lambda}
\big[\partial_{\mu'}\tau^{+0}_{s'}(\mu,\mu^{\prime})\big]_{\mu^{\prime}=\mu}.
\end{split}
\label{eqn:K-kernel_eig}
\end{equation}

\subsubsection{Local operator expansion.}
An important practical advantage of the present formulation is that  $X_{s}(\lambda)$ can be readily expanded in terms of local operators. This step is of main interest in applications where evaluation of local correlation functions 
plays the primary role.
To see how this works, we consider the resolution of operators $X_{s}(\lambda)$ with respect to local clusters of $r$ adjacent spins (\ref{clusterbasis}), by summing over all projections onto the finite sublattices of length $\Lambda$,
\begin{equation}
X_{s}(\lambda) = \lim_{\Lambda \to \infty}\lim_{N\to \infty}\sum_{r=1}^{\Lambda}\sum_{x=0}^{N-1}
\underbrace{\sum_{\ul{\alpha}}(\sigma^{\ul{\alpha}}_{[1,r]},X_{s}(\lambda))\,
\sigma^{\ul{\alpha}}_{[x,x+r-1]}}_{\hat{\cal S}^x(d_r(\lambda))}.
\label{Xexp}
\end{equation}
Of course the `limits' have to be understood in the sense as discussed in Sec.~\ref{subsect:pseudolocal}.
Here operators $d_{r}(\lambda)$ represent projections of $X_{s}(\lambda)$ onto local densities with support size (order) $r$,
where by virtue of Eq.~\eqref{eqn:on1} the HS norms $\|d_{r}(\lambda)\|_{\rm HS}$ decay exponentially with $r$.
We note that strictly local charges $H^{(k)}$ are, ignoring irrelevant constant prefactors, just the Taylor series coefficients generated by expanding $X_{1/2}(\lambda)$ around $\lambda=0$.

Thanks to the factorizability of the leading eigenvalue and the corresponding eigenvector, all $k$-point amplitudes 
$(\sigma^{\ul{\alpha}}_{[1,k]},X_{j}(\lambda))$ can be efficiently computed by introducing
a set of \emph{auxiliary vertex operators},
\begin{equation}
\mathbb{X}_{s}^{\alpha}(\lambda) = \mathbb{L}^{+\alpha}_{s}(\lambda),
\label{eqn:vertex_operators}
\end{equation}
one for each $\alpha \in \mathcal{J}$. This allows us to write a matrix product representation
\begin{equation}
(\sigma^{\ul{\alpha}}_{[1,k]},X_{s}(\lambda))=
\bra{\psi^{\rm L}_{\alpha_{1}}(\lambda)}\mathbb{X}^{\alpha_{2}}(\lambda)\cdots
\mathbb{X}^{\alpha_{k-1}}(\lambda)\ket{\psi^{\rm R}_{\alpha_{k}}(\lambda)}.
\label{eqn:X_expansion}
\end{equation}
This formula is exact in the thermodynamic limit ($N\to\infty$, see Eq.~\eqref{Xexp}) while in finite lattices there are corrections which vanish exponentially in $N$ and can be estimated in terms of subleading eigenvalues of $\mathbb{T}_{s,s}$.
The boundary vectors in Eq.~\eqref{eqn:X_expansion} are set as
\begin{equation}
\ket{\psi^{\rm R}_{\alpha}(\lambda)} = \mathbb{L}^{+\alpha}_{s}(\lambda)\ket{\psi_{0}},\quad
\bra{\psi^{\rm L}_{\alpha}(\lambda)} = \bra{\psi_{0}}[-\ii \partial_{\mu}\mathbb{L}^{+\alpha}_{s}(\lambda,\mu)]_{\mu=\lambda}.
\label{eqn:boundary_vectors_general}
\end{equation}
Here we wish to note that, in order to produce a non-vanishing amplitude, the $\mu$-derivative which is included in the definition 
of $X_{s}(\lambda)$ (cf. Eq.~\eqref{eqn:X_product_from}) must necessarily act on the \emph{first} site in the matrix product representation of operators $X_{j}(\lambda)$ (see Eq.~\eqref{eqn:X_as_MPO}) due to Eq.~\eqref{eqn:hit}.

\subsubsection{Computation of Hilbert-Schmidt kernel.}
Quasilocal charges $X_{s}(\lambda)$ are linearly independent, but not manifestly orthogonal with respect to HS inner product.
Below we show how to obtain explicit expressions for kernels $K_{s,s^{\prime}}$, and subsequently use them to carry out
the `Gram--Schmidt orthogonalization'.
For simplicity we restrict our discussion to the isotropic point $\Delta=1$, where we find
\begin{equation}
\bra{\psi_0}\vec{\mathbb{L}}^{-}_s(\lambda) = 2{\mathcal{N}}_{s}(\lambda)\bra{\psi_0}\vec{\mathbf{S}}_{1},\quad
\vec{\mathbb{L}}^{+}_s(\lambda)\ket{\psi_0} = -2{\mathcal{N}}_{s}(\lambda)\vec{\mathbf{S}}_{1}\ket{\psi_0},
\label{eqn:boundary_action_XXX}
\end{equation}
while boundary vectors given in Eq.~\eqref{eqn:boundary_vectors_general} can now be chosen symmetrically and take the form
\begin{equation}
\ket{\psi_{\alpha}} = \sqrt{2} \mathcal{N}_{s} \mm{S}_{1}^\alpha\ket{\psi_0}.
\label{eqn:X_boundary_vector}
\end{equation}
A direct route to evaluate HSK $K_{s,s^{\prime}}(\lambda,\mu)$ as defined in Eq.~\eqref{eqn:K-kernel_eig} is to rewrite
the initial representation~\eqref{eqn:K-kernel_def} in terms of the resolvent of the auxiliary transfer matrix (see Ref.~\cite{IMP15} for details)
which can be rewritten in terms of a geometric series
\begin{equation}
K_{s,s^{\prime}}(\lambda,\mu) =
\bra{\Psi}\left(\one - \mathbb{T}_{s,s^{\prime}}(\lambda,\mu)\right)^{-1}\ket{\Psi} = 
\sum_{k=0}^{\infty}\bra{\Psi}[\mathbb{T}_{s,s^{\prime}}(\lambda,\mu)]^{k}\ket{\Psi},
\label{KTinv}
\end{equation}
where $\ket{\Psi}=\sum_{\alpha \in \{\x,\y,\z\}}\ket{\psi_{\alpha}}\otimes \ket{\psi_{\alpha}}$. In the above sum, each term
$\bra{\Psi}[\mathbb{T}_{s,s^{\prime}}(\lambda,\mu)]^{k}\ket{\Psi}$ actually corresponds to a contribution of an order-$k$ density
$d_{k}(\lambda)$, which is finite since it obeys the quasilocality condition.
A key point in this calculation is to recognize that the leading eigenvalues reside in an invariant singlet subspace
$\mathcal{V}_{0}\subset \mathcal{V}^{\otimes 2}_{s}\otimes \mathcal{V}^{\otimes 2}_{s^{\prime}}$ spanned by 
a convenient basis $\mathcal{V}_0 = {\rm lsp}\{\ket{j}; j=0,1,2,\ldots,2s\}$, where $\ket{0}\equiv \ket{\Psi_0}$, $\ket{1}\equiv \ket{\Psi}$.
Noticing that $\mathbb{F}_{s,s'}$ does not couple $\ket{\Psi_{0}}$ to the remaining states from $\mathcal{V}_{0}$ allows to cast
Eq.~\eqref{eqn:K-kernel_eig} expressed as Eq.~\eqref{KTinv} in terms of a solution to a linear system of $2s$ equations,
\begin{equation}
\mathbb{F}^{(0)}_{s,s^{\prime}}(\lambda,\mu)\ket{\Xi}=\ket{\Psi},\qquad K_{s,s^{\prime}}(\lambda,\mu) = \braket{\Psi}{\Xi},
\label{eqn:K-kernel_system}
\end{equation}
introducing the restriction of $\mathbb{F}_{s,s^{\prime}}$ to subspace $\mathcal{V}_{0}$
denoted by $\mathbb{F}^{(0)}_{s,s^{\prime}}$. The solution to Eq.~\eqref{eqn:K-kernel_system} is given in a closed form~\cite{IMP15}
\begin{align}
K_{s,s^{\prime}}(\lambda,\mu) &= \mathcal{N}_{s}(\lambda)\mathcal{N}_{s^{\prime}}(\mu)\kappa_{s,s^{\prime}}(\lambda-\mu),
\label{eq:explicit}
\\ \kappa_{s,s^{\prime}}(\lambda) &= \sum_{k=1}^{{\rm dim}\mathcal{V}_{0}-1}k(k+2|s^{\prime}-s|)
\frac{(2s+1)(2s^{\prime}+1)-2k|s^{\prime}-s|-k^{2}}{(2s+1)(2s^{\prime}+1)}a_{2|s^{\prime}-s|+2k}(\lambda),
\end{align}
where $a_{2s}(\lambda)=s/(s^{2}+\lambda^{2})$ are Cauchy--Lorentz kernels.
Kernels $a_{2s}$ play the central role as quasi-particle scattering phase shifts of the underlying scattering theory, as briefly explained in Sec.~\ref{sec:string-charge}.

\subsubsection{Orthogonalization procedure.}
The aim here is to construct mutually orthogonal families of quasilocal operators $\widetilde{X}_s(\lambda)$.
By considering a generic charge with $s>\half$ we set
\begin{equation}
\widetilde{X}_{s}(\lambda) =
X_{s}(\lambda) - \sum_{s^{\prime}}^{s^{\prime}<s}\int_{-\infty}^{\infty} \dd \mu f_{s,s^{\prime}}(\lambda,\mu)X_{s^{\prime}}(\mu),
\end{equation}
and minimize the inner product by solving the following variational problem:
\begin{equation}
\frac{\delta}{\delta f_{s,s^{\prime}}(\lambda,\mu)}\left(\widetilde{X}_{s}(\lambda),\widetilde{X}_{s}(\lambda)\right)=0.
\end{equation}
This yields a linear system of $2s-1$ coupled Fredholm integral equations,
\begin{equation}
\sum_{s^{\prime \prime}}^{s^{\prime \prime}<s}\int_{-\infty}^{\infty}\dd \nu K_{s^{\prime},s^{\prime \prime}}(\mu,\nu)
f_{s,s^{\prime \prime}}(\lambda,\nu) = K_{s^{\prime},s}(\mu,\lambda),\qquad \forall s^{\prime}<s,
\end{equation}
which can be reduced to a linear convolution system, using the explicit representation for the HSK \eref{eq:explicit},
\begin{equation}
\sum_{s^{\prime \prime}}^{s^{\prime \prime}<s}\kappa_{s^{\prime},s^{\prime \prime}}\star \widetilde{f}_{s^{\prime \prime},s} =
\kappa_{s^{\prime},s},
\label{eqn:orthogonalization_system}
\end{equation}
after rescaling the functions
$\widetilde{f}_{s^{\prime},s}(\mu-\lambda)=(\mathcal{N}_{s^{\prime}}(\mu)/\mathcal{N}_{s}(\lambda))f_{s,s^{\prime}}(\lambda,\mu)$.
The convolution operation is defined as $(f\star g)(\lambda)=\int_{-\infty}^{\infty}\dd \mu f(\lambda-\mu)g(\mu)$.
Explicit results for the solutions of Eq.~\eqref{eqn:orthogonalization_system} can be found in Ref.~\cite{IMP15}.

\subsection{Fusion hierarchy approach}
\label{subsection:fusionapproach}

We have previously highlighted the meaning of the inversion identity Eq.~\eqref{eqn:inversion_formula} and learned about its
importance for identifying quasilocal conserved quantities. In this section, we explore a different route and show how to consistently 
retrieve the inversion formula from Eq.~\eqref{eqn:inversion_formula} by resorting to an algebraic diagonalization of higher-spin 
operators $T_{s}(\lambda)$.

In Sec.~\ref{subsect:hirota} we explained how the entire set of canonical $T$-operators can be simultaneously diagonalized 
by means of Baxter's $Q$-operator. Assuming that the large-$N$ behaviour of Eq.~\eqref{eqn:Hirota_solution} can be read from the 
$N$-dependent scalars $\zeta_{2s,k}$, the sum is dominated by the highest term at index $k=2s$,
\begin{equation}
\frac{T^{+}_{s}(\lambda)}{T^{[2s+1]}_{0}(\lambda)} \stackrel{N\to \infty}{\longrightarrow}
\frac{Q^{[-2s]}(\lambda)}{Q^{[2s]}(\lambda)}.
\label{eqn:leading_term}
\end{equation}
This manifestly produces the inversion formula~\eqref{eqn:inversion_formula} on the level of operators. We
therefore expect that the formula~\eqref{eqn:leading_term} also makes sense on the level of typical eigenvalues and
can therefore be used to obtain the action of $X_{j}(\lambda)$ on (Bethe) eigenstates.

In view of Eq.~\eqref{eqn:leading_term} we, in addition, conclude that the
`quasilocality domain' can be analytically continued from the real axis to the whole `physical strip' in the complex plane,
\begin{equation}
\mathcal{P}_{\eta}=\{\lambda\in \CC;|{\rm Im}(\lambda)|< \tfrac{\ii \eta}{2}\}.
\label{eqn:physical_domain}
\end{equation}
We note that the charges $X_{s}(\lambda)$ are Hermitian for $\lambda \in \RaR$, but they become
non-Hermitian for ${\rm Im}(\lambda)\neq 0$.

As a consequence of Eq.~\eqref{eqn:leading_term}, the general version (for arbitrary anisotropy $\Delta$) of the unitary quasilocal charges from Eq.~\eqref{eqn:X_definition} 
admits a useful compact representation in terms of the $Q$-operator
\begin{equation}
X_{s}(\lambda) = -\ii \partial_{\lambda}\log\,\frac{Q^{[-2s]}(\lambda)}{Q^{[2s]}(\lambda)},\qquad \lambda \in \mathcal{P}_{\eta}.
\label{eqn:Q-representation}
\end{equation}
The charges $X_{s}(\lambda)$ can now be effectively diagonalized using the fact that eigenvalues of the Baxter's $Q$-operator
(denoted by $\mathcal{Q}(\lambda)$) are $q$-deformed polynomials with zeros coinciding with the set of Bethe roots $\{\lambda_j\}$ \footnote{Here we ignore a 
subtle fact that Baxter's $Q$-operator becomes singular in the presence of periodic boundary condition and requires to be regularized 
in some way~\cite{shortcut}.
In our formulae, $Q$-s always appear in certain ratios which are always well-behaved. Apart from this, we do not rely on an
operatorial construction of $Q$-operator, but merely use its spectrum which pertains to Bethe string configurations.},
\begin{equation}
\mathcal{Q}(\lambda)= c \prod_{j=1}^{M}\sin{(\lambda-\lambda_j)},
\label{eqn:Q_spectrum}
\end{equation}
where $c$ is an inessential scalar prefactor.
At this point the identification with the spectrum of the model has been made, which shall play a central role in the subsequent 
discussion of applications in the area of `quantum quenches'. Further details are presented in Sec.~\ref{subsect:quenches}.

\subsection{Gapless regime}
\label{gapless} 

In this section we generalize the results for the isotropic and gapped cases derived in the previous section to the gapless regime.
Without loss of generality we restrict our considerations to the positive side of the critical interval $\Delta \in (0,1)$. For 
technical reasons we exclude the non-interacting point at $\Delta = 0$, which due to the
exceptional degeneracy requires a special treatment.

In the gapless regime we introduce a three-parametric family of conserved operators
\begin{equation}
X_{(s,u)}(\lambda) =
-\ii \partial_{\lambda}\log \frac{T^{+}_{(s,u)}(\lambda)}{T^{[j+1]}_{(0,u)}(\lambda)},\quad s=\half,1,\ldots
\label{eqn:X_def_gapless}
\end{equation}
An important difference with respect to the family of charges used in the gapped regime is that $T$-operators now acquire another
quantum label, the so-called (string) parity number $u\in \{\pm 1\}$. The latter merely represents a $\pi/2$ displacement
of the spectral parameter in the imaginary direction, namely
\begin{equation}
T^{[\pm k]}_{(s,u)}(\lambda) = T_{s}\left(\lambda \pm k\tfrac{\ii \eta}{2}+(1-u)\tfrac{\ii \pi}{4}\mp \ii 0^{+}\right)\qquad
{\rm for}\quad \lambda \in \mathcal{P}_{\eta}.
\label{eqn:T_def_gapless}
\end{equation}
It is important to
stress that operators from Eq.~\eqref{eqn:X_def_gapless} do not automatically inherit quasilocality from the gapped counterparts.
Even though in the present case the structural form of the solution Eq.~\eqref{eqn:Hirota_solution} to the Hirota equation
remains unaffected, the scalar functions undergo the following modification
\begin{equation}
\zeta_{(s,u),k}(\lambda) = \frac{\sinh{(\lambda+(2(k-s)+1)\tfrac{\ii \eta}{2}+(1-u)\tfrac{\ii \pi}{4})}}
{\sinh{(\lambda+(2s+1)\tfrac{\ii \eta}{2}+(1-u)\tfrac{\ii \pi}{4})}},\quad k=0,1,\ldots 2s.
\end{equation}
For the inversion identity to hold, the following condition should be satisfied
\begin{equation}
|\zeta_{(s,u),k}(\lambda)|<1 \qquad {\rm for}\quad k=0,1,\ldots 2s-1.
\end{equation}

In stark contrast to the gapped (and isotropic) case, given a root of unity deformation $q=\exp{(\ii \pi l/m)}$, only a \emph{finite} 
number of (linearly) independent charges with quantum labels $(s,u)$ can satisfy this condition.
For instance, for the simple roots of the form $\eta/\pi=1/m$, there are precisely $m-1$ charges with labels $(s,+)$
for $s=\tfrac{1}{2},1,\ldots \tfrac{m-1}{2}$.
On the other hand, at generic roots of unity identifying the complete set of charges becomes more involved~\cite{StringCharge}.
To give a flavour, at $\eta/\pi = 3/7$, we have four independent families of charges corresponding to the set
\begin{equation}
\{X_{(\frac{1}{2},+)},X_{(1,+)},X_{(2,-)},X_{(3,+)}\}.
\label{Xcontent}
\end{equation}
While the total number of quasilocal charges at a given value of $\eta$ and their associated quantum labels might seem a bit 
arbitrary at a first glance, it is explained below in Sec.~\ref{subsect:quenches}, that the labels can be matched to the known and 
well-established quasi-particle thermodynamic content of the model.

\section{Quasilocal charges from non-unitary representations}
\label{sec:non-unitary}

Here we turn our attention to the construction of quasilocal conserved charges, using non-unitary representations of
$\mathcal{U}_{q}(\mathfrak{sl}(2))$. In the first part we consider the highest-weight representations as elaborated on in Ref.~\cite{ProsenNPB14} (see also \cite{Pereira14}), building on previous results \cite{ProsenPRL106,PI13}. This construction yields conserved operators which break the spin reversal symmetry of the model and which are used for establishing the 
ballistic transport property of the high-temperature anisotropic Heisenberg model. The second part discusses an analogous construction, 
this time with semi-cyclic representations which, interestingly, break even the $U(1)$ symmetry of the model, following Ref.~\cite{Zadnik16}.

\subsection{Charges from highest-weight representations}
\label{subsect:hw}

Let us remain in the gapless regime and keep the root of unity parametrization of the anisotropy given as
$\Delta=\cos(\eta)$, or $q=e^{\ii \eta}$, with $\eta = \pi l/m$, and $l,m\in\ZZ_+$ co-prime. In what follows, the basic building block 
of our construction is a reparametrized Lax operator Eq.~\eqref{eqn:Lax_operator_deformed}, where for our convenience (and to comply 
with Refs.~\cite{PI13,ProsenNPB14}) we perform a rescaling by a factor $\sinh{(\eta)}/\sin{(\lambda)}$ and subsequently make a 
substitution $\eta\to -\ii\eta$ (but refraining from substituting $\lambda\to-\ii\lambda$ as in Sec.~\ref{subsect:lax}). This results in a trigonometric form of the Lax operator:
\begin{equation}
\mathbf{L}_{s}(\lambda) = \frac{1}{\sin{(\lambda)}}
\begin{pmatrix}
\sin(\lambda+\eta\,\mathbf{s}^{\z}_{s}) & \sin{(\eta)}\,\mathbf{s}^{-}_{s} \cr
\sin{(\eta)}\,\mathbf{s}^{+}_{s} & \sin(\lambda-\eta\,\mathbf{s}^{\z}_{s})
\end{pmatrix}.
\label{eqn:lax}
\end{equation}
Considering the $m$-dimensional highest-weight auxiliary space representation~\eqref{eqn:noncompact_irrep},
the commuting transfer operators are given in accordance with the standard prescription
\begin{equation}
T^{\rm hw}_{s}(\lambda) = {\rm Tr}_{\rm a}\left\{\mathbf{L}_{s}(\lambda)^{\otimes N}\right\}.
\end{equation}

Without further ado, we define the following family of commuting operators by
differentiating $T^{\rm hw}_{s}(\lambda)$ with respect to continuous spin $s$,
\begin{equation}
Z(\lambda) =  \frac{\sin{(\lambda})^{2}}{2\eta\sin{(\eta)}}\,\partial_{s} T^{\rm hw}_{s}(\lambda)\vert_{s=0}-
\frac{\sin{(\lambda)}\cos{(\lambda)}}{2 \sin{(\eta)}}\,M.
\label{eqn:Zdef0}
\end{equation}
Note that in this way the contribution of the magnetization $M=\sum_{x \in \Lambda}\sigma^{\z}_{x}$ cancels from $Z(\lambda)$,
and hence, by construction, only the operator terms acting non-trivially on two or more sites remain. With the aid of Lax operator
components  
\begin{equation}
\mathbf{L}(\lambda)\equiv \mathbf{L}_{0}(\lambda)=\sum_{\alpha\in \mathcal{J}}\mathbf{L}^{\alpha}(\lambda)\sigma^{\alpha},\qquad
\widetilde{\mathbf{L}}(\lambda) \equiv \partial_{s}\mathbf{L}_{s}(\lambda)\bigr|_{s=0},
\label{eqn:rescaled}
\end{equation}
we can, following the logic presented in Sec.~\ref{subsect:X}, expand the family of conserved operators $Z(\lambda)$ in the large-$N$ 
limit in terms of $r$-spin clusters,
\begin{equation}
Z(\lambda) = \lim_{\Lambda \to \infty}\lim_{N\to \infty}\sum_{r=1}^{\Lambda}\sum_{x=0}^{N-1}
\underbrace{\sum_{\ul{\alpha}}(\sigma^{\ul{\alpha}}_{[1,r]},Z(\lambda))\,
\sigma^{\ul{\alpha}}_{[x,x+r-1]}}_{\hat{\cal S}^x(d_r(\lambda))}.
\label{eqn:Zdef1}
\end{equation}
The amplitudes are now encoded as matrix product expressions
\begin{align}
(\sigma^{-}\otimes \sigma^{\alpha_{2},\ldots,\alpha_{r-1}}_{[2,r-1]}\otimes \sigma^{+},Z(\lambda)) =
\bra{\rm L}\mathbf{L}^{\alpha_{2}}(\lambda) \cdots
\mathbf{L}^{\alpha_{r-1}}(\lambda) \ket{\rm R},
\label{eqn:qp}
\end{align}
while the boundary vectors are given as $\bra{\rm L}\equiv \frac{\sin\lambda}{\sin\eta}\bra{0}\mathbf{L}^{-}$, $\ket{\rm R}\equiv \frac{\sin\lambda}{2\eta}\widetilde{\mathbf{L}}^{+}\ket{0}$ (in addition to that, $(\sigma^{-}\otimes\sigma^{+},Z(\lambda))=1$).
By inspecting the Lax components (cf. Eq.~\eref{eqn:laxcomponents_hw} below) we learn that all amplitudes which violate the selection 
rule $\alpha_{1}=-$ and $\alpha_{r}=+$ vanish. Another remark that we would like to make is that in any finite-$N$ lattice the 
expression for the conserved operators $Z(\lambda)$, as given by Eq.~\eqref{eqn:Zdef1} without taking the limits and setting $\Lambda=N$, in fact acquires a finite-size correction of the form
\begin{equation}
c(\lambda) = \sum_{x=0}^{N-1} \hat{\cal S}^x\left(\sum_{n=1}^{m-1} \bra{n}\mathbf{L}(\lambda)^{\otimes (N-1)} \otimes \widetilde{\mathbf{L}}(\lambda)\ket{n}\right),
\label{eqn:correction}
\end{equation}
which gets exponentially suppressed with $N$ with respect to HS norm (see Ref.~\cite{ProsenNPB14}).

\begin{figure}[htb]
\centering
\includegraphics[width=0.55\textwidth]{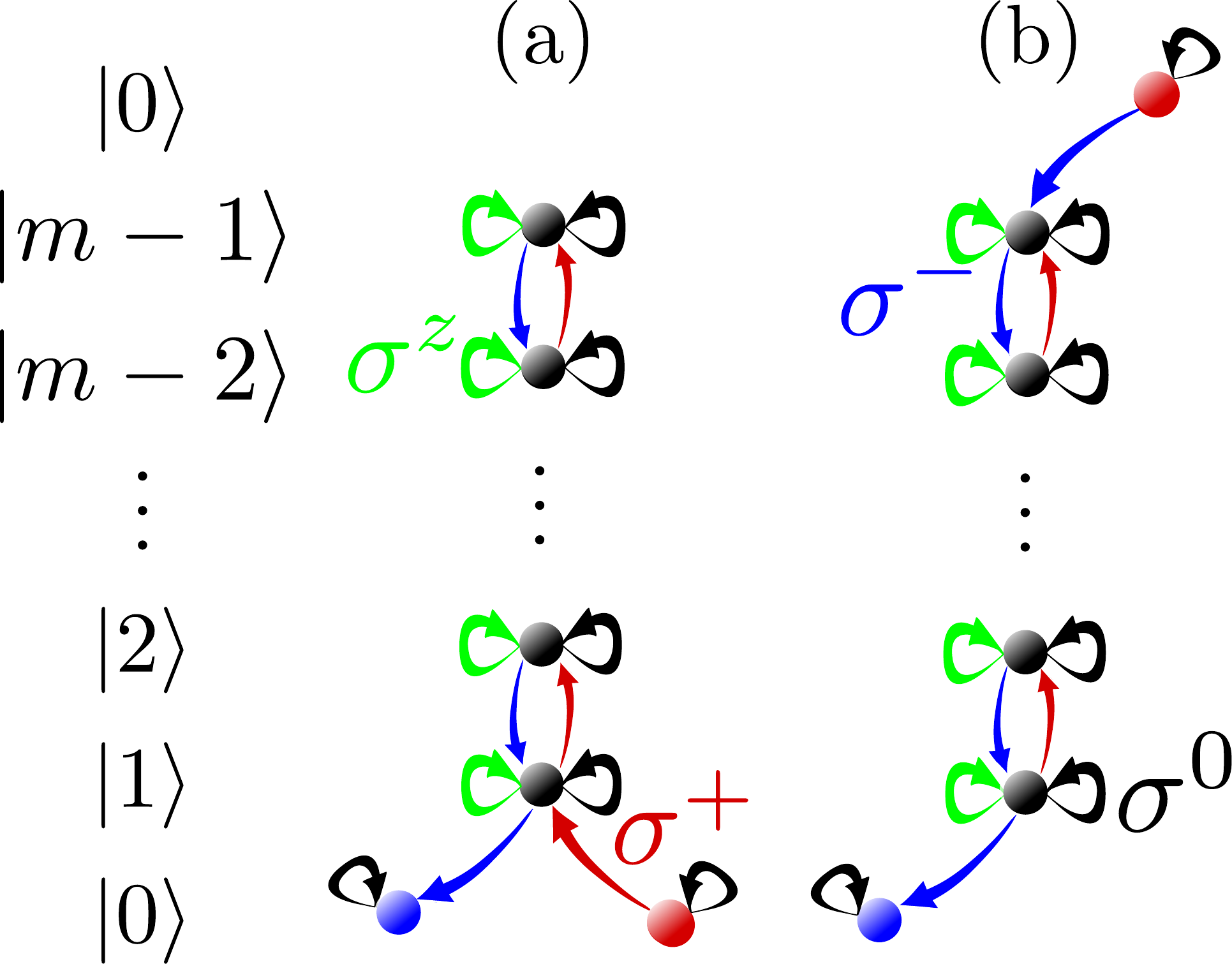}
\caption{Schematic depiction of the construction of a non-unitary quasilocal charge $Z(\lambda)$, for the highest-weight (a) and the
semi-cyclic (b) case. Each term in the local operator expansion Eq.~\eqref{eqn:Zdef1} corresponds to a distinct $N$-step walk in the directed graph, starting in the red node and ending in the blue node.
The vertical level $n$ of the node corresponds to a state in auxiliary space $\ket{n}$, while coloured arrows indicate
physical space operator (black $\sigma^0$, green $\sigma^\z$, red $\sigma^+$, blue $\sigma^-$) picked at $x-$th step of the walk, 
corresponding to the lattice site at position $x=1,2\ldots N$.
The amplitude of overall term is given by the product of matrix elements of the corresponding Lax operators between appropriate 
auxiliary states (vertical levels of the nodes, indicated on the left).}
\label{fig:Z-fig}
\end{figure}

Let us briefly comment on the technical part of what steps have been made to arrive at Eq.~\eqref{eqn:Zdef1}. Due to translational
invariance, each term in the operator expansion of Eq.~\eqref{eqn:Zdef0} has been rearranged so that the right-most position in the 
product of Lax operators always belongs to the 
differentiated Lax operator, $\widetilde{\mathbf{L}}(\lambda)$. The trace in $\partial_{s} T_{s}(\lambda)$ is then split into two 
parts, a sum over states $\ket{n\neq 0}$, producing the correction~\eqref{eqn:correction}, and the projection onto the `vacuum' $\ket{0}$ part which results in Eq.~\eqref{eqn:Zdef1}. Explicit form of the amplitudes given by 
Eq.~\eqref{eqn:qp} can be deduced from the Lax components, Eq.~\eqref{eqn:rescaled}, reading 
\begin{align}
\mathbf{L}^{0}(\lambda) &= \sum_{n=0}^{m-1} \cos{(n\eta)}\ket{n}\!\bra{n},&\quad
\widetilde{\mathbf{L}}^{0}(\lambda) &= \eta \sum_{n=1}^{m-1} \sin{(n\eta)} \ket{n}\!\bra{n},\nonumber\\
\mathbf{L}^{\z}(\lambda) &=-\cot(\lambda)\sum_{n=1}^{m-1} \sin{(n\eta)}\ket{n}\!\bra{n},&\quad
\widetilde{\mathbf{L}}^{\z}(\lambda) &= \eta \cot{(\lambda)} \sum_{n=0}^{m-1} \cos{(n\eta)} \ket{n}\!\bra{n},\nonumber\\
\mathbf{L}^+(\lambda) &=-\frac{1}{\sin(\lambda)}\sum_{n=1}^{m-2} \sin{(n\eta)} \ket{n+1}\!\bra{n},&\quad
\widetilde{\mathbf{L}}^{+}(\lambda) &= \frac{2\eta}{\sin{(\lambda)}} \sum_{n=0}^{m-2} \cos{(n\eta)} \ket{n+1}\!\bra{n},\nonumber\\
\mathbf{L}^-(\lambda) &=\frac{1}{\sin(\lambda)}\sum_{n=0}^{m-2} \sin{((n+1)\eta)} \ket{n}\!\bra{n+1},&\quad
\widetilde{\mathbf{L}}^{-}(\lambda) &= 0.
\label{eqn:laxcomponents_hw}
\end{align}
For a diagrammatic illustration of explicit construction of the highest-weight $Z$-charges~\eqref{eqn:Zdef1},
see Fig.~\ref{fig:Z-fig} (panel (a)).

\subsubsection{Quasilocality of Z-charges.}
Considering the HS inner product of an arbitrary pair of non-unitary quasilocal charges from Eq.~\eqref{eqn:Zdef1}, one again defines 
the HSK as
\begin{equation}
K(\lambda,\mu)=\lim_{N\to\infty}\frac{1}{N}(Z(\lambda),Z(\mu))= \frac{1}{4}\bra{1}(\one-\mathbb{T}(\lambda,\mu))^{-1}\ket{1}.
\label{eqn:innerproductT}
\end{equation}
The associated \emph{auxiliary transfer matrix} $\mathbb{T}$ is an operator on the reduced auxiliary space
${\rm lsp}\{\ket{n}\simeq\ket{n}\otimes\ket{n};n=1,...,m-1\}$ of the form\footnote{The exact bijective correspondence, used to produce 
this symmetrized matrix form is $\ket{n}\otimes\ket{n} \leftrightarrow |\sin{(n\eta)}|\ket{n}$, $\bra{n}\otimes\bra{n} \leftrightarrow |\sin{(n\eta)}|^{-1}\bra{n}$.}
\begin{align}
\mathbb{T}(\lambda,\mu) &= \sum_{n=1}^{m-1} (\cos{(n\eta)}^{2} +
\cot{(\lambda)}\cot{(\mu)}\sin{(n\eta)^{2}})\ket{n}\!\bra{n} \nonumber \\
&+ \sum_{n=1}^{m-2}\frac{|\sin{(n\eta)}\sin{((n+1)\eta)}|}{2\sin{(\lambda)}\sin{(\mu)}}\left(\ket{n}\!\bra{n\!+\!1} + \ket{n\!+\!1}\!\bra{n}\right).
\label{eqn:Tmatrix}
\end{align}
This matrix is \emph{contracting} when parameters $\lambda$ and $\mu$ lie inside the strip
\begin{equation}
\mathcal{D}_{m}=\left\{\lambda \in \mathbb{C}; \left|\,{\rm Re}(\lambda) - \frac{\pi}{2}\right| < \frac{\pi}{2m} \right\}.
\end{equation}
Quasilocality of conserved operators from Eq.~\eqref{eqn:Zdef1} is then an immediate consequence of this statement~\cite{ProsenNPB14}.
The reader has to be reminded that we have disregarded the correction term \eqref{eqn:correction}.
Using an equivalent procedure to the one described above, it can be shown that the contribution of this term to HSK is 
exponentially suppressed in the system size~\cite{ProsenNPB14}. In order to see that, one must examine the action of
$\mathbb{T}$ on invariant subspaces of $\mathcal{V}^{(m)}_{s}\otimes \mathcal{V}^{(m)}_{s}$ which are spanned by elements 
$\ket{n}\otimes\ket{n+k}$, for different fixed $k$. Such a decomposition reduces the auxiliary transfer matrix into the 
block diagonal form. One then proceeds by proving that each block itself is a contracting matrix.

Evaluating Eq.~\eqref{eqn:innerproductT} amounts to solving the linear equation
\begin{equation}
(\one-\mathbb{T}(\lambda,\mu))\ket{\psi}=\ket{1},
\label{linHW}
\end{equation}
for the components $\psi_j = \braket{j}{\psi}$ of $\ket{\psi}$. The final result is
\begin{equation}
K(\lambda,\mu)=\frac{1}{4}\psi_{1}=
-\frac{\sin{(\lambda)}\sin{(\mu)}\,\sin((m-1)(\lambda+\mu))}{2\sin^2{(\eta)}\sin{(m(\lambda+\mu))}}.
\label{HSKhw}
\end{equation}

The construction from above can also be applied to the case of \emph{twisted boundary conditions}. The Hamiltonian then consists of an open boundary part and a two-site term, acting on the first and the last site of the chain 
\begin{equation}
2e^{i\phi}\sigma^-\otimes\one_{2^{N-2}}\otimes\sigma^{+} + 2e^{-i\phi}
\sigma^{-}\otimes\one_{2^{N-2}}\otimes\sigma^{+} + \Delta\sigma^{z}\otimes\one_{2^{N-2}}\otimes\sigma^{z},
\end{equation}
introducing a flux parameter $\phi$, such that the $\phi=0$ case corresponds to the Hamiltonian with periodic boundary conditions.
The transfer operator in case of twisted boundary conditions takes the following form,
\begin{equation}
T_{s}(\lambda;\phi) = {\rm \Tr}_{\rm a}\left\{e^{-\ii \phi\,\mathbf{s}^{\z}_{s}}\mathbf{L}_{s}(\lambda)^{\otimes N}\right\},
\end{equation}
while the conserved charges are generated similarly as in Eq.~\eqref{eqn:Zdef1}, with the prescription~\eref{eqn:Zdef0}, but using 
a modified $s$-derivative, $\partial_{s}\to\partial_{s}+\ii\phi$. In this case the HS kernel from Eq.~\eqref{eqn:innerproductT} 
remains independent of $\phi$ and hence quasilocality is preserved. This concludes the review of highest-weight conserved charges.

\subsection{Charges from semi-cyclic representations}
\label{scch}

After having discussed how to obtain quasilocal charges from the highest-weight auxiliary modules, we now turn our attention to 
another family of  representations of $\mathcal{U}_q(\mathfrak{sl}(2))$ at roots of unity -- the semi-cyclic representations.
To this end we retain the $m$-dimensional auxiliary spaces, $\mathcal{V}^{(m)}_{s}={\rm lsp}\{\ket{n};k=0,...,m-1\}$, but
modify the algebra generators as defined in Eq.~\eqref{eqn:noncompact_irrep} by an addition of an extra coupling:
\begin{equation}
\begin{split}
\mathbf{s}^{\z}_{s} &= \sum_{n=0}^{m-1} (s-n)\ket{n}\bra{n}, \\
\mathbf{s}^{+}_{s} &= \sum_{n=0}^{m-2} [n+1]_q\ket{n}\bra{n+1}+\alpha \ket{m-1}\bra{0},\\
\mathbf{s}^{-}_{s} &= \sum_{n=0}^{m-2} [2s-n]_q\ket{n+1}\bra{n}. 
\end{split}
\label{eqn:generators_sc}
\end{equation}
Here we have introduced the `\emph{coupling}' parameter $\alpha$, linking the first and the last basis states.\footnote{In our 
notation, the dependence on additional parameter $\alpha$ will not be explicitly written. One should nevertheless bear this 
dependence in mind.} Since the action of ladder operators is periodic only in one 
direction, such a representation is referred to as \emph{semi-cyclic}.
The algebraic relations \eqref{eqn:q-deformed_commutation} are still satisfied.

There are other possible alterations of the representation of the algebra generators, all of them resulting in a certain kind of periodicity~\cite{KasselBook}. 
In the following we will, for the sake of simplicity, only consider the above example. Since all other semi-cyclic representations generate the same quasilocal charges, up to trivial transformations, this means no loss of generality~\cite{Zadnik16}.

As we will see, the coupling of the lowest and highest-weight vectors in $\mathcal{V}^{(m)}_{s}$ results in a family of conserved 
charges which do not conserve the total magnetization $M$ (i.e. they break the $U(1)$ symmetry). Apart from this, the charges considered here only exist for \emph{odd} 
dimensions $m$. 
While non-conservation of magnetization is obvious from the explicit expressions, non-existence of these charges for even $m$ 
stems from the mismatch between the canonical $\mathcal{U}_q(\mathfrak{sl}(2))$ relations \eqref{eqn:q-deformed_commutation} and slightly modified 
relations which directly imply commutativity of the transfer operators with the Hamiltonian, see Refs.~\cite{Zadnik16,Korff03}.
The allowed values of anisotropy parameter are:
\begin{equation}
\eta = \frac{2 l}{2k-1}\pi \quad \text{or} \quad \eta = \pi-\frac{2 l}{2k-1}\pi, \qquad {\rm for}\quad k,l\in\mathbb{N},\,l<k.
\end{equation}

\subsubsection{Constructing the semi-cyclic charges.}
The same transfer matrix as in the case of highest-weight representations can be used, but this time
we differentiate it with respect to the coupling parameter $\alpha$, at $\alpha=0$ and $s=0$.
We now put
\begin{equation}
\mathbf{L}(\lambda)=\mathbf{L}_{0}(\lambda)\bigr|_{\alpha=0},\qquad
\widetilde{\mathbf{L}}(\lambda)=\partial_{\alpha}\mathbf{L}_{0}(\lambda)\bigr|_{\alpha=0},
\end{equation}
with the only non-trivial component of $\widetilde{\mathbf{L}}$ being
\begin{equation}
\widetilde{\mathbf{L}}^{-}(\lambda) =\frac{\sin{(\eta)}}{\sin{(\lambda)}}\ket{m-1}\bra{0}.\nonumber
\end{equation}
The conserved charges are this time defined as
\begin{equation}
Z(\lambda)=\frac{\sin^{2}{(\lambda)}}{\sin^{2}{(\eta)}}\partial_{\alpha} T^{\rm sc}_{s}(\lambda)\bigr|_{\alpha=0,s=0},
\label{eqn:cyclicconserved}
\end{equation}
where $T^{\rm sc}_{s}(\lambda)$ is the semi-cyclic transfer matrix defined with auxiliary space generators \eqref{eqn:generators_sc}. Once again the formula~\eqref{eqn:Zdef1} applies, thereby the amplitudes can be expressed in a 
canonical way
\begin{align}
(\sigma^{-}\otimes \sigma^{\alpha_{2},\ldots,\alpha_{r-1}}_{[2,r-1]}\otimes \sigma^{-},Z(\lambda)) =
\bra{\rm L}\mathbf{L}^{\alpha_{2}}(\lambda) \cdots
\mathbf{L}^{\alpha_{r-1}}(\lambda) \ket{\rm R},
\label{eqn:qp2}
\end{align}
with $\bra{\rm L}\equiv\frac{\sin\lambda}{\sin\eta} \bra{0}\mathbf{L}^{-}$, $\ket{\rm R}\equiv\frac{\sin\lambda}{\sin\eta} \widetilde{\mathbf{L}}^{-}\ket{0}$. The remaining string of Lax components in the LHS of Eq.~\eref{eqn:qp2} must connect $\bra{1}$ to $\ket{m-1}$ so the second sum in the expansion \eqref{eqn:Zdef1} actually starts at $r=m$. Because each term of $Z(\lambda)$ consists of a surplus of exactly $m$ operators
$\sigma^{-}$ over operators $\sigma^{+}$, these charges do not conserve magnetization $M$.
A diagrammatic presentation of semi-cyclic $Z$-charges is shown in Fig.~\ref{fig:Z-fig} (panel (b)).

\subsubsection{Quasilocality.}
What remains to be done is to derive the quasilocality property. The latter follows from a slightly modified calculation
with respect to the situation which we had previously with the highest-weight charges.
A careful inspection shows that the same auxiliary transfer matrix as given by Eq.~\eqref{eqn:Tmatrix} for a highest-weight 
representation, can be used to express the semi-cyclic HSK as
\begin{equation}
K(\lambda,\mu)=\lim_{N\to\infty}\frac{1}{N}(Z(\lambda),Z(\mu))= \frac{1}{4}\bra{1}(\one-\mathbb{T}(\lambda,\mu))^{-1}\ket{m-1}.\label{eqn:K_sc}
\end{equation}
Again, a solution of a simple tridiagonal system \eref{linHW} of equations yields an explicit expression
\begin{equation}
K(\lambda,\mu)=\frac{1}{4}\psi_{m-1}=\frac{\sin{(\lambda)}\sin{(\mu)}\sin{(\lambda+\mu)}}{2\sin^{2}{(\eta)}\sin{(m(\lambda+\mu))}}.
\end{equation}
To produce $\psi_{m-1}$ as defined previously in Sec.~\ref{subsect:hw}, the states $\ket{1}$ and $\ket{m-1}$ have to be exchanged. To 
this end  we conjugate Eq.~\eqref{eqn:K_sc} and recall that $\mathbb{T}(\lambda,\mu)$ is symmetric.

\section{Applications}
\label{sec:applications}

Let us finally focus on various physical applications of quasilocal conserved charges
in the domain of non-equilibrium quantum physics. Here both classes considered above, i.e. unitary and non-unitary charges, will
be examined. We shall begin with non-unitary $Z$-charges and show how they directly relate to non-equilibrium states with currents.
On the flip side, unitary $X$-charges will play an instrumental role for understanding equilibration in quantum quenches.
But before heading on, we need to clarify an important role of the spin reversal parity symmetry and its breaking.

\paragraph{Spin reversal and CPT symmetry of generic transfer matrices.}
We wish to elaborate on an important $\ZZ_{2}$ symmetry of all finite-dimensional unitary representations of the 
quantum group $\mathcal{U}_{q}(\mathfrak{sl}({2}))$, and consequently of the $XXZ$ Hamiltonian itself, which is manifestly broken for non-unitary representations. This symmetry breaking has some remarkable physical implications which shall be presented in
the following.

The $\ZZ_{2}$ symmetry under scrutiny is a parity generated by the spin-reversal canonical transformation
\begin{equation}
\label{spin-reversal}
s^{\z} \to -s^{\z}\quad s^{\pm} \to s^{\mp}.
\end{equation}
In fundamental representation the latter amounts to applying the product of $\sigma^{\x}$,
\begin{equation}
A\to P A P^{-1}:\qquad P = \prod_{x=1}^{N}\sigma^{\x}_{x} = P^{-1},
\end{equation}
where $A$ can be any observable on the entire Hilbert space $\mathcal{H}$.
It is easy to show that all transfer matrices belonging to finite-dimensional irreducible unitary representations are
manifestly $P$-invariant,
\begin{equation}
P T_{s}(\lambda) P^{-1} = T_{s}(\lambda), \quad s\in\half \mathbb{Z}_{+},
\label{eq:PTP}
\end{equation}
implying the same property also for the corresponding local and quasilocal charges,
\begin{equation}
[H^{(k)},P]=0,\quad [X_s(\lambda),P]=0.
\end{equation}

For the root of unity deformations $q=\exp(\ii\pi l/m)$ there exists another class of irreducible representations. These are
non-unitary $m$-dimensional highest-weight representations of $\mathcal{U}_{q}(\mathfrak{sl}({2}))$ discussed previously
in Sec.~\ref{sec:non-unitary}.
They are distinguished by the property, which can be readily verified, that {\em no} similarity transformation $\mm{x} \to \mm{G} \mm{x} \mm{G}^{-1}$ of the auxiliary space representation of the algebra exists 
which would generate the spin-reversal canonical transformation  \eref{spin-reversal}. These non-unitary representations \eref{eqn:noncompact_irrep} 
are labelled by a complex-spin parameter $s\in\CC$ and are henceforth not
$P$-invariant. We note that existence of an invertible $\mm{G}$, such that $\mm{G}\mm{s}^{\z} \mm{G}^{-1} = -\mm{s}^{\z}$, 
$\mm{G}\mm{s}^\pm \mm{G}^{-1} = \mm{s}^\mp$, is equivalent to a spin-reversal symmetry of the Lax operator \eqref{eqn:lax} 
 $P \mm{L}_s(\lambda) P^{-1} = \mm{G} \mm{L}_s(\lambda) \mm{G}^{-1}$, where $P$ acts nontrivially only on the physical space and $\mm{G}$ only on the auxiliary space, and consequently implies
 Eq.~\eref{eq:PTP}.

The highest-weight transfer matrices for complex spins and the quasilocal charges they generate instead exhibit a weaker symmetry,
\begin{equation}
P T^{\rm hw}_{s}(\lambda) P^{-1} = (T^{\rm hw}_{s}(\pi-\lambda))^{T}, \quad P Z(\lambda) P^{-1} = (Z(\pi-\lambda))^{T},\quad s\in\CC.
\label{eq:CPT}
\end{equation}
As the transposition can be associated with time-reversal operation, while reflection of the spectral parameter $\lambda \to \pi-\lambda$ 
can be thought of as the `charge conjugation' (after a suitable rotation and a shift of the spectral parameter it would correspond to 
$\lambda\to \bar{\lambda}$), the relation~\eqref{eq:CPT} can in fact be interpreted as a CPT symmetry of a generic highest-weight transfer matrix. The fact that complex-spin transfer matrices $T^{\rm hw}_{s}(\lambda)$ break spin-reversal symmetry can be 
fruitfully explored for the analysis of ballistic spin transport in anisotropic Heisenberg chains as will be demonstrated
in Sec.~\ref{sec:mazur}.

An equivalent CPT symmetry (\ref{eq:CPT}) holds also for the semi-cyclic transfer matrices and the corresponding quasilocal charges as discussed in Sec.~\ref{scch}.

\subsection{Mazur bounds on Drude weights}
\label{sec:mazur}

\subsubsection{Ballistic linear response.}
The main motivation for constructing pseudolocal conservation laws originated from the idea of using
such objects to estimate the ballistic contribution to transport coefficients,
such as Drude weights or, more generally, zero frequency dynamical susceptibilities~\cite{ProsenJPA98,ProsenPRE99}. 
It is perhaps worth noticing that related indicators of ballistic transport are nowadays directly experimentally
accessible~\cite{Hild,Ronzheimer,Schneider,Xia}.

By considering an extensive current $J = \sum_x \hat{\mathcal{S}}^x(j)$ with a local density $j$, say the
spin/particle/energy/etc. current, the Kubo linear response formula for the non-dissipative (real) part of the respective conductivity
is of the form
\begin{equation}
\sigma'(\omega) = \lim_{t\to\infty}\lim_{N\to\infty}\frac{\beta}{N}
\int_{0}^{t} \dd t' e^{\ii\omega t'} (J(t'),J(0))_{\beta},
\label{eqn:Kubo}
\end{equation}
Here the time-evolution reads $J(t) = e^{\ii H t} J  e^{-\ii H t}$, and
\begin{equation}
(A,B)_\beta = \beta^{-1}Z^{-1}_{\beta} \int_0^\beta \dd \lambda\,{\rm \Tr}
\left( A^\dagger e^{-\lambda H} B e^{-(\beta-\lambda) H}\right),
\label{eqn:Kubo-Mori}
\end{equation}
is the Kubo--Mori bracket with $Z_{\beta}={\rm Tr}(\,e^{-\beta H})$ denoting the partition function.
Note that the proper order of limits in Eq.~\eqref{eqn:Kubo}, namely firstly the thermodynamic limit $N\to\infty$ and then $t\to\infty$, which is in general important.
When either $A$ or $B$ is a conserved operator, Eq.~\eqref{eqn:Kubo-Mori} simplifies to a thermal state
$(A,B)_\beta = \omega_\beta(A^\dagger B)$, whereas at high temperatures $\beta \to 0$, the overlap $(A,B)_\beta\equiv (A,B)$
reduces to Hilbert--Schmidt inner product \eqref{eqn:HS}. The real part of the spin conductivity is normally split as
\begin{equation}
\sigma'_J(\omega) = 2\pi D_{J}\delta(\omega) + \sigma^{\rm reg}_J(\omega),
\end{equation}
where $\sigma^{\rm reg}_{J}$ is the regular part and $D_{J}$ is the singular contribution called the Drude weight.
The latter can be expressed by means of the linear response formula~\eqref{eqn:Kubo},
\begin{equation}
D_{J} = \lim_{t\to\infty}\lim_{N\to\infty}\frac{\beta}{2t N} \int_0^t \dd t' (J(t'), J(0))_\beta.
\end{equation}
Under certain mild assumptions on analyticity of local correlation functions, which are discussed in Ref.~\cite{IP13}, the order of 
the limits for $D_{J}$ can in fact be reversed and using time-invariance of the thermal state $\omega_{\beta}$ the Drude weight gets expressed in terms of time-averaged current as
\begin{align}
D_{J} &= \lim_{N\to\infty} \frac{\beta}{2N} \omega_\beta\left( \bar{J}^2 \right),
\label{eqn:DJ} \\
\bar{J} &= \lim_{t\to\infty} \frac{1}{t} \int_{0}^{t} \dd t' J(t').
\end{align}
A nontrivial value of the Drude weight $D_{J} > 0$ signals the ballistic (ideal) DC transport and is equivalent
(cf.  Eq.~\eqref{eqn:volscaling}) to the statement that the time-averaged current is a pseudolocal operator with respect
to the Gibbs state $\omega_{\beta}$ (see also Ref.~\cite{Doyon15}). 
We have thus related pseudolocality of time-averaged observables to ballistic linear response.

\subsubsection{Mazur bound.}
Computing time-averages of current operators seems a highly nontrivial task in interacting models. One can instead estimate the Drude 
weight from below using a bound due to Mazur~\cite{Mazur69} and Suzuki~\cite{Suzuki71} in terms of some conserved Hermitian operator $I=I^\dagger$, $[H,I]=0$. 
We start by writing out the expectation value of a \emph{nonnegative} operator $(\bar{J} - \alpha I)^{2}$, where $\alpha\in\RaR$ is a free parameter,
\begin{equation}
\omega_\beta(\bar{J}^{2}) - 2\alpha \omega_{\beta}(J I) + \alpha^{2} \omega_{\beta}(I^{2}) \ge 0.
\label{eqn:bound1}
\end{equation}
We used the fact that $\omega_{\beta}(\bar{J}I) = \omega_{\beta}(J I)$, which is due to the time-invariance of $\omega_{\beta}$ and 
conservation of $I$. After optimizing Eq.~\eqref{eqn:bound1} with respect to $\alpha$, we obtain
\begin{equation}
\omega_\beta(\bar{J}^2) \geq \frac{ \left(\omega_\beta( J I )\right)^{2}}{\omega_{\beta}(I^{2})}.
\end{equation}
Dividing by $2N$ and taking the limit $N\to\infty$, we produce the Mazur bound on the Drude weight, which has first been pointed out 
in Ref.~\cite{Zotos97},
\begin{equation}
D_{J} \geq \lim_{N\to\infty} \frac{\left(\omega_{\beta}(j I)\right)^{2}}{2N\omega_{\beta}(I^{2})}.
\end{equation}
In summary, a conserved pseudolocal operator $I$ which satisfies $\omega_{\beta}(j I) \neq 0$ implies ballistic 
transport and consequently allows to put a strict lower bound on the Drude weight. For example, by taking a translationally invariant 
extensive conserved operator $I = \sum_x \mathcal{S}^x(q)$, with density $q$ satisfying $\omega_{\beta}(q^{2}) < \infty$, one finds 
$D_J > 0$ if $\sum_x \omega_\beta(j \hat{\cal S}^x(q)) \neq 0$, where the last sum always converges due to exponential clustering of 
Gibbs states in one dimension~\cite{Araki69}.

In addition, as a consequence of an effective causality on the locally interacting lattice (i.e. Lieb--Robinson bounds~\cite{BRBook})
it can be shown that the above Mazur bound holds even when $I$ is not exactly conserved on any finite lattice with open boundaries but
the commutator $[H,I]$ contains terms localized near the boundary sites~\cite{IP13}.

When dealing with a larger set of pseudolocal conserved operators, say a countable set $\{I_k,k=1,2\ldots\}$, the Mazur 
bound can be further improved. To see how this works, we study the operator $(\bar{J} - \sum_{k} \alpha_{k} I_{k})^{2}$,
which after repeating the above reasoning results in
\begin{equation}
D_{J} \geq \frac{\beta}{2} \sum_{k,l} \omega_\beta(j I_k) (K_\beta^{-1})_{k,l} \omega_\beta(j I_l),
\end{equation}
where $K_\beta$ is a positive-definite overlap matrix $(K_\beta)_{k,l} = \lim_{N\to\infty} \frac{1}{N} \omega_{\beta} (I_{k} I_{l})$.
In this sense, if the above bound gets saturated for \emph{all} local currents $j$, it would be meaningful to regard 
the set of pseudolocal charges $\{I_{k}\}$ as being \emph{complete}. It is presently not known if such complete sets of pseudolocal 
conserved operators can be systematically identified in interacting models.

In previous sections we have defined and discussed certain continuous families (rather than discrete sequences) of pseudolocal 
charges which were referred to as quasilocal (cf. Eq.~\eqref{eqn:on1}). They comprise the charges $X_{s}(\lambda)$ and $Z(\lambda)$ which are analytic in $\lambda \in \mathbb{C}$ and become quasilocal when restricted to suitable domains ${\cal D}\subset\CC$. Since all $X_{s}(\lambda)$ are \emph{even} under spin-reversal transformation, while the spin current is \emph{odd},
\begin{equation}
P j P^{-1} = -j,
\end{equation}
we immediately conclude that all the charges coming from unitary representations are irrelevant for the Drude weight, namely 
$\omega_\beta(j X_{s}(\lambda)) \equiv 0$. For this reason we subsequently consider only the set
$\{Z(\lambda);\lambda\in \mathcal{D}\}$.
Similarly as in the previously considered discrete case, we start by studying the following operator
\begin{equation}
B = \bar{J} - \int_{\cal D} \dd^2 \lambda f(\lambda) Z(\lambda),
\label{eq:B}
\end{equation}
where the integration is over the quasilocality domain ${\cal D}$. It is worth stressing that in general $Z(\lambda)$ are not Hermitian.
Nevertheless, the expectation value of $B^\dagger B$ is always nonnegative
\begin{align}
0 \le \frac{1}{2N}\omega_\beta(B^\dagger B) &= \frac{1}{\beta}D_J \nonumber \\ 
&- \frac{1}{2N}\int_{\mathcal{D}}\dd^2\lambda f(\lambda) \omega_{\beta} (J Z(\lambda))
 - \frac{1}{2N}\int_{\mathcal{D}}\dd^2\lambda \overline{f(\lambda)}\omega_{\beta}(Z(\lambda)^\dagger J)\nonumber \\
&+ \frac{1}{2N}\int_{\mathcal{D}}\dd^2\lambda\int_{\mathcal{D}}\dd^2\lambda' 
\overline{f(\lambda)}f(\lambda') \omega_\beta(Z(\lambda)^\dagger Z(\lambda')).
\label{eqn:lastterm}
\end{align}
We proceed by defining the overlap coefficients of an extensive observable $J$ along the conserved operators in terms
of the \emph{holomorphic} function
\begin{equation}
Z_{J}(\lambda) = \lim_{N\to\infty} \frac{1}{N}\omega_{\beta}(J Z(\lambda)) = \lim_{N\to\infty} \omega_{\beta}(j Z(\lambda)),
\end{equation}
assuming the limit $N\to\infty$ exists. For infinite temperature $\beta\to 0$ the existence of the limit and consequently holomorphicity of $Z(\lambda)$ 
simply follow from the explicit matrix product 
operator expression \eref{eqn:qp}. The limit in the last term of Eq.~\eqref{eqn:lastterm} exists as well, due to pseudolocality of $Z(\lambda)$, 
and can be written in terms of a Hermitian kernel
\begin{equation}
\kappa(\lambda,\lambda') = \lim_{N\to\infty} \frac{1}{N} \omega_\beta(Z(\lambda)^\dagger Z(\lambda')) =
\overline{\kappa(\lambda',\lambda)}, \quad \lambda,\lambda' \in \mathcal{D}.
\end{equation}
Therefore $D_{J}$ should satisfy the inequality
\begin{equation} 
\frac{1}{\beta}D_J \ge F[f] = \int_{{\cal D}}\!\!\dd^2\lambda\, {\rm Re}(Z_J(\lambda)f(\lambda)) - \frac{1}{2} \int_{\mathcal{D}}\dd^2\lambda\int_{\mathcal{D}}\dd^2\lambda'\,\kappa(\lambda,\lambda')\overline{f(\lambda)}f(\lambda'),
\label{eqn:estimate}
\end{equation}
for any $f$. Optimization of the right hand-side with respect to $f$
\begin{equation}
\delta F[f] = {\rm Re}\int\dd^2\lambda\,\overline{\delta\!f(\lambda)}\left\{ \overline{Z_J(\lambda)} -  \int\!\dd^2\lambda' \kappa(\lambda,\lambda')f(\lambda')\right\} = 0,
\end{equation} 
results in the complex Fredholm equation of the first kind for the unknown function $f$,
\begin{equation}
\int_{{\cal D}}\!\dd^2\lambda' \kappa(\lambda,\lambda')f(\lambda') = \overline{Z_J(\lambda)}.
\label{eqn:Fredholm}
\end{equation}
The solution of the above equation can be plugged back to the estimate \eqref{eqn:estimate}, yielding the final
Mazur--Suzuki lower bound
\begin{equation}
D_{J} \ge \frac{\beta}{2}\!\int_{{\cal D}}\!\dd^2\lambda\,f(\lambda) Z_J(\lambda). 
\label{eqn:DAb}
\end{equation}
The bound is manifestly real due to the hermiticity of the kernel.

\subsubsection{Spin Drude weight in gapless $XXZ$ chain.}
\label{spindrude}
The recipe explained above can be readily demonstrated on a paradigmatic example of the high-temperature spin Drude weight for the 
spin current $j=\ii(\sigma^{+} \otimes \sigma^{-} - \sigma^{-}\otimes \sigma^{+})$ in the gapless regime of $XXZ$ model at roots of 
unity anisotropies. There the expression for the kernel reads $\kappa_0(\lambda,\lambda')=K(\bar{\lambda},\lambda')$, with the Hilbert--Schmidt kernel given by Eq.~\eqref{HSKhw}.
The expression for the spin current and matrix product formula for the densities of
$Z(\lambda)$ Eqs.~(\ref{eqn:Zdef1},\ref{eqn:qp}) yield a constant overlap function $Z_J(\lambda) = \ii/4$ and the integral equation \eqref{eqn:Fredholm} can be solved, remarkably, by a simple function
\begin{equation}
f(\lambda) = -\frac{\ii}{\pi} m \sin^2(\pi/m) \frac{1}{|\sin\lambda|^4}.
\label{f}
\end{equation}
Another elementary integral then yields the lower bound \cite{PI13} $D_{J} \geq D_{K}/4$ with
\begin{equation}
D_{K} = \frac{\beta}{4} \frac{\sin^{2}{(\pi l/m)}}{\sin^{2}{(\pi/m)}}
\left(1 - \frac{m}{2\pi}\sin{\left(\frac{2\pi}{m}\right)}\right).
\label{eqn:DK}
\end{equation}
It is noteworthy that the lower bound \eqref{eqn:DK} agrees exactly with the Thermodynamic Bethe Ansatz (TBA) calculation 
\cite{Zotos99,Benz05} at the special (isolated) points of anisotropy at $\eta = \pi/m$, corresponding to $q$-deformation at 
\emph{simple} roots of unity ($l=1$). Since TBA calculation for other values of $l$ seems to be highly
nontrivial and has not yet been performed, we can only conjecture that the bound (\ref{eqn:DK}) is in fact saturating the exact 
value of high-temperature spin Drude weight for a dense set of commensurate anisotropies $\Delta = \cos{(\pi l/m)}$. 
Such a conclusion can also be based on the comparison with numerical results of the state-of-the-art density matrix renormalization group (DMRG) methods 
\cite{Karrasch12,Karrasch13} which indicate no significant deviations from the lower bound $D_{K}$~\cite{KMprivate}.
One obtains similarly good agreement by comparing to exact real-time dynamical simulations with random initial wave-function sampling on smaller systems and perform appropriate finite size scaling analysis\cite{RobinTypicality}.
 
\begin{figure}[htb]
\centering
\includegraphics[width=0.75\textwidth]{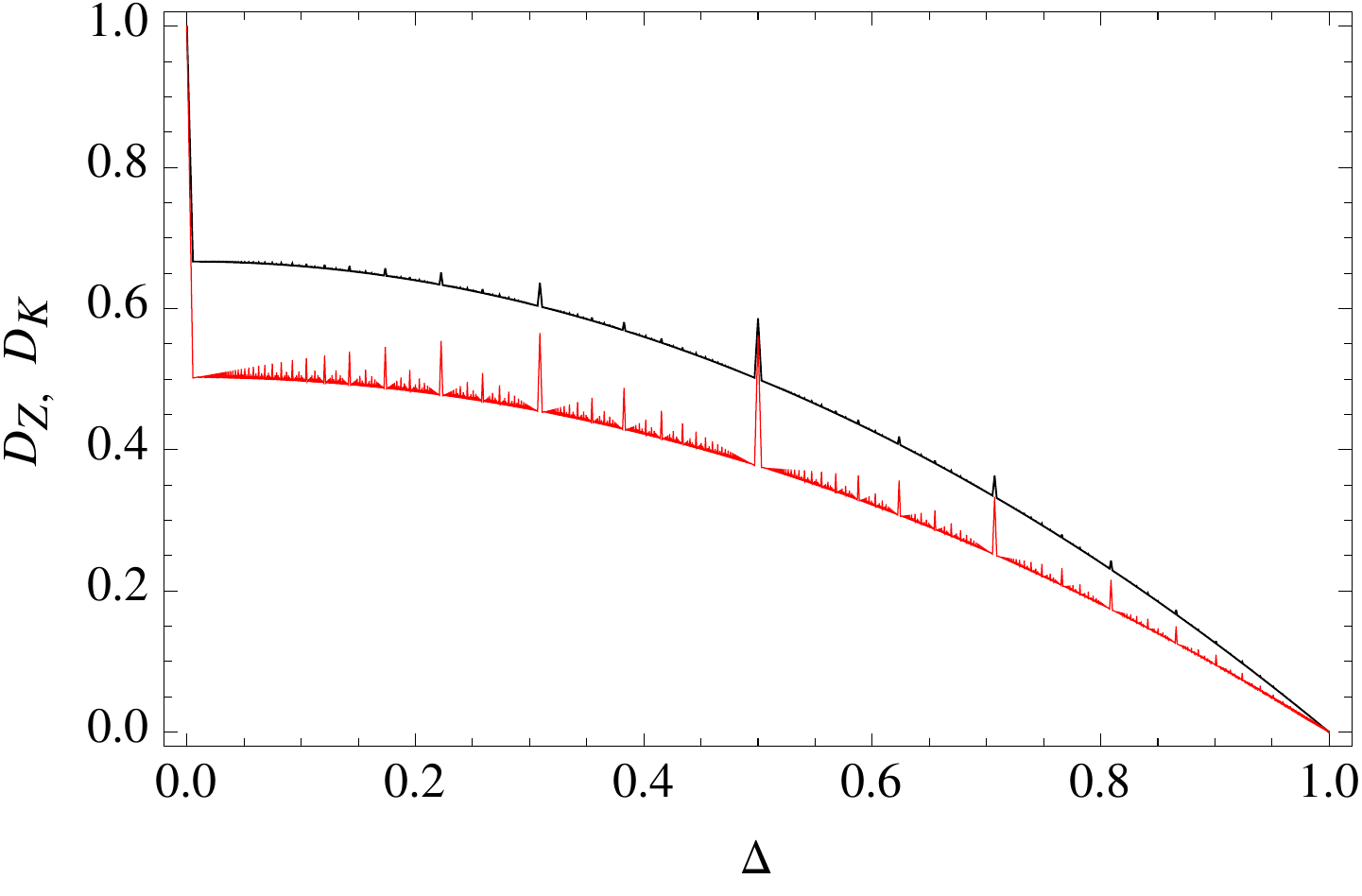}
\caption{Lower bound on the spin Drude weight $D_{K}$ (black, cf. Eq.~\eqref{eqn:DK} as computed in Ref.~\cite{PI13}.
In comparison we show (in red) the bound optimized for a single charge obtained initially in Ref.~\cite{ProsenPRL106}.
In either case the bound exhibits a pronounced fractal-like (nowhere continuous) dependence on parameter $\Delta$.}
\label{fig:Drude}
\end{figure}

\subsubsection{Operator time averaging.}
Saturation of the spin Drude weight bound suggests even a stronger conjecture, namely that the set of quasilocal 
conserved charges $\{Z(\lambda)\}$ is \emph{complete} for a class of local observables that are \emph{odd} under spin reversal for 
an arbitrary root of unity anisotropy. This would imply that an exact equality should be reached in Eq.~\eqref{eqn:lastterm} for the 
optimal weight function $f(\lambda)$ which solves the Fredholm equation \eqref{eqn:Fredholm}, namely 
$\lim_{N\to\infty} (1/N)\omega_\beta(B^\dagger B) = 0$. In a weak sense (with respect to a thermal state $\omega_\beta$) this
statement is equivalent to
\begin{equation}
\bar{J} = \int_{\cal D} \dd^2 \lambda f(\lambda) Z(\lambda).
\label{tav}
\end{equation}

Note that one can use the concept of operator time averaging to formally describe the steady state of $XXZ$ model pierced with a flux $\phi$ and undergoing a small flux quench $\phi \to \phi + \delta_\phi$, namely starting from a thermal density matrix 
$\varrho_\beta$, one may show~\cite{Marcin} that after-quench current carrying steady state is given by the density operator 
\begin{equation}
\bar{\varrho} = \varrho_\beta (\one - \delta_\phi \beta \bar{J}) + {\cal O}(\delta_\phi^2).
\end{equation}

Furthermore, the concept of time-averaged extensive local operators has been used to implement a useful numerical algorithm to 
search for unknown quasilocal charges of an arbitrary locally-interacting lattice model~\cite{MPP15}. One should simply recall that 
for any operator $O$, which is an extensive translational invariant sum of traceless local operators, $\bar{O}$ is by construction a 
pseudolocal conserved operator, or it vanishes in a suitable norm if $O$ is ergodic. Taking a maximal linearly independent set of such 
local extensive operators $\{O_n\}$ up to some maximal order of locality $M\ll N$, enumerated with $n=1\ldots {\cal M}$, ${\cal M} \sim (d^2)^M$, one can define a nonnegative definite HS kernel as the matrix $K_{n,n'} = ( \bar{O}_n, \bar{O}_{n'}) = ( O_n, \bar{O}_{n'})$. The number of {\em independent} pseudolocal conserved operators $\bar{O}_n$, with effective support not larger than $M$, can thus be determined 
as an effective rank of the matrix $K$ with eigenvectors yielding the quasilocal charges expanded in $\{O_n\}$.
Implementation of this method in the case of isotropic $XXX$ model~\cite{MPP15}  gave the first constructive empirical evidence on existence 
of unitary quasilocal charges $X_s$~\cite{IMP15}.

\subsection{Quantum quenches}
\label{subsect:quenches}

Motivated by recent experimental progress in optical 
lattices~\cite{Kinoshita04,Kinoshita06,Hofferberth_Nature07,Gring_Science12,Trotzky_Nature12,Cheneau_Nature12,Langen_Science15,Schmiedmayer_review} and 
a plethora of numerical simulations of strongly correlated matter in low dimensions, a very popular setup studied over the last 
decade is the problem of a `quantum
quench'~\cite{CalabreseCardy,MWNM07,KLA07,IC09,RSMS09,CalabresePRL106,Barmettler09,Calabrese12I,Calabrese12II,CK12}:
at initial time, an ideally isolated (closed) system is prepared in an initial state $\ket{\Psi}$, and subsequently, by a sudden 
change of interactions, let to evolve according to a unitary evolution generated by a post-quench Hamiltonian $H$. The situation which 
is particularly appealing from  the theoretical viewpoint is when $H$ is integrable. Many aspects regarding quantum quenches, ranging 
from classical field theories \cite{DeLuca_review}, conformal field theories \cite{Bernard_review,Calabrese_review}, disordered systems \cite{Vasseur_review}, Luttinger model \cite{Cazalilla_review}, to integrable lattice systems \cite{Vidmar_review,Essler_review,Caux_review} are discussed in the reviews of the present volume.

\subsubsection{Complete Generalized Gibbs Ensembles.}

One of the pivotal questions is to understand the process of equilibration from the microscopic perspective~\cite{Deutsch91,Polkovnikov_colloquium}.
In homogeneous quantum systems with generic interactions the relaxation towards canonical Gibbs 
ensemble is typically explained in the framework of the {\em eigenstate thermalization 
hypothesis}~\cite{Deutsch91,Srednicki94,RigolNature}, which states that eigenstates which are close in energy give approximately the 
same values of local correlation functions. The situation
with integrable interactions is however different as time-evolution is severely constrained due to the existence of
a macroscopic number of local (and quasilocal) conserved quantities.

It has been conjectured in Refs.~\cite{Rigol_PRA06,Rigol_PRL07} that statistical properties of local quantities in many-particle 
quantum systems which possess an `extensive number' of conserved local charges $I_n$ should comply with predictions of a
Generalized Gibbs Ensemble~\cite{Rigol11,KWE11,CalabresePRL106,Calabrese12I,Fioretto12,CK12,Vidmar_review} , given by a formal expansion
\begin{equation}
\rho_{\rm GGE}\sim \exp{\left(-\sum_{n}\beta_{n}I_{n}\right)}.
\label{eqn:GGE}
\end{equation}
The `GGE conjecture' asserts that the ergodic average of an operator $A$ with a finite support  
\begin{equation}
\langle A\rangle_\psi=\lim_{T\to\infty}\frac{1}{T}\int_0^T \langle\psi(t)|A|\psi(t)\rangle dt,
\label{eqn:time_avg_observable}
\end{equation}
can be reproduced by tracing with respect to an appropriate GGE of the form \eqref{eqn:GGE}, with the `chemical potentials' $\beta_{m}$ being determined from expectation values of the charges with respect to the initial state.

A great body of work has already been devoted to applicability of the GGE in
non-interacting models~\cite{CalabresePRL106,Calabrese12I,Calabrese12II,Fagotti13,Fagotti14}, and a closely related
phenomenon of prethermalization~\cite{KWE11,Gring_Science12,EKMR14,BF15,Bertini15}.

Explicit verification of the GGE paradigm in a truly interacting quantum integrable models required a bit more effort though.
Initial studies focused on Heisenberg $XXZ$ chain and compared predictions of truncated GGEs made of hitherto known local charges
against numerical results for the time-evolved local observables~\cite{FE13,Pozsgay2013,FCEC14}.
First exact results have been obtained for the case of the Lieb--Liniger model in Ref.~\cite{DeNardisPRA14} by resorting
to the so-called {\em quench action} method, developed previously in \cite{CE13} (cf. \cite{Caux_review} for a review).
In this approach, a generalized free energy functional is constructed which incorporates the restrictions imposed by the
initial condition in  the form of an exact overlap coefficient. By employing TBA
framework~\cite{YY69,Takahashi71,TS72,TakahashiBook}, the saddle-point of such a functional yields the sought for steady-state
ensemble via coupled non-linear integral equations for a set of variational variables. These thermodynamic variables are, as we shall shortly discuss, a set of analytic functions representing distributions of Bethe strings.

Sometimes, e.g. for certain simple product states, the overlap formulas which enter as an input to quench action method can be 
evaluated explicitly~\cite{Pozsgay_overlaps}. Two independent studies~\cite{WoutersPRL,Pozsgay_GGE} unambiguously demonstrated
that GGEs composed from only the hitherto known local charges fail to recover the exact results
(see also Refs.~\cite{Brockmann14,Mestyan14}). The failure has been related to the fact that strictly local charges 
Eq.~\eqref{eqn:ultralocal} do not provide enough information to determine the distributions of the bound states which are present
in an initial state~\cite{GA14}. The results of these studies hinted on the presence of additional (sufficiently local) conservation 
laws in the unitary (or spin-reversal symmetric) sector.

\subsubsection{String-charge duality.}
\label{sec:string-charge}

Here we explain, following Ref.~\cite{StringCharge}, the connection between the spectra of quasilocal charges $X_{s}$ and 
distributions of Bethe strings. The latter should be interpreted as thermodynamic particle content of an integrable lattice theory.
Hence, the main task shall be to extract the large-$N$ behaviour of eigenvalues of $T$-operators.
A convenient tool to achieve this is
to employ the Baxter $Q$-operator~\cite{Baxter73,BaxterBook,shortcut} and exploit the fact that its eigenvalues are
given by a (deformed) polynomial with zeros coinciding with Bethe roots. Below we present the main steps by 
specializing to the \emph{gapped} regime.

\paragraph{Bethe equations and string hypothesis.}
To set the stage we need to briefly describe how to characterize the spectra of integrable lattice models in the
thermodynamic regime. The elementary building block of an integrable model is the single-particle $S$-matrix $S_{1}$ which
for the $XXZ$ model reads
\begin{equation}
S_{1}(\lambda,\mu)\equiv S_{1}(\lambda-\mu) = \frac{\sin{(\lambda-\mu-\tfrac{\ii \eta}{2})}}{\sin{(\lambda-\mu+\tfrac{\ii \eta}{2})}}.
\label{eqn:scattering_matrix}
\end{equation}
From a scattering theory point of view, the spectral parameters $\lambda$ and $\mu$ pertain to rapidities of the two quasi-particles 
involved in a scattering event. For composite objects which
consist of $j$ excitations -- commonly referred to as the $j$-strings -- a set of fused scattering matrices $S_{j}$ are introduced
\begin{equation}
S_{j}(\lambda)=\frac{\sin{(\lambda-j\tfrac{\ii \eta}{2})}}{\sin{(\lambda+j\tfrac{\ii \eta}{2})}},\qquad j=1,2,\ldots
\end{equation}
Scattering among two different types of strings is governed by string-to-string scattering matrices
\begin{equation}
S_{j,k}(\lambda)=S_{|j-k|}(\lambda)S_{j+k}(\lambda)\prod_{i=1}^{{\rm min}(j,k)-1}S^{2}_{|j-k|+2i}(\lambda).
\end{equation}
With the aid of scattering matrices, the Bethe Ansatz equations, representing a quantization condition for quasi-particle
rapidities $\lambda_{j}$ in a periodic system, are cast in the form
\begin{equation}
e^{\ii p(\lambda_j)N}\prod_{k=1}^{M}S_{1,1}(\lambda_{j}-\lambda_{k})=-1,\qquad j=1,2,\ldots M.
\label{eqn:Bethe_equations}
\end{equation}
Here $M$ is the number of Bethe roots (related to the magnetization of the eigenstate) and $p(\lambda)$ encodes the 
momentum of an elementary excitation on top of a ferromagnetic vacuum state,
\begin{equation}
e^{\ii p(\lambda)}=\frac{\sin{(\lambda+\tfrac{\ii \eta}{2})}}{\sin{(\lambda-\tfrac{\ii \eta}{2})}}.
\label{eqn:momentum_def}
\end{equation}

The \emph{string hypothesis}~\cite{Takahashi71,Gaudin71,TS72,TakahashiBook} states that in the large-$N$ limit the Bethe roots
(i.e. solutions $\lambda_{j}$ to Eq.~\eqref{eqn:Bethe_equations}) for a typical eigenstate become equidistantly displaced in the 
imaginary direction in the rapidity complex-plane,
\begin{equation}
\{\lambda^{k,j}_{\alpha}\}\equiv \{\lambda^{k}_{\alpha}+(k+1-2j)\tfrac{\ii \eta}{2}|j=1,2,\ldots k\}.
\label{eqn:strings}
\end{equation}
Such string formations physically correspond to bound states of magnons.
By partitioning the Bethe roots in terms of strings, Bethe equations \eqref{eqn:Bethe_equations} can be rewritten in terms of string 
centres $\lambda^{k}_{\alpha}\in \RaR$. Thus, taking their logarithmic form and considering the thermodynamic limit when
string centres get smoothly distributed along the real axis, we arrive at the following non-linear coupled integral 
equations~\cite{Takahashi71,Gaudin71,TakahashiBook}
\begin{equation}
\rho_{j}(\lambda)+\rhoh_{j}(\lambda) = a_{j}(\lambda) - \sum_{k}\int_{-\pi/2}^{\pi/2}\frac{\dd \mu}{2\pi}\,a_{j,k}(\lambda-\mu)\rho_{k}(\mu),
\label{eqn:string_Bethe_equations}
\end{equation}
known as the Bethe--Yang equations for the strings. The integral kernels in Eq.~\eqref{eqn:string_Bethe_equations} 
are given by the derivatives of scattering phase shifts and the corresponding string-to-string phase shifts
\begin{equation}
a_{j}(\lambda)=-\ii \partial_{\lambda}\log\,S_{j}(\lambda),\quad 
a_{j,k}(\lambda)=-\ii \partial_{\lambda}\log\,S_{j,k}(\lambda),
\label{eqn:a_kernels}
\end{equation}
in the respective order. One of the advantages of Eq.~\eqref{eqn:string_Bethe_equations} in comparison to the finite-volume counterpart
is that we no longer have to deal with a complicated set of quantized quasi-momenta (encoded by Bethe roots $\lambda_{j}$).
Instead, now quasi-momenta take values in the continuum which allows us to cast the description in terms of analytic distributions 
$\rho_{j}(\lambda)$ which count the number of Bethe strings whose centres occupy an interval $[\lambda,\lambda+ \dd\lambda]$. 
Similarly, $\rhoh_{j}(\lambda)$ denote the complementary variables, parametrizing distributions of Bethe holes (the positions of 
string centres which are in principle available, but remain unoccupied).

\paragraph{Thermodynamic spectra.}
To obtain the spectra of charges $X_s$ we make use of representation~\eqref{eqn:Q-representation}. By neglecting the contributions
which are subleading in $N$ we have
\begin{equation}
\bra{\{\lambda_{j}\}}X_{s}(\lambda)\ket{\{\lambda_{j}\}} =
-\ii \partial_{\lambda}\log\,\frac{\mathcal{Q}^{[-2s]}(\lambda)}{\mathcal{Q}^{[2s]}(\lambda)},
\label{eqn:X_leading_term}
\end{equation}
where $\ket{\{\lambda_j\}}$ denote a Bethe eigenstate parametrized by a set of roots $\{\lambda_j\}$.
Working under the `string hypothesis' (cf. Sec.~\ref{sec:string-charge}), the spectra of quasilocal charges $\mathcal{X}_{s}$, 
\begin{equation}
\mathcal{X}_{s}(\lambda)=\lim_{N\to \infty}\frac{1}{N}\bra{\{\lambda_j\}}X_{s}(\lambda)\ket{\{\lambda_j\}},
\end{equation}
can be readily expressed in terms of densities of string centers $\rho_{j}(\lambda)$. Specifically,
by plugging the expression for the spectrum (cf. Eq.~\eqref{eqn:Q_spectrum} in Eq.~\eqref{eqn:X_leading_term}),
we arrive at~\cite{StringCharge}
\begin{equation}
\mathcal{X}_{s}(\lambda) = \sum_{k}\int_{-\pi/2}^{\pi/2}\frac{\dd \mu}{2\pi}\,G_{2s,k}(\lambda-\mu)\rho_{k}(\mu).
\label{eqn:charge-string}
\end{equation}
The set of kernels $G_{2s,k}$ can be expressed using scattering matrices among the strings
\begin{equation}
G_{2s,k}(\lambda) = \sum_{j=1}^{k}-\ii \partial_{\lambda}\log\,S_{2s}(\lambda+(k+1-2j)\tfrac{\ii \eta}{2})
= \sum_{j=1}^{{\rm min}(2s,k)}a_{|2s-k|-1+2j}(\lambda).
\label{eqn:Green_function}
\end{equation}

Let us introduce a \emph{discrete d'Alembert operator} $\square$, whose action on any set of objects
$f_{s}\equiv f_s(\lambda)$ (with $s=\tfrac{1}{2}\ZZ_{+}$) which are \emph{analytic} inside the physical strip $\mathcal{P}_{\eta}$ is prescribed by
\begin{equation}
\square f_{s} = f^{+}_{s} + f^{-}_{s} - f_{s-\frac{1}{2}} - f_{s+\frac{1}{2}}.
\label{eqn:dAlembert}
\end{equation}
By acting with the d'Alembertian on the kernel functions from Eq.~\eqref{eqn:Green_function} we conclude that
\begin{equation}
\square G_{j,k}(\lambda)= \delta_{j,k}\,\delta(\lambda),\qquad j=2s \in \mathbb{N}.
\end{equation}
This result allows us to interpret $G_{j,k}$ as a discrete 2D \emph{Green's function} of the `wave operator' $\square$.
The relation Eq.~\eqref{eqn:charge-string} can be readily inverted, enabling to express the entire set of density functions
$\rho_{j}(\lambda)$ in terms of eigenvalues of the charges $X_{s}(\lambda)$ as~\cite{StringCharge}
\begin{equation}
\rho_{2s}(\lambda) = \square \mathcal{X}_{s}(\lambda).
\label{rho2s}
\end{equation}
The distributions of holes $\rhoh_{2s}(\lambda)$ can be obtained in a similar fashion~\cite{Ilievski_GGE,StringCharge},
\begin{equation}
\rhoh_{2s}=a_{2s}-\mathcal{X}^{+}_{s}-\mathcal{X}^{-}_{s}.
\label{rho2sb}
\end{equation}
In the scope of quantum quench applications, a set of densities $\rho_{2s}$ provides a complete description of local
correlation functions (cf.~\cite{Mestyan14,Mestyan15}).

Finally, let us make a brief account on the \emph{gapless regime} as well. Although the string hypothesis in the $|\Delta|<1$ regime
can still be formulated, taking the deformation parameter $q=e^{\ii\eta}$ from the unit circle makes the analysis
rather cumbersome and technically involved. The string content in the gapless regime for an arbitrary value of anisotropy has been 
derived in Ref.~\cite{TS72}. Due to limited space we do not attempt to review it here. We nevertheless wish to point out the three 
principal differences in comparison with the situation in the gapped case: (i) string configurations acquire (beside the string length) an 
additional parity label $u\in\{\pm 1\}$ (see Sec.~\ref{gapless}),
(ii) the allowed string lengths depend strongly (and discontinuously) on $\eta$, and (iii) at root of unity value of $q$ the number of 
allowed distinct string types is always \emph{finite}.
Moreover, in the spirit of string-charge duality, the number of (dynamical) strings should still be in a bijective correspondence with the number of quasilocal charges, as discussed in Sec.~\ref{gapless}. To complete our example for  $\eta=3\pi/7$, where the charge content is given by a set \eqref{Xcontent}, we provide the corresponding string 
content: $${(1,+),(1,-),(3,+),(5,-)}.$$

Below we explain a computational scheme to determine the densities of Bethe roots from the eigenvalues of $X_{s}(\lambda)$.
This can be done, in contrast to a more common practice, without ever resorting to the variational approach based on a generalized 
free energy functional. The manifest locality of quasilocal charges $X_{s}(\lambda)$ in the spin basis (cf. Eq.~\eqref{eqn:X_product_from}) greatly simplifies this task and allows us to resort to rather standard techniques.

\subsubsection{Evaluation of charges.}

In this section we address the problem of computing expectation values of the quasilocal charges $X_{s}$
with respect to a generic\footnote{Strictly speaking, we are implicitly assuming that our reference state is `local', i.e.
is compatible with cluster decomposition principle~\cite{WeinbergBook,SC14}. In this case we are able to express $\ket{\Psi}$ in the
thermodynamic limit as a single macrostate (a state given by prescribing distributions of Bethe strings).}
pure state $\ket{\Psi}$. While performing this task in full generality remains out of reach at the moment,
we make a restriction to a class of matrix product states where an efficient implementation is possible. In what follows we 
essentially recast the results of Refs.~\cite{FE13,FCEC14} in the present language.

In order to keep the level of technicality at a minimum, we shall in addition restrict ourselves only
to \emph{periodic product states}
\begin{equation}
\ket{\Psi}=\ket{\psi}^{\otimes N/N_{p}},
\end{equation}
where $\ket{\psi}$ is a state on the block of $N_p$ spins and $N_{p}\in \NN$ is the periodicity of the state.
Our aim is to compute
\begin{equation}
\mathcal{X}^{\Psi}_{s}(\lambda) = \lim_{N\to \infty}\frac{1}{N}\bra{\Psi}X_{s}(\lambda)\ket{\Psi}.
\label{eqn:X_eigenvalue}
\end{equation}
Due to the product structure of $\ket{\Psi}$ we can make use of standard transfer matrix techniques.
The first step is to introduce a \emph{boundary partition function}
\begin{equation}
\mathcal{Z}^{\Psi}_{s}(\lambda,\mu)=\lim_{N\to \infty}\frac{1}{N}\Tr_{\rm a}\left\{\UU^{\Psi}_{s}(\lambda,\mu)^{N/N_p}\right\},
\label{eqn:boundary_partition_function}
\end{equation}
which is given by iterating a one-step auxiliary propagator,
\begin{equation}
\UU^{\Psi}_{s}(\lambda,\mu) = \bra{\psi}\LL_{s}(\lambda,\mu)^{\otimes N_p}\ket{\psi}.
\label{eqn:U_propagator}
\end{equation}
Subsequently we evaluate
\begin{equation}
\mathcal{X}^{\Psi}_{s}(\lambda)=-\ii \partial_{\mu}\mathcal{Z}^{\Psi}_{s}(\lambda,\mu)|_{\mu=\lambda}.
\label{eqn:X_on_state}
\end{equation}
We note that partition functions given by Eq.~\eqref{eqn:boundary_partition_function} are in essence merely the contracted
quantum transfer operators $X_{s}(\lambda,\mu)$ from Eq.~\eqref{eqn:X_as_MPO} (depicted in Fig.~\ref{fig:X-fig})
where in the vertical direction we project onto components determined by the reference state $\ket{\psi}$. Such a
contraction over one period $N_{p}$ yields the propagator from Eq.~\eqref{eqn:U_propagator}.

The construction sketched above can be adapted for general translational invariant matrix product states
(see Refs.~\cite{FE13,FCEC14,StringCharge}).

\subsubsection{Closed-form results.}
\label{subsect:closedform}

In Sec.~\ref{subsect:hirota} we already mentioned that higher-spin $T$-operators constitute the \emph{canonical} solution to
Hirota difference equations (alias the $T$-system). However, Hirota difference equations admit different solutions as well. 
Remarkably, there exists a class of initial conditions which relax to equilibrium steady states
(specified by a collection of density functions $\rho^{\Psi}_{j}$) which can be cast as distinct solutions of
the Hirota equations. Below we mention two particular examples, which have been previously studied in the literature, when 
equilibrium states admit simple representative product states: (i) a spin-singlet dimerized state
$\ket{\rm D}=\tfrac{1}{\sqrt{2}}(\ket{\ua \da}-\ket{\da \ua})^{\otimes N/2}$ and (ii)
N\'eel state $\ket{\rm N}=\ket{\ua \da}^{\otimes N/2}$. In other words, these two states can be understood as members
of a basin of attraction for equilibrium states which assume parametrizations in terms of non-canonical solutions
to the functional relation of the $T$-system.

In the following, we use a small font to explicitly distinguish \emph{non-canonical} $t$-functions and
$q$-functions, $t_s(\lambda),q(\lambda)$, from the canonical objects, i.e. fused transfer matrices $T_s(\lambda)$ and
Baxter $Q$-operator $Q(\lambda)$ defined in Sec.~\ref{subsect:hirota}.
By relaxing the constraint $t_{0}=\varphi^{-}$, the linear auxiliary problem associated to the Hirota equation takes the form
\begin{equation}
t_{s+\frac{1}{2}}q^{[2s]}-t^{-}_{s}q^{[2s+2]}=\varphi^{[2s]}\ol{q}^{[-2s-2]},
\end{equation}
and can be explicitly solved as
\begin{equation}
t_{s}=t^{[-2s]}_{0}\frac{q^{[2s+1]}}{q^{[-2s+1]}}+q^{[2s+1]}\ol{q}^{[-2s-1]}
\sum_{k=1}^{2s}\frac{\varphi^{[2(k-s)-1]}}{q^{[2(k-s)-1]}q^{[2(k-s)+1]}}.
\end{equation}
In the present case, $q$-functions can be considered as auxiliary complex-valued functions which contain the information about
the equilibrium state at hand, closely related to auxiliary functions which enter in non-linear integral equations in the scope
of the Quantum Transfer Matrix (QTM) method~\cite{Klumper92,Klumper93,Klumper02}.

To keep things simple, we specialize below only to the isotropic point $\Delta=1$.
For the dimerized state $\ket{\rm D}$ the solution is remarkably simple and reads $q^{\rm D}(\lambda)=\lambda^{2}$.
These results generate the entire tower of $t$-functions
\begin{equation}
t^{\rm D}_{s}(\lambda) = (2s+1)\lambda,\qquad s\in \tfrac{1}{2}\mathbb{Z}_{+},
\end{equation}
which can, in turn, be mapped to $y$-functions $y^{\Psi}_{j} = \rhoh^{\Psi}_{j} / \rho^{\Psi}_{j}$,
\begin{equation}
y^{\rm D}_{j}(\lambda) = \frac{((j+1)^{2}-1)\lambda^{2}}{(\lambda+(j+1)\tfrac{\ii}{2})(\lambda-(j+1)\tfrac{\ii}{2})},\qquad
j\in \mathbb{Z}_{+}.
\end{equation}
As we have already explained (cf. Eqs.\eqref{rho2s}, \eqref{rho2sb}), the $y$-functions can be related to 
expectation values of the charges on the state $\ket{\rm D}$,
\begin{equation}
\mathcal{X}^{\rm D}_{\frac{1}{2}}(\lambda) = \frac{5+2\lambda^{2}}{4(1+\lambda^{2})^{2}},\quad
\mathcal{X}^{\rm D}_{1}(\lambda) = \frac{4(17+4\lambda^{2})}{(9+4\lambda^{2})^{2}},\quad
\mathcal{X}^{\rm D}_{\frac{3}{2}}(\lambda) = \frac{3(13+2\lambda^{2})}{4(4+\lambda^{2})^{2}}.
\end{equation}
We remark that in practice one should work in the opposite direction: by computing a few initial values of
the charges and employing the string-charge relationship one can explicitly check whether the $y$-functions fulfil the $Y$-system
hierarchy. It is not clear if a general systematic procedure exists to directly determine which states admit a
description in the $Y$-system format. The analogous expressions for the N\'eel state (including the expressions for the gapped case) 
are provided in Ref.~\cite{StringCharge}.

To conclude this section, let us stress that the unitary charges $X_{s}$ from the compact sector cannot be
sufficient for characterizing \emph{non-equilibrium} steady states, i.e. states which exhibit particle currents.
In this situation, the quasilocal charges $Z(\lambda)$ which break the spin-reversal invariance have to be 
included~\cite{Marcin,NMO16}. To the best of our knowledge, it remains presently unknown how the $Z$-charges act on Bethe eigenstates.

\subsection{Steady states of boundary-driven chains}
\label{sec:ness}

Quantum transport is typically studied in the framework of the linear response theory.
An alternative way is to adopt an open system perspective.
A simple effective setup for that is to use the approach of non-unitary evolution equations which are commonly referred to as
\emph{quantum master equations}. A central concept here is a Markovian (time-local) evolution
\begin{equation}
\varrho(t)=e^{\hat{\mathcal{L}}t}\varrho(0),
\label{eqn:non-unitary_evolution}
\end{equation}
which preserves the \emph{trace} and \emph{positivity} of density operators $\varrho$ at any time. The generator $\hat{\mathcal{L}}$ is of 
Lindblad form and acts linearly on density matrices as
\begin{equation}
\hat{\mathcal{L}}\varrho = -\ii[H,\varrho] +  \sum_{k}\left(2A_{k}\varrho A^{\dagger}_{k} -
\{A^{\dagger}_{k}A_{k},\varrho\}\right),
\label{eqn:Lindblad_generator}
\end{equation}
where $H$ encompasses all interactions attributed to the unitary part of the 
process, and the set $\{ A_{k} \}$ contains the Lindblad `jump operators' which are used to model dissipative processes.
For a comprehensive introduction on the Lindblad equation formalism we refer the reader to Refs.~\cite{BreuerBook,HuelgaBook}.

In Refs.~\cite{ZnidaricPRL11,ZnidaricJSTAT11,ProsenPRL106,ProsenPRL107,KPS13,IZ14,IP14} Lindblad equation has been used to `drive' a 
quantum many-body system far from equilibrium. Two common scenarios describe the situations where the Lindblad bath operators simulate (a) 
dephasing noise due to uncontrolled degrees of freedom in the bulk, or (b) particle/magnetic reservoirs with different chemical potentials/magnetizations attached at the system's boundaries.

General instances can be studied by adapting a time-dependent DMRG technique to the Liouville dynamics~\cite{PZ09}. On the flip side, 
certain interesting situations permit an exact analytic description, the most notable example being non-interacting particles 
experiencing Gaussian noise which can be treated in a unified manner within the formalism of `third 
quantization'~\cite{3Q,Zunkovic10}.
While deriving exact solutions for the full Liouvillian dynamics of an interacting system remains an open challenge up to date, 
certain steady state density operators, i.e. fixed points of Liouvillian dynamics, which allow for an efficient matrix product form 
have been found and investigated (the first non-trivial example being perhaps the situation of noninteracting particles with bulk dephasing noise~\cite{ZnidaricJSTAT10,ZnidaricJPA10}). In some sense, one can understand these as quantum counterparts of their more 
popular classical cousins known as asymmetric simple exclusion processes~\cite{Derrida93,Derrida07}.

For the Heisenberg spin-$1/2$ chain, a model under scrutiny in this review, driven by incoherent in/out boundary processes: $A_1=\sqrt{\Gamma}\sigma^+_1$, 
$A_2=\sqrt{\Gamma}\sigma^-_2$, the steady state in the weak-coupling limit has been constructed first in Ref.~\cite{ProsenPRL106}, and later on extended to the non-perturbative regime in Ref.~\cite{ProsenPRL107}.
What is remarkable, and perhaps somewhat surprising as well, is that the density operator of the {\em current carrying} steady state found in Ref.~\cite{ProsenPRL106}
is a fully mixed state perturbed with an operator of the non-unitary quasilocal family, namely 
\begin{equation}
\varrho_\infty = \varrho(t\to\infty) \sim \one + \frac{\ii \Gamma}{2}\left(Z^{\rm vac}\left(\pi/2\right)-Z^{\rm vac}\left(\pi/2\right)^\dagger\right) + {\cal O}(\Gamma^2).
\label{eq:rhoinf}
\end{equation}
The only distinction from the conserved operators given by Eq.~\eqref{eqn:Zdef0} is that instead of taking the trace over the auxiliary 
space the adequate transfer matrix is now defined as an expectation value in the highest-weight state (vacuum)
\begin{equation}
T^{{\rm vac}}_{s}(\lambda)=\left(\frac{\sin{(\lambda)}}{\sin{(\lambda + s\eta)}}\right)^N \bra{0}\mathbf{L}_{s}(\lambda)^{\otimes N}\ket{0},
\end{equation}
where the Lax operator is taken from Eq.~\eref{eqn:lax}.
Consequently, the local operator expansion of the open boundary charge $Z^{\rm vac}$ is given with the same formula as before, Eq.~\eqref{eqn:Zdef1}, where the shift $\hat{\cal S}^{x}$ no longer acts periodically (meaning that the sum over $x$ runs only up to $N-r$). While the vacuum transfer matrices and the derived quasilocal charges still mutually commute, $[T^{{\rm vac}}_{s}(\lambda),T^{{\rm vac}}_{s'}(\lambda')]=0$, and
$[Z^{\rm vac}(\lambda),Z^{\rm vac}(\lambda')]=0$, the manifest absence of translational invariance breaks the conservation property,
\begin{equation}
\begin{split}
[H,T^{{\rm vac}(N)}_{s}(\lambda)] &= \frac{2 \sin(\eta)}{\sin(\lambda+s\eta)}\left( b \otimes T^{{\rm vac}(N-1)}_{s}(\lambda) + T^{{\rm vac}(N-1)}_{s}(\lambda) \otimes b\right), \\
[H,Z^{\rm vac}(\pi/2)] &= \sigma^{\z}_{1} - \sigma^{\z}_{N}.
\end{split}
\label{eqn:comm_HV}
\end{equation}
where the Hamiltonian of the anisotropic Heisenberg chain is now taken with \emph{open} boundary conditions and $b=\sigma^\z \sin(\lambda)\sin(\eta s) - \sigma^0 \cos(\lambda)\cos(\eta s)$ is a boundary operator.
The first identity follows straightforwardly from the RLL relation \eqref{eqn:RLL_relation}, while the second one follows from the first one after taking the derivative $\partial_s|_{s=0,\lambda=\pi/2}$.
Note that the second line of Eq.\eqref{eqn:comm_HV} has a form of a conservation law, i.e. time-derivative of $Z^{\rm vac}$ in a finite volume equals net 
surface currents, where the spin density $\sigma^{\z}_{x}$ plays the role of the formal `current'.
In spite of `almost-conservation' in a finite volume, it has been rigorously shown in Ref.~\cite{IP13}, resorting to
quasilocality and Lieb--Robinson causality bounds, that Eq.~\eqref{eqn:comm_HV} yields a conserved quantity in the
thermodynamic limit and in effect provides an equivalent set of quasilocal conservation laws as those introduced in 
Sec.~\ref{subsect:hw}.

Moreover, it can be shown (see \cite{ProsenPRL107,KPS13,PI13}, and \cite{Prosen_review,Ilievski_thesis} for a review) that the vacuum 
transfer matrix generates an exact, non-perturbative steady state density operator via the purification ansatz
\begin{equation}
\varrho_\infty = \frac{\Omega \Omega^\dagger}{\Tr(\Omega \Omega^\dagger)},
\quad
\Omega(\lambda) = (T^{\rm vac}_s(\lambda))^T,
\label{eq:pur}
\end{equation}
if one sets the spectral parameter and identifies the noise strength $\Gamma$ with a complex auxiliary spin $s$, in either one of the following two ways
\begin{equation}
\lambda = \frac{\pi}{2}, \quad  \tan (\eta s) = \frac{\ii \Gamma}{2\sin(\eta)} \qquad {\rm or}\qquad
\lambda = 0, \quad \cot (\eta s) = \frac{\ii \Gamma}{2\sin(\eta)}.
\end{equation}
These two assignments yield identical steady-state density operator \eref{eq:pur}.
In light of the fact that in the canonical $\sigma^\z_x$--basis the amplitude operator $\Omega$ becomes a strictly upper-triangular matrix~\cite{ProsenPRL107}, the ansatz \eqref{eq:pur} can also be understood as a \emph{many-body Cholesky factorization}.
The ansatz~\eqref{eq:pur} in fact exactly solves a much larger set of boundary-driven Lindblad equations, namely taking an arbitrary 
pair of asymmetric (left/right) noise strengths $\Gamma_{\rm L,R}$ and adding arbitrary boundary magnetic fields in $z-$direction 
$h_{\rm L,R}$ uniquely parametrizing two complex variables $s,\lambda$ (see Ref.~\cite{Prosen_review}).
We note that the notation of this section was adapted for the regime $|\Delta| < 1$ where
quasilocal $Z$-charges have an effect and the corresponding transport is ballistic. To find the gapped counterparts one has to make a substitution $\eta \to \ii \eta$, or replace trigonometric functions with the corresponding hyperbolic functions.

It is also perhaps instructive to stress that the vacuum charges $Z^{\rm vac}(\lambda)$ are manifestly nondiagonalizable objects with
a nontrivial Jordan structure. For example, the spectrum of $Z^{\rm vac}(\pi/2)$ only consists of $\{0\}$, hence the operator is
nilpotent for any finite volume, but nevertheless generates a highly nontrivial steady state.
The approach of generating quasilocal almost-conserved charges as perturbative solutions to boundary-driven Lindblad equations
may be useful also in other integrable models (see Ref.~\cite{Prosen_review} for a review) and should perhaps be further explored in 
future.

\section{Discussion}
\label{discussion}

\subsection{Future prospectives}

\paragraph{Spin chains.}
Even though applications of quasilocal conservations laws which we covered in this review have been fully
concentrated on the paradigmatic case of the Heisenberg $XXZ$ model, it is natural to expect
that analogous quantities exist for a much broader class of integrable models (see e.g. Ref.~\cite{PV16} for a recent application to 
gapless spin-$1$ chains).
The simplest extensions should involve quantum lattice models associated with the so-called fundamental
solutions to Yang--Baxter equation, with underlying symmetry algebras based on Lie algebras of higher rank and their quantizations 
(deformations). Additionally, supersymmetric cousins (e.g. $t-J$ model, EKS model)
shall be of interest in paving the way towards the celebrated Hubbard model~\cite{Hubbard63,Shastry88,HubbardBook}.
Note that a novel family of transfer matrices which violate particle-hole symmetry and correspond to non-unitary auxiliary 
representations has recently been proposed for the Hubbard model~\cite{PP15}, based on preserving the integrability of the associated 
boundary driven master equation~\cite{PPRL14}. A possibility of generating quasilocal conserved quantities remains to be 
explored.

For all models mentioned above it is well-known that thermodynamic spectra can be partitioned into Bethe
root compositions (strings) which pertain to bound states of elementary excitations.
In order to ensure that macrostates (e.g. thermal states and their generalizations) are mutually distinguishable, the number of
distinct particle types (see Refs.~\cite{Zamolodchikov91,KN92,KNS94,KSZ08,Gromov09,KL10}) has to be matched with the number of 
independent families of (quasi)local charges.

Another example of an integrable theory which has recently drawn a great deal of attention
due to its experimental significance is a Bose gas with $\delta$-like repulsive interactions,
known better as the Lieb--Liniger model~\cite{LL63} (Nonlinear Schr{\"o}dinger equation in the language of
second quantization). Yet, in spite of its wide popularity, the second-quantized form of the
entire tower of local charges have not been obtained explicitly so far~\cite{Davies90,DK11}.
Besides, there also exist certain obstructions which are intimately related to pathological UV divergences as discussed in 
Refs.~\cite{KormosPRB13,DeNardisPRA14}. In Ref.~\cite{EMP15} the authors attempted to overcome
the difficulties by `mildly' relaxing the conventional form of locality. Alternatively, there is an option
to employ a suitable integrable regularization (e.g. by introducing a UV cutoff)
allowing to treat the lattice counterparts in the thermodynamic limit first, then construct/compute the observables,
and take the continuum limit only at the end (see e.g. Ref.~\cite{KormosPRB14}).
The effectively local, or quasilocal conserved charges could then be derived using the methods presented in this review.

\paragraph{Integrability in AdS/CFT correspondence.}
One of the hallmarks of theoretical physics of the last two decades is the discovery of the gauge-gravity 
duality~\cite{Maldacena,GPK98}. The most prominent example of this is the celebrated $\mathcal{N}=4$ superconformal Yang--Mills theory 
which is conjectured to be dual to a certain type of the superstring theory \cite{AdSCFT_review}.
One of its surprising features is that the Heisenberg spin Hamiltonian arises in the scalar sector as 
the one-loop approximation of the dilatation operator. The scattering matrix behind the
scenes has an exceptional structure and turns out to be tightly related to the famous Fermi--Hubbard model and some other related models of strongly-correlated electrons dubbed as the Hubbard--Shastry model~\cite{FQ12}. Constructions and physical applications of 
quasilocal charges have not yet been explored in this context.

\paragraph{Correlation functions.}
In this work we have not devoted any attention to the problem of calculating static and dynamic correlation 
functions of local observables, a task which typically represents an ultimate goal of any successful computational framework.
A systematic procedure to encompass a wide range of interacting integrable theories in a universal and robust language still awaits to be developed. In this review, we have only addressed the problem of determining
Bethe root distributions which parametrize a (non-thermal) equilibrium state. A mapping between the string
densities and local correlators for the gapped regime of the $XXZ$ model has been conjectured
in Refs.~\cite{Mestyan14,Pozsgay_failure,Mestyan15}. An alternative route is to follow the Quantum Transfer Matrix
approach~\cite{Klumper93,Boos07,Boos08} which was pursued in Ref.~\cite{FCEC14}.

\subsection{Beyond quasilocality}

We have discussed at length the implication of pseudolocality of conserved quantities on several observable physical 
properties, such as ballistic (ideal) high-temperature transport and equilibration to non-thermal states. However, in some
other rudimentary integrable models, a normal, {\em diffusive} spin or particle transport has been observed by numerical simulations,
e.g. in the gapped Heisenberg model~\cite{PZ09,ZnidaricPRL11,Robin,ZnidaricJSTAT11,Fabian},
or half-filled Fermi-Hubbard model~\cite{PZPRB12,KKM14,Jin}.

Diffusive high-temperature transport in an integrable model can be considered as an indication of the absence of {\em relevant} 
pseudolocal charges, i.e. linearly extensive charges with non-vanishing overlap with a current operator.
In the opposite case, the Mazur bound is strictly positive, implying ballistic conductivity.
Even then, however, one may obtain other interesting bounds employing conserved operators with different volume-scaling properties.
For example, if there exists a conservation law $Q$ with \emph{quadratic volume scaling} of the Hilbert-Schmidt
norm $\|Q\|_{\rm HS}^2 = q N^2 + {\cal O}(N)$, then a rigorous derivation~\cite{ProsenPRE14}, in spirit very similar to
the proof of Mazur bound for quantum spin lattice systems~\cite{IP13} but with appropriately balanced limits $N\to\infty,t\to\infty$, 
yields a rigorous lower bound on the diffusion constant 
\begin{equation}
D_{\rm diff} \ge \frac{|(j,Q)|^2}{8 v q}.
\end{equation}
Here $j$ is a local current operator and $v$ is the Lieb-Robinson velocity~\cite{LR72,BRBook}. 
Simple examples of such bounds have been elaborated in Ref.~\cite{ProsenPRE14} for the $XXX$ and Fermi--Hubbard models,
where $Q$ in fact corresponds to a level-$1$ generator of Yangian symmetry~\cite{Bernard}.
Systematic exploration of quadratically extensive conserved charges in integrable systems and their applications to diffusive 
transport and quantum relaxation has not yet been undertaken.

\subsection{Conclusions}
\label{sec:conclusions}

This review is devoted to certain types of effective localities in the context of quantum integrable lattice models termed
pseudolocality and quasilocality. The notion of locality indisputably plays a monumental role in the foundations of statistical
mechanics, both on the classical and quantum level.
We have presented and exemplified the meaning of quasilocal conserved quantities by discussing various applications of non-ergodic 
phenomena in a paradigmatic interacting system, the anisotropic Heisenberg model.
Specifically, we have elaborated on the importance of quasilocal charges in the description of generalized (non-thermal) equilibria on 
one hand, and their vital role in understanding certain anomalous transport characteristics such as divergent high-temperature spin 
conductivity on the other hand.

A key observation is that statistical ensembles, given by reduced density matrices which emerge in the steady-state limit after a 
relaxation process starting from any `physical' initial state, are, due to effective dephasing, only capable of retaining a part of
information about the initial condition which is encoded in local and pseudolocal conservation laws. Identification of a complete set 
of such charges provides us with a complete description of local correlation functions in generalized equilibria.

This naturally brings us to an elusive question which has been posted at the beginning, namely a controversial issue of
the proper counting of degrees of freedom in an integrable lattice model. 
As we have explained, spectra of integrable lattice models in the thermodynamic limit organize in
an astounding way and permit to cast our description in terms of stable quasi-particles~\cite{YY69,Gaudin71,Takahashi71,ZZ79}.
This picture is in principle valid for any equilibrium state and even for elementary quasi-particle excitations on top of 
them~\cite{QF13}. Quasi-particles are labelled by a representation label (auxiliary spin in our example) and a continuous rapidity 
variable (corresponding to quasi-momentum). Having this in mind it should not be difficult to understand why higher-spin transfer 
operators, despite fulfilling a system of functional identities, are nonetheless \emph{linearly} independent variables.
Therefore, the naive proposal of matching the number of degrees of freedom with the number of local Hilbert spaces of the lattice 
system cannot be correct.

We have furthermore discussed an interesting (although somewhat atypical) situation when the above picture is incomplete and needs to 
be appropriately extended. This happens when the underlying symmetry algebra becomes enlarged which implies extra 
degeneracies in the spectrum. Perhaps the simplest example of that occurs in the gapless regime of the $XXZ$ model, governed by 
$\mathcal{U}_{q}(\mathfrak{sl}({2}))$ quantum symmetry at root of unity deformations where an enriched symmetry led to the discovery of 
an extra family of quasilocal charges pertaining to non-unitary representations of the quantum group in the auxiliary space.
In this review we exposed some of their implications on non-decaying currents and associated anomalous transport properties and 
presented a rigorous non-trivial bound on the singular contribution to the spin conductivity (Drude weight).

The last type of applications which we presented briefly were integrable spin chains subjected to Markovian dissipative boundaries.
The time evolution in such cases is governed by a non-unitary process described by Lindblad master equation and generally leads
to a unique steady state which is far from canonical thermal equilibrium. We owe to stress however that such `integrable instances' 
which emerge as a consequence of an effective evolution describing an open system can be profoundly different from the conventional 
non-equilibrium settings in the scope of isolated systems which evolve according to the unitary evolution law and consequently
the relevant class of symmetries which become important might be quite different. In addition, we notice that
dissipation processes are strictly only well-defined in a finite volume while studies of equilibration in isolated systems
typically deal with extended systems.

Aside of several novel theoretical insights which have been outlined in this review, it is worth mentioning
that our formulation can also prove advantageous from a practical computational standpoint.
An obvious example of that are explicit matrix product representations of quasilocal charges $X_{s}(\lambda)$ and $Z(\lambda)$ which 
do not only admit a unified abstract representation but also enable a direct and efficient computation using methods from the standard 
statistical mechanics toolbox. In essence, this lifts the Bethe ansatz concepts to operator level right away in the thermodynamic 
regime, circumventing a long-standing challenge of achieving this by pursuing the programme of algebraic Bethe ansatz,
see e.g.~\cite{AL03,KM10}.

In conclusion, apart from a few successful physical applications in the realm of quantum quenches and quantum transport,
much of the formal origin and group-theoretic interpretation is still missing at the moment.
A notable example is the question of completeness of the $Z$-charges and their reconciliation with the spectrum and the quasi-particle 
content. We hope that this review can provide a source of inspiration for the ongoing investigation of open directions.

\section*{Acknowledgements} 
TP and EI enjoyed useful discussions and/or fruitful collaboration on related problems with I. Affleck, M. Brockmann, J.-S. Caux,
B. Doyon, F. H. L. Essler, J. De Nardis, A. Kl\" umper, M. Mierzejewski, V. Popkov, P. Prelov\v sek, E. Quinn, J. Sirker,
and B. Wouters. The work has been supported by Programme P1-0044, and Grants J1-5439 and N1-0025 of Slovenian Research Agency (ARRS).

\section*{References}

\bibliographystyle{iopart-num}
\bibliography{qlrev_revised}

\providecommand{\newblock}{}
\begin{thebibliography}{100}
\expandafter\ifx\csname url\endcsname\relax
  \def\url#1{{\tt #1}}\fi
\expandafter\ifx\csname urlprefix\endcsname\relax\def\urlprefix{URL }\fi
\providecommand{\eprint}[2][]{\url{#2}}

\bibitem{LSM61}
Lieb E, Schultz T and Mattis D 1961 {\em Annals of Physics\/} {\bf 16}
  407–466

\bibitem{AreaLaws}
Eisert J, Cramer M and Plenio M~B 2010 {\em Rev. Mod. Phys.\/} {\bf 82}(1)
  277--306

\bibitem{Drinfeld89}
Drinfeld V~G 1989 {Quasi-Hopf algebras and Knizhnik-Zamolodchikov equations}
  {\em Problems of modern quantum field theory\/} (Springer) pp 1--13

\bibitem{FRT90}
Faddeev L, Reshetikhin N~Y and Takhtajan L 1990 {\em Yang-Baxter Equation In
  Integrable Systems. Series: Advanced Series in Mathematical Physics, ISBN:
  978-981-02-0120-3. WORLD SCIENTIFIC, Edited by Michio Jimbo, vol. 10, pp.
  299-309\/} {\bf 10} 299--309

\bibitem{KasselBook}
Kassel C 2012 {\em Quantum groups\/} vol 155 (Springer Science \& Business
  Media)

\bibitem{SierraBook}
G{\'o}mez C, Ruiz-Altaba M and Sierra G 2005 {\em Quantum groups in
  two-dimensional physics\/} (Cambridge University Press)

\bibitem{KorepinBook}
Korepin V~E, Bogoliubov N~M and Izergin A~G 1997 {\em Quantum inverse
  scattering method and correlation functions\/} (Cambridge university press)

\bibitem{BaxterBook}
Baxter R~J 2007 {\em Exactly solved models in statistical mechanics\/} (Courier
  Corporation)

\bibitem{Schutz01}
Sch{\"u}tz G~M 2001 {\em Phase transitions and critical phenomena\/} {\bf 19}
  1--251

\bibitem{FaddeevBook}
Faddeev L and Takhtajan L 2007 {\em {Hamiltonian methods in the theory of
  solitons}\/} (Springer Science \& Business Media)

\bibitem{BBTBook}
Babelon O, Bernard D and Talon M 2003 {\em Introduction to classical integrable
  systems\/} (Cambridge University Press)

\bibitem{Prosen_review}
Prosen T 2015 {\em Journal of Physics A: Mathematical and Theoretical\/} {\bf
  48} 373001

\bibitem{Bernard_review}
Bernard D and Doyon B 2016 {\em arXiv preprint arXiv:1603.07765\/}

\bibitem{Calabrese_review}
Calabrese P and Cardy J 2016 {\em arXiv preprint arXiv:1603.02889\/}

\bibitem{Caux_review}
Caux J~S 2016 {\em arXiv preprint arXiv:1603.04689\/}

\bibitem{Cazalilla_review}
Cazalilla M and Chung M~C 2016 {\em arXiv preprint arXiv:1603.04252\/}

\bibitem{DeLuca_review}
De~Luca A and Mussardo G 2016 {\em arXiv preprint arXiv:1603.08628\/}

\bibitem{Essler_review}
Essler F~H and Fagotti M 2016 {\em arXiv preprint arXiv:1603.06452\/}

\bibitem{Vasseur_review}
Vasseur R and Moore J~E 2016 {\em arXiv preprint arXiv:1603.06618\/}

\bibitem{Vidmar_review}
Vidmar L and Rigol M 2016 {\em arXiv preprint arXiv:1604.03990\/}

\bibitem{Schmiedmayer_review}
Langen T, Gasenzer T and Schmiedmayer J 2016 {\em arXiv preprint
  arXiv:1603.0938\/}

\bibitem{Bloch_review}
Bloch I, Dalibard J and Zwerger W 2008 {\em Reviews of Modern Physics\/} {\bf
  80} 885

\bibitem{Kinoshita04}
Kinoshita T, Wenger T and Weiss D~S 2004 {\em Science\/} {\bf 305} 1125--1128

\bibitem{Kinoshita06}
Kinoshita T, Wenger T and Weiss D~S 2006 {\em Nature\/} {\bf 440} 900--903

\bibitem{Hofferberth_Nature07}
Hofferberth S, Lesanovsky I, Fischer B, Schumm T and Schmiedmayer J 2007 {\em
  Nature\/} {\bf 449} 324--327

\bibitem{Gring_Science12}
Gring M, Kuhnert M, Langen T, Kitagawa T, Rauer B, Schreitl M, Mazets I, Smith
  D~A, Demler E and Schmiedmayer J 2012 {\em Science\/} {\bf 337} 1318--1322

\bibitem{Trotzky_Nature12}
Trotzky S, Chen Y~A, Flesch A, McCulloch I~P, Schollw{\"o}ck U, Eisert J and
  Bloch I 2012 {\em Nature Physics\/} {\bf 8} 325--330

\bibitem{Cheneau_Nature12}
Cheneau M, Barmettler P, Poletti D, Endres M, Schau{\ss} P, Fukuhara T, Gross
  C, Bloch I, Kollath C and Kuhr S 2012 {\em Nature\/} {\bf 481} 484--487

\bibitem{Langen_Science15}
Langen T, Erne S, Geiger R, Rauer B, Schweigler T, Kuhnert M, Rohringer W,
  Mazets I~E, Gasenzer T and Schmiedmayer J 2015 {\em Science\/} {\bf 348}
  207--211

\bibitem{Hild}
Hild S, Fukuhara T, Schau\ss{} P, Zeiher J, Knap M, Demler E, Bloch I and Gross
  C 2014 {\em Phys. Rev. Lett.\/} {\bf 113}(14) 147205

\bibitem{Ronzheimer}
Ronzheimer J~P, Schreiber M, Braun S, Hodgman S~S, Langer S, McCulloch I~P,
  Heidrich-Meisner F, Bloch I and Schneider U 2013 {\em Phys. Rev. Lett.\/}
  {\bf 110}(20) 205301

\bibitem{Schneider}
Schneider U, Hackerm{\"u}ller L, Ronzheimer J~P, Will S, Braun S, Best T, Bloch
  I, Demler E, Mandt S, Rasch D {\em et~al.\/} 2012 {\em Nature Physics\/} {\bf
  8} 213--218

\bibitem{Xia}
Xia L, Zundel L~A, Carrasquilla J, Reinhard A, Wilson J~M, Rigol M and Weiss
  D~S 2015 {\em Nature Physics\/} {\bf 11} 316--320

\bibitem{AABook}
Arnolʹd V~I and Avez A 1968 {\em Ergodic problems of classical mechanics\/}
  vol~9 (Benjamin)

\bibitem{Prosen00}
Prosen T 2000 {\em Progress of Theoretical Physics Supplement\/} {\bf 139}
  191--203

\bibitem{BCK15}
Brandino G, Caux J~S and Konik R 2015 {\em Physical Review X\/} {\bf 5} 041043

\bibitem{BabbittThomas79}
Babbitt D and Thomas L 1979 {\em Journal of Mathematical Analysis and
  Applications\/} {\bf 72} 305--328

\bibitem{Zotos97}
Zotos X, Naef F and Prelovsek P 1997 {\em Physical Review B\/} {\bf 55} 11029

\bibitem{Zotos99}
Zotos X 1999 {\em Physical Review Letters\/} {\bf 82} 1764

\bibitem{WoutersPRL}
Wouters B, De~Nardis J, Brockmann M, Fioretto D, Rigol M and Caux J~S 2014 {\em
  Physical Review Letters\/} {\bf 113} 117202

\bibitem{Pozsgay_GGE}
Pozsgay B, Mesty{\'a}n M, Werner M, Kormos M, Zar{\'a}nd G and Tak{\'a}cs G
  2014 {\em Physical Review Letters\/} {\bf 113} 117203

\bibitem{ProsenPRL106}
Prosen T 2011 {\em Physical Review Letters\/} {\bf 106} 217206

\bibitem{FHM03}
Heidrich-Meisner F, Honecker A, Cabra D and Brenig W 2003 {\em Physical Review
  B\/} {\bf 68} 134436

\bibitem{HPZ11}
Herbrych J, Prelov{\v{s}}ek P and Zotos X 2011 {\em Physical Review B\/} {\bf
  84} 155125

\bibitem{Karrasch12}
Karrasch C, Bardarson J and Moore J 2012 {\em Physical Review Letters\/} {\bf
  108} 227206

\bibitem{Karrasch13}
Karrasch C, Hauschild J, Langer S and Heidrich-Meisner F 2013 {\em Physical
  Review B\/} {\bf 87} 245128

\bibitem{SPA09}
Sirker J, Pereira R and Affleck I 2009 {\em Physical Review Letters\/} {\bf
  103} 216602

\bibitem{SPA11}
Sirker J, Pereira R and Affleck I 2011 {\em Physical Review B\/} {\bf 83}
  035115

\bibitem{PI13}
Prosen T and Ilievski E 2013 {\em Physical Review Letters\/} {\bf 111} 057203

\bibitem{ProsenNPB14}
Prosen T 2014 {\em Nuclear Physics B\/} {\bf 886} 1177--1198

\bibitem{Pereira14}
Pereira R, Pasquier V, Sirker J and Affleck I 2014 {\em Journal of Statistical
  Mechanics: Theory and Experiment\/} {\bf 2014} P09037

\bibitem{Zadnik16}
Zadnik L, Medenjak M and Prosen T 2016 {\em Nuclear Physics B\/} {\bf 902}
  339--353

\bibitem{PV16}
Piroli L and Vernier E 2016 {\em arXiv preprint arXiv:1601.07289\/}

\bibitem{GA14}
Goldstein G and Andrei N 2014 {\em Physical Review A\/} {\bf 90} 043625

\bibitem{IMP15}
Ilievski E, Medenjak M and Prosen T 2015 {\em Physical Review Letters\/} {\bf
  115} 120601

\bibitem{Ilievski_GGE}
Ilievski E, De~Nardis J, Wouters B, Caux J~S, Essler F~H and Prosen T 2015 {\em
  Physical Review Letters\/} {\bf 115} 157201

\bibitem{Bethe31}
Bethe H 1931 {\em Zeitschrift f{\"u}r Physik\/} {\bf 71} 205--226

\bibitem{Faddeev_arxiv}
Faddeev L 1996 {\em arXiv preprint hep-th/9605187\/}

\bibitem{Sklyanin_arxiv}
Sklyanin E 1992 {\em arXiv preprint hep-th/9211111\/}

\bibitem{GaudinBook}
Gaudin M 1983 {\em {La fonction d'onde de Bethe}\/} vol~1 (Elsevier Masson)

\bibitem{BRBook}
Bratteli O and Robinson D~W 2012 {\em {Operator Algebras and Quantum
  Statistical Mechanics: Volume 1: C*-and W*-Algebras. Symmetry Groups.
  Decomposition of States}\/} (Springer Science \& Business Media)

\bibitem{Prosen98}
Prosen T 1998 {\em Physical Review Letters\/} {\bf 80} 1808

\bibitem{ProsenJPA98}
Prosen T 1998 {\em Journal of Physics A: Mathematical and General\/} {\bf 31}
  L645

\bibitem{ProsenPRE99}
Prosen T 1999 {\em Physical Review E\/} {\bf 60} 3949

\bibitem{Doyon15}
Doyon B 2015 {\em arXiv preprint arXiv:1512.03713\/}

\bibitem{McGuire64}
McGuire J~B 1964 {\em Journal of Mathematical Physics\/} {\bf 5} 622--636

\bibitem{Yang67}
Yang C~N 1967 {\em Physical Review Letters\/} {\bf 19} 1312

\bibitem{ZZ79}
Zamolodchikov A~B and Zamolodchikov A~B 1979 {\em Annals of physics\/} {\bf
  120} 253--291

\bibitem{FK95}
Faddeev L and Korchemsky G 1995 {\em Physics Letters B\/} {\bf 342} 311--322

\bibitem{Derkachov01}
Derkachov S, Karakhanyan D and Kirschner R 2001 {\em Nuclear Physics B\/} {\bf
  618} 589--616

\bibitem{Karakh02}
Karakhanyan D, Kirschner R and Mirumyan M 2002 {\em Nuclear Physics B\/} {\bf
  636} 529--548

\bibitem{BLZII}
Bazhanov V~V, Lukyanov S~L and Zamolodchikov A~B 1997 {\em Communications in
  Mathematical Physics\/} {\bf 190} 247--278

\bibitem{KLWZ97}
Krichever I, Lipan O, Wiegmann P and Zabrodin A 1997 {\em Communications in
  Mathematical Physics\/} {\bf 188} 267--304

\bibitem{Zabrodin97}
Zabrodin A 1997 {\em International Journal of Modern Physics B\/} {\bf 11}
  3125--3158

\bibitem{KSZ08}
Kazakov V, Sorin A and Zabrodin A 2008 {\em Nuclear physics B\/} {\bf 790}
  345--413

\bibitem{Gromov09}
Gromov N, Kazakov V and Vieira P 2009 {\em Journal of High Energy Physics\/}
  {\bf 2009} 060

\bibitem{KL10}
Kazakov V and Leurent S 2010 {\em arXiv preprint arXiv:1007.1770\/}

\bibitem{KP92}
Kl{\"u}mper A and Pearce P~A 1992 {\em Physica A: Statistical Mechanics and its
  Applications\/} {\bf 183} 304--350

\bibitem{KNS94}
Kuniba A, Nakanishi T and Suzuki J 1994 {\em International Journal of Modern
  Physics A\/} {\bf 9} 5215--5266

\bibitem{BR90}
Bazhanov V and Reshetikhin N 1990 {\em Journal of Physics A: Mathematical and
  General\/} {\bf 23} 1477

\bibitem{Baxter73}
Baxter R 1973 {\em Annals of Physics\/} {\bf 76} 1--24

\bibitem{shortcut}
Bazhanov V~V, {\L}ukowski T, Meneghelli C and Staudacher M 2010 {\em Journal of
  Statistical Mechanics: Theory and Experiment\/} {\bf 2010} P11002

\bibitem{BFLM11}
Bazhanov V~V, Frassek R, {\L}ukowski T, Meneghelli C and Staudacher M 2011 {\em
  Nuclear Physics B\/} {\bf 850} 148--174

\bibitem{DM06}
Derkachov S~{\'E} and Manashov A 2006 {\em Journal of Physics A: Mathematical
  and General\/} {\bf 39} 4147

\bibitem{Pearce87}
Pearce P~A 1987 {\em Physical Review Letters\/} {\bf 58} 1502

\bibitem{Klumper89}
Kl{\"u}mper A, Schadschneider A and Zittartz J 1989 {\em Zeitschrift f{\"u}r
  Physik B Condensed Matter\/} {\bf 76} 247--258

\bibitem{StringCharge}
Ilievski E, Quinn E, De~Nardis J and Brockmann M 2015 {\em arXiv preprint
  arXiv:1512.04454\/}

\bibitem{Korff03}
Korff C 2003 {\em Journal of Physics A: Mathematical and General\/} {\bf 36}
  5229

\bibitem{IP13}
Ilievski E and Prosen T 2013 {\em Communications in Mathematical Physics\/}
  {\bf 318} 809--830

\bibitem{Mazur69}
Mazur P 1969 {\em Physica\/} {\bf 43} 533--545

\bibitem{Suzuki71}
Suzuki M 1971 {\em Physica\/} {\bf 51} 277--291

\bibitem{Araki69}
Araki H 1969 {\em Communications in Mathematical Physics\/} {\bf 14} 120--157

\bibitem{Benz05}
Benz J, Fukui T, Kl{\"u}mper A and Scheeren C 2005 {\em Journal of the Physical
  Society of Japan\/} {\bf 74} 181--190

\bibitem{KMprivate}
Karrasch C and Moore J~E 2014 {\em Private communication\/}

\bibitem{RobinTypicality}
Steinigeweg R, Gemmer J and Brenig W 2014 {\em Phys. Rev. Lett.\/} {\bf
  112}(12) 120601

\bibitem{Marcin}
Mierzejewski M, Prelov{\v{s}}ek P and Prosen T 2014 {\em Physical Review
  Letters\/} {\bf 113} 020602

\bibitem{MPP15}
Mierzejewski M, Prelov{\v{s}}ek P and Prosen T 2015 {\em Physical Review
  Letters\/} {\bf 114} 140601

\bibitem{CalabreseCardy}
Calabrese P and Cardy J 2006 {\em Phys. Rev. Lett.\/} {\bf 96} 136801

\bibitem{MWNM07}
Manmana S~R, Wessel S, Noack R~M and Muramatsu A 2007 {\em Physical Review
  Letters\/} {\bf 98} 210405

\bibitem{KLA07}
Kollath C, L{\"a}uchli A~M and Altman E 2007 {\em Physical Review Letters\/}
  {\bf 98} 180601

\bibitem{IC09}
Iucci A and Cazalilla M 2009 {\em Physical Review A\/} {\bf 80} 063619

\bibitem{RSMS09}
Rossini D, Silva A, Mussardo G and Santoro G~E 2009 {\em Physical Review
  Letters\/} {\bf 102} 127204

\bibitem{CalabresePRL106}
Calabrese P, Essler F~H~L and Fagotti M 2011 {\em Phys. Rev. Lett.\/} {\bf 106}
  227203

\bibitem{Barmettler09}
Barmettler P, Punk M, Gritsev V, Demler E and Altman E 2009 {\em Physical
  Review Letters\/} {\bf 102} 130603

\bibitem{Calabrese12I}
Calabrese P, Essler F~H~L and Fagotti M 2012 {\em J. Stat. Mech.: Th. Exp.\/}
  {\bf 2012} P07016

\bibitem{Calabrese12II}
Calabrese P, Essler F~H~L and Fagotti M 2012 {\em J. Stat. Mech.: Th. Exp.\/}
  {\bf 2012} P07022

\bibitem{CK12}
Caux J~S and Konik R~M 2012 {\em Physical Review Letters\/} {\bf 109} 175301

\bibitem{Deutsch91}
Deutsch J 1991 {\em Physical Review A\/} {\bf 43} 2046

\bibitem{Polkovnikov_colloquium}
Polkovnikov A, Sengupta K, Silva A and Vengalattore M 2011 {\em Reviews of
  Modern Physics\/} {\bf 83} 863

\bibitem{Srednicki94}
Srednicki M 1994 {\em Physical Review E\/} {\bf 50} 888

\bibitem{RigolNature}
Rigol M, Dunjko V and Olshanii M 2008 {\em Nature\/} {\bf 452} 854--858

\bibitem{Rigol_PRA06}
Rigol M, Muramatsu A and Olshanii M 2006 {\em Physical Review A\/} {\bf 74}
  053616

\bibitem{Rigol_PRL07}
Rigol M, Dunjko V, Yurovsky V and Olshanii M 2007 {\em Physical Review
  Letters\/} {\bf 98} 050405

\bibitem{Rigol11}
Cassidy A~C, Clark C~W and Rigol M 2011 {\em Physical Review Letters\/} {\bf
  106} 140405

\bibitem{KWE11}
Kollar M, Wolf F~A and Eckstein M 2011 {\em Physical Review B\/} {\bf 84}
  054304

\bibitem{Fioretto12}
Fioretto D and Mussardo G 2010 {\em New J. Phys.\/} {\bf 12} 055015

\bibitem{Fagotti13}
Fagotti M and Essler F~H~L 2013 {\em Phys. Rev. B\/} {\bf 87}(24) 245107

\bibitem{Fagotti14}
Fagotti M 2014 {\em Journal of Statistical Mechanics: Theory and Experiment\/}
  {\bf 2014} P03016

\bibitem{EKMR14}
Essler F, Kehrein S, Manmana S and Robinson N 2014 {\em Physical Review B\/}
  {\bf 89} 165104

\bibitem{BF15}
Bertini B and Fagotti M 2015 {\em Journal of Statistical Mechanics: Theory and
  Experiment\/} {\bf 2015} P07012

\bibitem{Bertini15}
Bertini B, Essler F~H, Groha S and Robinson N~J 2015 {\em Physical Review
  Letters\/} {\bf 115} 180601

\bibitem{FE13}
Fagotti M and Essler F~H 2013 {\em Journal of Statistical Mechanics: Theory and
  Experiment\/} {\bf 2013} P07012

\bibitem{Pozsgay2013}
Pozsgay B 2013 {\em Journal of Statistical Mechanics: Theory and Experiment\/}
  {\bf 2013} P07003

\bibitem{FCEC14}
Fagotti M, Collura M, Essler F~H and Calabrese P 2014 {\em Physical Review B\/}
  {\bf 89} 125101

\bibitem{DeNardisPRA14}
De~Nardis J, Wouters B, Brockmann M and Caux J~S 2014 {\em Phys. Rev. A\/} {\bf
  89}(3) 033601

\bibitem{CE13}
Caux J~S and Essler F~H 2013 {\em Physical Review Letters\/} {\bf 110} 257203

\bibitem{YY69}
Yang C~N and Yang C 1969 {\em Journal of Mathematical Physics\/} {\bf 10}
  1115--1122

\bibitem{Takahashi71}
Takahashi M 1971 {\em Progress of Theoretical Physics\/} {\bf 46} 401--415

\bibitem{TS72}
Takahashi M and Suzuki M 1972 {\em Progress of theoretical physics\/} {\bf 48}
  2187--2209

\bibitem{TakahashiBook}
Takahashi M 2005 {\em Thermodynamics of one-dimensional solvable models\/}
  (Cambridge University Press)

\bibitem{Pozsgay_overlaps}
Pozsgay B 2014 {\em Journal of Statistical Mechanics: Theory and Experiment\/}
  {\bf 2014} P06011

\bibitem{Brockmann14}
Brockmann M, De~Nardis J, Wouters B and Caux J~S 2014 {\em Journal of Physics
  A: Mathematical and Theoretical\/} {\bf 47} 145003

\bibitem{Mestyan14}
Mesty{\'a}n M and Pozsgay B 2014 {\em Journal of Statistical Mechanics: Theory
  and Experiment\/} {\bf 2014} P09020

\bibitem{Gaudin71}
Gaudin M 1971 {\em Physical Review Letters\/} {\bf 26} 1301

\bibitem{Mestyan15}
Mesty{\'a}n M, Pozsgay B, Tak{\'a}cs G and Werner M 2015 {\em Journal of
  Statistical Mechanics: Theory and Experiment\/} {\bf 2015} P04001

\bibitem{WeinbergBook}
Weinberg S 1996 {\em The quantum theory of fields\/} vol~2 (Cambridge
  university press)

\bibitem{SC14}
Sotiriadis S and Calabrese P 2014 {\em Journal of Statistical Mechanics: Theory
  and Experiment\/} {\bf 2014} P07024

\bibitem{Klumper92}
Kl{\"u}mper A 1992 {\em Annalen der Physik\/} {\bf 504} 540--553

\bibitem{Klumper93}
Kl{\"u}mper A 1993 {\em Zeitschrift f{\"u}r Physik B Condensed Matter\/} {\bf
  91} 507--519

\bibitem{Klumper02}
Kl{\"u}mper A and Sakai K 2002 {\em Journal of Physics A: Mathematical and
  General\/} {\bf 35} 2173

\bibitem{NMO16}
Nakagawa Y~O, Misguich G and Oshikawa M 2016 {\em arXiv preprint
  arXiv:1601.06167\/}

\bibitem{BreuerBook}
Breuer H~P and Petruccione F 2002 {\em The theory of open quantum systems\/}
  (Oxford University Press on Demand)

\bibitem{HuelgaBook}
Rivas {\'A} and Huelga S~F 2011 {\em {Open quantum systems: An introduction}\/}
  (Springer Science \& Business Media)

\bibitem{ZnidaricPRL11}
{\v{Z}}nidari{\v{c}} M 2011 {\em Physical Review Letters\/} {\bf 106} 220601

\bibitem{ZnidaricJSTAT11}
{\v{Z}}nidari{\v{c}} M 2011 {\em Journal of Statistical Mechanics: Theory and
  Experiment\/} {\bf 2011} P12008

\bibitem{ProsenPRL107}
Prosen T 2011 {\em Physical Review Letters\/} {\bf 107} 137201

\bibitem{KPS13}
Karevski D, Popkov V and Sch{\"u}tz G 2013 {\em Physical Review Letters\/} {\bf
  110} 047201

\bibitem{IZ14}
Ilievski E and {\v{Z}}unkovi{\v{c}} B 2014 {\em Journal of Statistical
  Mechanics: Theory and Experiment\/} {\bf 2014} P01001

\bibitem{IP14}
Ilievski E and Prosen T 2014 {\em Nuclear Physics B\/} {\bf 882} 485--500

\bibitem{PZ09}
Prosen T and {\v{Z}}nidari{\v{c}} M 2009 {\em Journal of Statistical Mechanics:
  Theory and Experiment\/} {\bf 2009} P02035

\bibitem{3Q}
Prosen T 2008 {\em New Journal of Physics\/} {\bf 10} 043026

\bibitem{Zunkovic10}
Prosen T and {\v{Z}}unkovi{\v{c}} B 2010 {\em New Journal of Physics\/} {\bf
  12} 025016

\bibitem{ZnidaricJSTAT10}
{\v{Z}}nidari{\v{c}} M 2010 {\em Journal of Statistical Mechanics: Theory and
  Experiment\/} {\bf 2010} L05002

\bibitem{ZnidaricJPA10}
{\v{Z}}nidari{\v{c}} M 2010 {\em Journal of Physics A: Mathematical and
  Theoretical\/} {\bf 43} 415004

\bibitem{Derrida93}
Derrida B, Evans M, Hakim V and Pasquier V 1993 {\em Journal of Physics A:
  Mathematical and General\/} {\bf 26} 1493

\bibitem{Derrida07}
Derrida B 2007 {\em Journal of Statistical Mechanics: Theory and Experiment\/}
  {\bf 2007} P07023

\bibitem{Ilievski_thesis}
Ilievski E 2014 {\em arXiv preprint arXiv:1410.1446\/}

\bibitem{Hubbard63}
Hubbard J 1963 Electron correlations in narrow energy bands {\em Proceedings of
  the Royal Society of London A: Mathematical, Physical and Engineering
  Sciences\/} vol 276 (The Royal Society) pp 238--257

\bibitem{Shastry88}
Shastry B~S 1988 {\em Journal of statistical physics\/} {\bf 50} 57--79

\bibitem{HubbardBook}
Essler F~H, Frahm H, G{\"o}hmann F, Kl{\"u}mper A and Korepin V~E 2005 {\em
  {The one-dimensional Hubbard model}\/} (Cambridge University Press)

\bibitem{PP15}
Popkov V and Prosen T 2015 {\em Physical Review Letters\/} {\bf 114} 127201

\bibitem{PPRL14}
Prosen T 2014 {\em Phys. Rev. Lett.\/} {\bf 112}(3) 030603

\bibitem{Zamolodchikov91}
Zamolodchikov A~B 1991 {\em Physics Letters B\/} {\bf 253} 391--394

\bibitem{KN92}
Kuniba A and Nakanishi T 1992 {\em Modern Physics Letters A\/} {\bf 7}
  3487--3494

\bibitem{LL63}
Lieb E~H and Liniger W 1963 {\em Physical Review\/} {\bf 130} 1605

\bibitem{Davies90}
Davies B 1990 {\em Physica A: Statistical Mechanics and its Applications\/}
  {\bf 167} 433--456

\bibitem{DK11}
Davies B and Korepin V~E 2011 {\em arXiv preprint arXiv:1109.6604\/}

\bibitem{KormosPRB13}
Kormos M, Shashi A, Chou Y~Z, Caux J~S and Imambekov A 2013 {\em Phys. Rev.
  B\/} {\bf 88}(20) 205131

\bibitem{EMP15}
Essler F~H~L, Mussardo G and Panfil M 2015 {\em Phys. Rev. A\/} {\bf 91}(5)
  051602

\bibitem{KormosPRB14}
Kormos M, Collura M and Calabrese P 2014 {\em Phys. Rev. A\/} {\bf 89}(1)
  013609

\bibitem{Maldacena}
Maldacena J 1999 {\em International journal of theoretical physics\/} {\bf 38}
  1113--1133

\bibitem{GPK98}
Gubser S~S, Klebanov I~R and Polyakov A~M 1998 {\em Physics Letters B\/} {\bf
  428} 105--114

\bibitem{AdSCFT_review}
Beisert N, Ahn C, Alday L~F, Bajnok Z, Drummond J~M, Freyhult L, Gromov N,
  Janik R~A, Kazakov V, Klose T {\em et~al.\/} 2012 {\em Letters in
  Mathematical Physics\/} {\bf 99} 3--32

\bibitem{FQ12}
Frolov S and Quinn E 2012 {\em Journal of Physics A: Mathematical and
  Theoretical\/} {\bf 45} 095004

\bibitem{Pozsgay_failure}
Pozsgay B 2014 {\em Journal of Statistical Mechanics: Theory and Experiment\/}
  {\bf 2014} P09026

\bibitem{Boos07}
Boos H~E, G{\"o}hmann F, Kl{\"u}mper A and Suzuki J 2007 {\em Journal of
  Physics A: Mathematical and Theoretical\/} {\bf 40} 10699

\bibitem{Boos08}
Boos H~E, Damerau J, G{\"o}hmann F, Kl{\"u}mper A, Suzuki J and Wei{\ss}e A
  2008 {\em Journal of Statistical Mechanics: Theory and Experiment\/} {\bf
  2008} P08010

\bibitem{Robin}
Steinigeweg R 2011 {\em Physical Review E\/} {\bf 84} 11136

\bibitem{Fabian}
Karrasch C, Moore J~E and Heidrich-Meisner F 2014 {\em Phys. Rev. B\/} {\bf
  89}(7) 075139

\bibitem{PZPRB12}
Prosen T and {\v{Z}}nidari{\v{c}} M 2012 {\em Physical Review B\/} {\bf 86}
  125118

\bibitem{KKM14}
Karrasch C, Kennes D~M and Moore J~E 2014 {\em Phys. Rev. B\/} {\bf 90}(15)
  155104

\bibitem{Jin}
Jin F, Steinigeweg R, Heidrich-Meisner F, Michielsen K and De~Raedt H 2015 {\em
  Physical Review B\/} {\bf 92} 205103

\bibitem{ProsenPRE14}
Prosen T 2014 {\em Physical Review E\/} {\bf 89}(1) 012142

\bibitem{LR72}
Lieb E~H and Robinson D~W 1972 {\em Communications in Mathematical Physics\/}
  {\bf 28} 251--257

\bibitem{Bernard}
Bernard D 1993 {An introduction to Yangian symmetries} {\em Integrable quantum
  field theories\/} (Springer) pp 39--52

\bibitem{QF13}
Quinn E and Frolov S 2013 {\em Journal of Physics A: Mathematical and
  Theoretical\/} {\bf 46} 205001

\bibitem{AL03}
Alcaraz F~C and Lazo M~J 2003 {\em Journal of Physics A: Mathematical and
  General\/} {\bf 37} L1

\bibitem{KM10}
Katsura H and Maruyama I 2010 {\em Journal of Physics A: Mathematical and
  Theoretical\/} {\bf 43} 175003

\end{thebibliography}

\end{document}